\documentclass{article}

\usepackage{hyperref}       
\usepackage{tikz}

\usepackage{subcaption,xcolor,lipsum}

\usepackage{url}            
\usepackage{booktabs}       
\usepackage{amsfonts}       
\usepackage{nicefrac}       
\usepackage{microtype}      
\hypersetup{colorlinks,linkcolor={blue},citecolor={blue},urlcolor={blue}}
\usepackage{natbib}
\usepackage{amsfonts}
\usepackage{graphicx,amssymb,mathrsfs,amsmath,color,fancyhdr}
\usepackage{amsthm}
\usepackage{amsmath}
\usepackage{cases}
\usepackage{bbding}
\usepackage{pifont}
\usepackage{mathbbol}
\usepackage{geometry}
\usepackage{cite}
\usepackage{algorithm}
\usepackage{algpseudocode}

\usepackage{latexsym,amsopn,amstext,amsthm,amsxtra,bm,calc,ifpdf}
\usepackage{enumerate}
\usepackage{listings}
\usepackage{enumitem}
\usepackage{subcaption}

\usepackage{multirow}
\usepackage{threeparttable} 

\usepackage{indentfirst}
\usepackage[toc,page]{appendix}
\usepackage{authblk}
\newcommand{\indep}{\perp \!\!\! \perp}

\usepackage{caption}
\usepackage{subcaption}

\hypersetup{
	colorlinks=true,
	linkcolor=black
}
\newtheorem{thm}{Theorem}

\newtheorem{condition}{Condition}

\newtheorem{assumption}{Assumption}
\newtheorem{lemma}{Lemma}

\newtheorem{corollary}{Corollary}
\newtheorem{proposition}{Proposition}
\newtheorem{pf}{Proof}
\usepackage{caption}

\def \E {\mathbb{E}}

\title{Semiparametric Efficient Inference  for the \\ Probability of Necessary and Sufficient Causation
}

\author[1]{Zhaoqing Tian} 
\author[1]{Peng Wu}
\affil[1]{School of Mathematics and Statistics, Beijing Technology and Business University}

\begin{document}

\maketitle

\begin{abstract}  
Causal attribution, which aims to explain why events or behaviors occur, is crucial in causal inference and enhances our understanding of cause-and-effect relationships in scientific research. The probabilities of necessary causation (PN) and sufficient causation (PS) are two of the most common quantities for attribution in causal inference. While many works have explored the identification or bounds of PN and PS, efficient estimation remains unaddressed. To fill this gap, this paper focuses on obtaining semiparametric efficient estimators of PN and PS under two sets of identifiability assumptions: strong ignorability and monotonicity, and strong ignorability and conditional independence. We derive efficient influence functions and semiparametric efficiency bounds for PN and PS under the two sets of identifiability assumptions, respectively.  Based on this, we propose efficient estimators for PN and PS, and show their large sample properties. Extensive simulations validate the superiority of our estimators compared to competing methods. We apply our methods to a real-world dataset to assess various risk factors affecting stroke.  
  
\end{abstract}


\section{Introduction}

 Causal attribution, which aims to explain why events or behaviors occur, is crucial in causal inference and enhances our understanding of cause-and-effect relationships in scientific research~\citep{antonakis2011counterfactuals,peters2017elements}.  For example, in medical research, causal attribution techniques can be used to ascertain whether a specific drug or treatment is the cause of a disease episode. This determination enables the development of more targeted and effective treatment strategies
 ~\citep{ankeny2014overlooked}. In economics, attribution techniques can be employed by policymakers to assess whether any observed improvements in macroeconomic conditions are attributable to the recent implementation of a tax reduction policy~\citep{gurevich2012role}. In public policy, attribution techniques help analyze the causes behind social phenomena, such as determining whether the implementation of load-shedding policies is responsible for a decrease in average student performance~\citep{antonakis2011counterfactuals}.

For attribution analysis, the probability of necessary causation (PN) and the probability of sufficient causation (PS), introduced by \citet{pearl2022probabilities}, are the two most common quantities in causal inference. PN measures the necessity of a cause for the occurrence of an event, whereas PS measures the sufficiency of a cause for an event. Unlike causal effect estimation, which studies how a particular treatment (or cause)  contributes to a specific outcome by focusing on the impact of that treatment on the outcome, attribution involves explaining and identifying the causes of events that have already occurred. As \citet{Dawid2022} and \citet{lu2023evaluating} put it, causal effect estimation is the ``effects of causes'', while attribution is the ``causes of effects''. Additionally, causal effect estimation is prospective, estimating (conditional) the mean of potential outcomes using observed covariates (pre-treatment variables). In contrast, causal attribution is retrospective, examining whether an event was caused by certain factors given the observed covariates and outcomes (post-treatment variables). 
Previous studies have discussed the identification of PN and PS under strongly ignorability assumption and monotonicity assumption~\citep{pearl2022probabilities,tian2000probabilities,pearl2009causal,cai2005variance} without covariates. 
The monotonicity assumption suggests that the treatment (or intervention or cause) has a consistent directional effect, either benefiting all individuals or having no adverse effect. However, this assumption may be too stringent in practice due to the heterogeneity of individuals.   
Thus, several methods focus on determining the bounds of PN and PS with only the strongly ignorability assumption, such as those proposed by \citet{pearl2022probabilities} and \citet{tian2000probabilities}. There is also literature on the application of PN and PS in fields such as medicine and engineering \citep{pearl2019sufficient}. Some recent studies have further narrowed the bounds of PN and PS by utilizing covariates or mediator variables under the strongly ignorability assumption \citep{dawid2017probability,  mueller2021causes}. Additionally, attention has been drawn to extending the concepts of PN and PS to cases involving multiple causes or multiple outcomes \citep{li2022learning, li2022probabilities, zhao2023conditional,  li2024probabilities, li2024retrospective}. 

Although many works have explored the identification or bounds of PN and PS, to the best of our knowledge, no literature has addressed their efficient estimation. To fill this gap, this paper focuses on obtaining the semiparametric efficient estimation of PN and PS under two sets of identifiability assumptions: strong ignorability~\citep{rubin1976inference} and monotonicity~\citep{manski1997monotone}, as well as strong ignorability and conditional independence~\citep{shen2013treatment}. The conditional independence assumption, which posits that potential outcomes are independent of each other given the covariates, is rarely discussed in the identification and estimation of PN and PS.

The main contributions of this article are summarized in three aspects. (1) We derive the efficient influence functions of PN and PS under two sets of identifiability assumptions and present the corresponding semiparametric efficiency bounds. 
In addition, we compare the semiparametric efficiency bounds under the two sets of identifiability assumptions and find that the semiparametric efficiency bound for PN under strong ignorability and conditional independence assumptions is smaller
than that under strong ignorability and monotonicity assumptions. We further investigate the impact of propensity scores on the estimation of PN and PS, by comparing the semiparametric efficiency bounds under the same identifiability assumptions with and without known propensity scores, and show that the efficiency bounds are lower when the propensity scores are known. 
(2) Based on the derived efficient influence functions, we propose new estimators for PN and PS under two sets of identifiability assumptions, respectively. We show that the proposed estimators are consistent, asymptotically normal, and locally efficient, meaning that the variance of the proposed estimators attain the corresponding semiparametric efficiency bound when all working models are correctly specified.  The proposed estimators under strong ignorability and monotonicity assumptions enjoy the property of double robustness; they are consistent if either the propensity score model or the outcome regression model is correctly specified. In contrast, the proposed estimators under strong ignorability and conditional independence assumptions are only single robust; they are consistent if the outcome regression model is correctly specified, regardless of whether the propensity score model is correctly specified.  (3) We conduct extensive simulation studies to empirically validate the superiority of the proposed estimators compared to the competing ones.  Moreover, we apply the proposed methods to assess different risk factors affecting stroke using the INTERSTROKE dataset. 
The remainder of this paper is organized as follows. In Section \ref{sec2}, we introduce the basic notation and identification assumptions. In Section \ref{sec3}, we present the efficient influence functions and the corresponding estimators for PN under two sets of identification assumptions. In Section \ref{sec4}, we establish the large sample properties of the proposed estimators. Section \ref{sec5} presents an extensive simulation study on PN to evaluate the finite sample performance of the proposed estimators. In Section \ref{sec6}, we demonstrate the proposed methods with a real-world dataset. In Section \ref{sec7}, we extend the previous analysis for PN to PS. A detailed discussion and conclusions are given in Section \ref{sec8}. All proofs can be found in the Supplementary Material.

\section{Setup} \label{sec2}


 



\subsection{Notation} 
We introduce the notation and definitions used in this paper.  Let $X$ represent the $p$-dimensional covariates, $A$ be a binary cause variable with $A = 1$ indicating the presence of the cause and $A = 0$ otherwise, and $Y$ denote the binary outcome, where $Y = 1$ indicates the occurrence of an event (e.g., disease)  and $Y = 0$ otherwise.  The observed data ${(X_i, A_i, Y_i): i = 1, \dots, n }$  consists of an independent and identically distributed sample of $n$ units from a superpopulation $\mathbb{P}$.     
To measure how likely $A$ is a cause of the outcome $Y$, 
we use the probability of necessary causation (PN) and the probability of sufficient causation (PS)~\citep{pearl2022probabilities}, which are the two most commonly used quantities. 
We define them using the potential outcome framework \citep{rubin1974estimating, Neyman1990}. Specifically,  
 let $(Y^0, Y^1)$ denote the potential outcome of $Y$ if  $A$ had been set to $a$ for $a = 0, 1$, respectively.   
 By the consistency assumption, the observed outcome $Y$ can be expressed as $Y = AY^1 + (1 - A)Y^0$. Then PN and PS are formally defined as
 \begin{align*} 
    \beta  = \mathbb{P}(Y^0 = 0| A = 1, Y = 1) \text{ and }  \gamma  = \mathbb{P}(Y^1 = 1| A = 0, Y = 0),
 \end{align*}
which measure how necessary $A$ is a cause of the outcome $Y$ and how sufficient $A$ is a cause of the outcome $Y$, respectively~\citep{pearl2009causality}.

Given that $\beta$ and $\gamma$ are structurally symmetric, 
for ease of presentation, we focus on the estimation of $\beta$ and defer the associated results for the estimation of $\gamma$ to Section \ref{sec7}.

\subsection{Identifiability Assumptions} 
For the identification of $\beta$ and $\gamma$, we first introduce the common strongly ignorability assumption. 

\begin{assumption}[Strongly Ignorability] (a) Unconfoundedness: $(Y^0, Y^1) \indep A \  | X$; (b) Overlap: $0 < e(x) := \mathbb{P}(A = 1| X = x) < 1$. \label{assump 1}
\end{assumption}
Assumption \ref{assump 1}(a) states that the cause $A$ is independent of the potential outcomes $(Y^0, Y^1)$ given the covariates $X$. This implies that $X$ includes all confounders affecting both $A$ and $Y$. Assumption \ref{assump 1}(b) states that units with any given values of the covariates have a positive probability of the presence of the cause. 
Assumption \ref{assump 1} is a standard assumption for identifying causal effects~\citep{rosenbaum1983central,Imbens-Rubin2015, Hernan-Robins2020}. Nevertheless, it is insufficient to identify $\beta$ and $\gamma$ \citep{pearl2009causality,pearl2019sufficient,peters2017elements}, because the definitions of PN and PS involve the joint distribution of potential outcomes $(Y^{0}, Y^{1})$.   
We then further impose Assumption \ref{assump 2} or \ref{assump 3} below.


\begin{assumption}[Monotonicity] \label{assump 2}
     $Y^0 \leq Y^1$. 
\end{assumption}

  Assumption  \ref{assump 2} indicates that 
  the presence of cause (i.e., $A=1$) 
  does not result in a decrease in the corresponding outcome \citep{angrist1996identification, pearl2009causality}. For example, consider a study on whether smoking (cause) leads to lung cancer (outcome).
  Let $A = 1$ represent smoking, $A = 0$ represent not smoking, $Y = 1$ represent having lung cancer, and $Y = 0$ represent not having lung cancer. If $Y^0$ and $Y^1$ denote the individual's lung cancer status when they are a non-smoker versus a smoker, respectively, then monotonicity implies that smoking will not reduce an individual's risk of developing lung cancer. 
  
\begin{assumption}[Conditional Independence] \label{assump 3}
     $Y^0 \indep Y^1 \ | \ X$.
\end{assumption}

Assumption \ref{assump 3} suggests conditional independence between the two potential outcomes $Y^0$ and $Y^1$ given the covariate $X$ \citep{shen2013treatment,yin2018assessing}. With this assumption, we can transform the joint distribution of $(Y^0, Y^1)$ given $X$ into the product of the marginal distributions of $Y^0$ and $Y^1$ given $X$, which is crucial to identify $\beta$ and $\gamma$.   
It is noteworthy that Assumption \ref{assump 2} and Assumption \ref{assump 3} do not imply each other, meaning they may provide different information for estimating $\beta$ and $\gamma$. 

The following Lemma \ref{lemma1} presents the identifiability of $\beta$ under two sets of assumptions. Let $\mu_a(x) = \E[Y = 1 | X = x, A = a]$,  $\mu_a = \E[\mu_a(X) \cdot e(X)]$ for $a = 0, 1$, and $\mu = \E[\mu_1(X) \mu_0(X) e(X)]$.

\begin{lemma}[Identifiability] \label{lemma1} It follows that 

(a) under Assumptions \ref{assump 1} and \ref{assump 2}, $\beta$ is identified as 
\begin{align*}
    \beta &= 1 - \frac{ \mathbb{E}[e(X)\mu_0(X)]}{\mathbb{E}[e(X)\mu_1(X)] } = 1 - \frac{\mu_0}{\mu_1}.
\end{align*}

(b)  under Assumptions  \ref{assump 1} and \ref{assump 3}, $\beta$ is identified as 
\begin{align*}
    \beta 
    &= 1 - \frac{\mathbb{E}[e(X)\mu_0(X)\mu_1(X)]}{\mathbb{E}[e(X)\mu_1(X)]} = 1 - \frac{\mu}{\mu_1}.
\end{align*}

\end{lemma}


Many previous studies have identified PN and PS under Assumptions \ref{assump 1} and \ref{assump 2}~\citep{pearl2022probabilities,tian2000probabilities,pearl2009causality,   dawid2017probability, mueller2021causes}. However, there is limited literature addressing the identification of PN and PS under Assumptions \ref{assump 1} and \ref{assump 3}. Lemma \ref{lemma1} presents the identifiability results for PN and PS under both sets of assumptions. 

 
%
Moreover, to the best of our knowledge, no literature has explored the efficiency of PN and PS. In this article, we aim to fill this gap by obtaining efficient estimators for PN and PS under Assumptions \ref{assump 1} and \ref{assump 2}, or Assumptions \ref{assump 1} and \ref{assump 3}. 

\section{Proposed Method} \label{sec3}
In this section, we first present the efficient influence functions (EIFs) $\beta$ under different identifiability assumptions and provide their corresponding semiparametric efficient bounds. Next, based on the derived EIFs, we construct semiparametric efficient estimators of $\beta$. 

\subsection{Efficiency Bounds of PN}

To fully utilize the collected data and the identifiability assumptions, we apply semiparametric efficiency theory to obtain efficient estimators of $\beta$. An efficient estimator, often regarded as the optimal, achieves the semiparametric efficiency bound, which represents the smallest possible asymptotic variance among all regular estimators given the observed data under the constraints imposed by the identifiability assumptions~\citep{ newey1990semiparametric, vdv-1998,tsiatis2006semiparametric}.

Let $\eta = \eta(X) := (e(X), \mu_0(X), \mu_1(X))$ denote the nuisance parameters. As an intermediate step in constructing efficient estimators, we calculate the EIFs and the semiparametric efficiency bounds of $\beta$, as demonstrated in Theorem \ref{thm1}. 



\begin{thm}[Efficiency Bounds]  \label{thm1} 
We have that 

(a) under Assumptions \ref{assump 1} and \ref{assump 2},  the efficient influence function of $\beta$ is 
\begin{align*}
    \phi_{\beta}(Y, A, X; \eta) &={} \frac{\mu_0}{\mu_1^2}A(Y-\mu_1(X)) - \frac{1}{\mu_1}\frac{1-A}{1-e(X)}(Y - \mu_0(X))e(X) \\
    &+ \frac{1}{\mu_1^2}[\mu_0\mu_1(X) - \mu_1\mu_0(X)](A - e(X)) 
    + \frac{1}{\mu_1^2}[\mu_0\mu_1(X) - \mu_1\mu_0(X)]e(X) \\
    &=\frac{(1-\beta)}{\mu_1} A Y - \frac{e(X)}{\mu_1} \frac{(1-A)}{1 - e(X)}(Y-\mu_0(X))  -   \frac{  \mu_0(X) }{\mu_1} A.   
\end{align*}
The associated semiparametric efficiency bound of $\beta$ is $\mathbb{V}(\phi_{\beta}(Y, A, X; \eta)) = \E[\phi_{\beta}^2(Y, A, X; \eta) ]$.

(b) under Assumptions \ref{assump 1} and \ref{assump 3},  the efficient influence function of $\beta$ is 
\begin{align*}
  \varphi_{\beta}(Y, A, X; \eta) &= \ A \cdot \dfrac{\mu - \mu_1\mu_0(X)}{\mu_1^2}(Y-\mu_1(X)) - \dfrac{e(X)}{\mu_1}\dfrac{1-A}{1-e(X)}(Y - \mu_0(X))\mu_1(X) \\
  &+ \dfrac{\mu - \mu_1\mu_0(X)}{\mu_1^2}\mu_1(X)(A - e(X)) + \dfrac{\mu - \mu_1\mu_0(X)}{\mu_1^2}\mu_1(X)e(X) \\
  &= \ A \cdot \dfrac{1 - \beta - \mu_0(X)}{\mu_1}Y - \dfrac{e(X)}{\mu_1}\dfrac{1-A}{1-e(X)}(Y - \mu_0(X))\mu_1(X).  
\end{align*}
The associated semiparametric efficiency bound of $\beta$ is $\mathbb{V}(\varphi_{\beta}(Y, A, X; \eta)) = \E[\varphi_{\beta}^2(Y, A, X; \eta) ]$. 
\end{thm}


Theorem \ref{thm1} presents the EIFs and the associated semiparametric efficiency bounds for $\beta$ under two different sets of identifiability assumptions. The EIFs of $\beta$ under Assumptions \ref{assump 1} and \ref{assump 2} differ from those under Assumptions \ref{assump 1} and \ref{assump 3}, reflecting the distinct information implied by Assumptions \ref{assump 2} and \ref{assump 3}. This is not surprising, as different identifiability assumptions restrict different model classes of the data-generating distribution, contributing different information to the efficiency of $\beta$. 


To examine the influence of Assumptions \ref{assump 2} and \ref{assump 3} on the estimation of $\beta$, we compare the semiparametric efficiency bounds given in Theorems \ref{thm1}(a) and \ref{thm1}(b).  

\begin{corollary}[Comparison of Bounds]\label{corollary 1}
Under Assumptions \ref{assump 1} and \ref{assump 2}, the semiparametric efficiency bound of $\beta$ is larger than that under Assumptions \ref{assump 1} and \ref{assump 3}. The magnitude of this difference is 
\begin{align*}
    \mathbb{E} \left [\frac{A\mu_0^2(X)}{\mu_1^2}(1 - Y^2) + \frac{(1 - A)e^2(X)}{\mu_1^2(1 - e(X))^2}(Y - \mu_0(X))^2(1 - \mu_1^2(X)) \right ]. 
\end{align*}
\end{corollary}

Corollary \ref{corollary 1} shows that the semiparametric efficiency bound for $\beta$ under Assumptions \ref{assump 1} and \ref{assump 3} is smaller than that under Assumptions \ref{assump 1} and \ref{assump 2}. This indicates that, under Assumptions \ref{assump 1}, Assumption \ref{assump 3} is more informative than Assumption \ref{assump 2} in terms of estimating $\beta$. This result is interesting given that Assumptions \ref{assump 2} and \ref{assump 3} do not imply each other.   

Motivated by \citet{hahn1998role}, we explore the role of propensity scores in the efficiency of $\beta$ by calculating the EIF for $\beta$ when the propensity score $e(X)$ is known, as shown in Theorem \ref{thm2}. 
\begin{thm}[Efficiency Bounds with Known Propensity Score]\label{thm2} When the propensity score $e(X)$ is known, then we have that 

(a) the efficient influence function of $\beta$ under Assumptions \ref{assump 1} and \ref{assump 2} is 
\begin{align*}
    \tilde \phi_{\beta}(Y, A, X; \eta) &= \frac{\mu_0}{\mu_1^2}A(Y-\mu_1(X)) - \frac{e(X)}{\mu_1}\frac{1-A}{1-e(X)}(Y - \mu_0(X)) + \frac{\mu_0}{\mu_1^2}\mu_1(X)e(X) - \frac{1}{\mu_1}\mu_0(X)e(X) \\
    &=\frac{1 - \beta}{\mu_1}A(Y-\mu_1(X)) - \frac{e(X)}{\mu_1}\frac{1-A}{1-e(X)}(Y - \mu_0(X)) + \frac{1 - \beta}{\mu_1}\mu_1(X)e(X) - \frac{1}{\mu_1}\mu_0(X)e(X). 
\end{align*}
The associated semiparametric efficiency bound of $\beta$ is $\mathbb{V}(\tilde \phi_{\beta}(Y, A, X; \eta))$. 

(b) the efficient influence function of $\beta$ under Assumptions \ref{assump 1} and \ref{assump 3} is 
\begin{align*}
  \tilde \varphi_{\beta}(Y, A, X; \eta) &= (\dfrac{\mu}{\mu_1^2} - \dfrac{\mu_0(X)}{\mu_1})A(Y-\mu_1(X)) - \dfrac{e(X)}{\mu_1}\dfrac{1-A}{1-e(X)}\mu_1(X)(Y - \mu_0(X)) \\
  &+ (\dfrac{\mu}{\mu_1^2} - \dfrac{\mu_0(X)}{\mu_1})\mu_1(X)e(X) \\
  &=(\dfrac{1 - \beta}{\mu_1} - \dfrac{\mu_0(X)}{\mu_1})A(Y-\mu_1(X)) - \dfrac{e(X)}{\mu_1}\dfrac{1-A}{1-e(X)}\mu_1(X)(Y - \mu_0(X)) \\
  &+ (\dfrac{1 - \beta}{\mu_1} - \dfrac{\mu_0(X)}{\mu_1})\mu_1(X)e(X).  
\end{align*}
The associated semiparametric efficiency bound of $\beta$ is $\mathbb{V}(\tilde \varphi_{\beta}(Y, A, X; \eta))$. 
\end{thm}



Parallel to Theorem \ref{thm1}, Theorem \ref{thm2} provides the EIFs of $\beta$ under two sets of identifiability assumptions when the propensity score is known. 
By comparing Theorem \ref{thm1} with Theorem \ref{thm2}, we observe that knowing the propensity score influences the form of the EIF of $\beta$, which is consistent with the previous literature on estimating average treatment effects on the treated~\citep{hahn1998role}. 

To measure the specific influence of the propensity score, we further compare the semiparametric efficiency bounds presented in Theorems \ref{thm1} and \ref{thm2}.  


\begin{proposition}[Efficiency Comparison] \label{proposition 1} We have that 
   
  (a) under Assumptions \ref{assump 1} and \ref{assump 2}, if the propensity score $e(X)$ is known, the semiparametric efficiency bound of $\beta$ is smaller compared to when the propensity score $e(X)$ is unknown. The magnitude of this difference is   
    \begin{align*}
        \mathbb{E} \left [\frac{1}{\mu_1^4}(\mu_0\mu_1(X) - \mu_1\mu_0(X))^2e(X)(1 - e(X)) \right ]. 
    \end{align*}

  (b)  under Assumptions \ref{assump 1} and \ref{assump 3}, if propensity score $e(X)$ is known, then the semiparametric efficiency bound of $\beta$ is lower compared to when the  propensity score $e(X)$ is unknown, and the magnitude of this difference is
    \begin{align*}
        \mathbb{E} \left [\frac{1}{\mu_1^4}(\mu\mu_1(X) - \mu_1\mu_1(X)\mu_0(X))^2e(X)(1 - e(X)) \right ]
    \end{align*}
\end{proposition}

Proposition \ref{proposition 1} quantifies the magnitude of the efficiency gain for $\beta$ when the propensity score is known. Under both sets of identifiability assumptions, the amount of the difference includes a term $e(X)(1 - e(X))$, indicating that as the propensity score approaches 1/2, the contribution of knowing the propensity score to the semiparametric efficiency bound for $\beta$ increases. This is intuitive because $e(X)(1 - e(X)) = \text{Var}(A|X)$ represents the uncertainty of $A$ conditional on $X$. The uncertainty reaches its maximum when $e(X) = 1/2$ and is zero when $e(X) = 0$ or 1.   

\subsection{Proposed Estimator} \label{sec3-2}
In this subsection, we develop efficient estimators for $\beta$. The proposed estimators rely on the estimation of nuisance parameters.  
 We utilize the technique of cross-fitting, a method widely employed in recent causal inference research \citep{chernozhukov2018double,wager2018estimation}, to estimate nuisance parameters $\eta$, the details are presented in Algorithm \ref{algorithm1}. We let $\hat{\eta}(X) = (\hat{e}(X), \hat{\mu}_0(X), \hat{\mu}_1(X))$ denote the estimators of nuisance parameters $\eta(X) = (e(X), \mu_0(X), \mu_1(X))$.  For simplicity, we let $Z = (Y, A, X)$ and $\phi_\beta(Z;  \eta) = \phi_\beta(A, X, Y;  \eta)$. Similarly, we can define  $\varphi_\beta(Z;  \eta)$, $\tilde \phi_\beta(Z;  \eta)$, and $\tilde \varphi_\beta(Z;  \eta)$.  

After estimating the nuisance parameters, we then construct the proposed estimators based on the EIFs outlined in Theorems \ref{thm1} and \ref{thm2}. It should be noted that the format of EIFs for $\beta$ outlined in Theorems \ref{thm1} and \ref{thm2} is more complex than the EIFs for average treatment effects (ATEs) under ignorability assumptions explored by \citet{hahn1998role} and \citet{hu2022identification}. Their EIFs are directly separable concerning the estimands of interest; that is, the EIFs can be written in the form $\xi(Z; \eta) - \tau$, involving a term only dependent on $(Z, \eta)$ minus the target estimand $\tau$. Thus, the estimator of $\tau$ can be directly constructed as $n^{-1}\sum_{i=1}^n \xi(Z_i; \hat \eta)$.    
In contrast, the EIFs in Theorems \ref{thm1} and \ref{thm2} do not exhibit direct separability, where the target estimand $\beta$ is intertwined with $(Z, \eta)$. 
Observing that the expectation of EIF is zero under the corresponding identifiability assumptions. Therefore, under Assumptions \ref{assump 1}-\ref{assump 2}, the estimator for $\beta$ can be derived naturally by solving the following sample estimating equation  
\begin{equation}\label{equa 1}
   \mathbb{E}_n [\phi_\beta(Z; \hat \eta)] = 0,  
\end{equation}
where $\mathbb{E}_n[U] = n^{-1}\sum_{i=1}^n U_i$ is the empirical expectation operator.  
Solving equation (\ref{equa 1}) yields that  
\begin{align}  \label{eq 2}
    \hat{\beta}_{mono} = \dfrac{1}{n}\sum\limits_{i=1}^n \Big \{ A_i(Y_i - \hat \mu_0(X_i)) -  \dfrac{(1-A_i)(Y_i - \hat \mu_0(X_i)) \hat e(X_i) }{1 - \hat e(X_i)}  \Big \}  \Big / \dfrac{1}{n} \sum\limits_{i=1}^n A_i Y_i. 
\end{align} 
Similarily, under Assumptions \ref{assump 1} and \ref{assump 3}, solving equation $\E_n[   \varphi_{\beta}(Z; \eta)  ]$ leads to that   
\begin{align}\label{eq 3}
    \hat \beta_{inde} = \dfrac{1}{n}\sum\limits_{i=1}^n A_i \cdot (1 - \hat \mu_0(X_i))Y_i - \hat e(X_i)\dfrac{1-A_i}{1-\hat e(X_i)}(Y_i - \hat \mu_0(X_i))\hat \mu_1(X_i) \Big / \dfrac{1}{n} \sum\limits_{i=1}^n A_i Y_i.  
\end{align}

Algorithm \ref{algorithm1} summarizes the detailed procedure for constructing the proposed estimators. 
\begin{algorithm}[h!]
\caption{Estimation Procedures of the Proposed Estimators} \label{algorithm1}
\begin{algorithmic}[1]

        \item[{\bf Step 1}] (sample splitting). We split the observed data into $K$ subsamples with equal size, each group containing $n/K$ observations (for simplicity, both $K$ and $n/K$ are assumed to be positive integers here) while defining $I_1$ to $I_K$ as the index numbers of the corresponding groups. Furthermore, let $I_k^C = \{I_1, \dots ,I_K\} \backslash I_k$ be the complement of $I_k$ for $k = 1,\dots,K$. 
        \medskip
        
        \item[{\bf Step 2}] (training nuisance parameters in every sub-sample). Let k rotate from 1 to K. In each round use the sub-sample $I_k^C$ to construct the estimators $\check{e}(X), \check{\mu}_0(X)$ and $\check{\mu}_1(X)$ for $e(X), \mu_0(X)$ and $\mu_1(X)$, and then obtain an predicted values of $\check{e}(X_i), \check{\mu}_0(X_i)$ and $\check{\mu}_1(X_i)$ for $i \in I_k$.
        \medskip
        
        \item[{\bf Step 3}] (get nuisance parameters). All the predicted values of $\check{e}(X_i), \check{\mu}_0(X_i)$ and $\check{\mu}_1(X_i)$ for $i \in \{1, \dots, n\}$ consist of the final estimates of $e(X), \mu_0(X)$ and $\mu_1(X)$, denote as $\hat{e}(X_i), \hat{\mu}_0(X_i)$ and $\hat{\mu}_1(X_i)$
        \medskip
        
       \item[{\bf Step 4}] (proposed estimators) obtain the proposed estimators of $\beta$ with equations \eqref{eq 2}-\eqref{eq 5}.
\end{algorithmic}
\end{algorithm}

In Algorithm \ref{algorithm1}, the full sample is sliced into many sub-samples, and the target parameter is estimated from each sub-sample, which is the same widely used method in the field of causal inference~\citep{wager2018estimation,kennedy2023towards}.

Similar to the construction of $\hat \beta_{mono}$ and $\hat \beta_{inde}$, when the propensity score $e(X)$ is known, the proposed estimator of $\beta$ under Assumptions \ref{assump 1}-\ref{assump 2} becomes  
\begin{align}\label{eq 4}
    \tilde{\beta}_{mono} =\dfrac{\dfrac{1}{n}\sum\limits_{i=1}^n \Big \{ A_i(Y_i-\hat\mu_1(X_i)) - e(X_i) \dfrac{1-A_i}{1- e(X_i)}(Y_i - \hat \mu_0(X_i)) + (\hat \mu_1(X_i) - \hat \mu_0(X_i)) e(X_i) \Big \}  }{ \dfrac{1}{n} \sum\limits_{i=1}^n A_i(Y_i-\hat\mu_1(X_i)) + \hat\mu_1(X_i) e(X_i) },
\end{align}
and the proposed estimator of $\beta$ under Assumptions \ref{assump 1} and \ref{assump 3} becomes 
  \begin{equation}
\begin{aligned}\label{eq 5}
    \tilde \beta_{inde} = &\frac{ \dfrac{1}{n}\sum\limits_{i=1}^n \Big \{ ( 1 - \hat \mu_0(X_i))A_i(Y_i - \hat \mu_1(X_i)) + ( 1 - \hat \mu_0(X_i)) \hat \mu_1(X_i)  e(X_i)  \Big \}}{ \dfrac{1}{n} \sum\limits_{i=1}^n \Big \{ A_i (Y_i - \hat \mu_1(X_i)) + \hat \mu_1(X_i) e(X_i) \Big \} } \\
    &- \frac{ \dfrac{1}{n}\sum\limits_{i=1}^n \Big \{  \dfrac{(1-A_i)(Y_i-\hat \mu_0(X_i)) e(X_i)}{1 - e(X_i)} \hat \mu_1(X_i) \Big \}}{ \dfrac{1}{n} \sum\limits_{i=1}^n \Big \{ A_i (Y_i - \hat \mu_1(X_i)) + \hat \mu_1(X_i) e(X_i) \Big \} },
\end{aligned}
\end{equation}
which are the solutions $   \mathbb{E}_n [\tilde \phi_\beta(Z; \hat \eta)] = 0$ and $   \mathbb{E}_n [\tilde \varphi_\beta(Z; \hat \eta)] = 0$, respectively.




Observing Equations \eqref{eq 2}-\eqref{eq 5} reveals that whether the propensity score is known or not significantly affects the form of the estimator for $\beta$ across different sets of identifiability assumptions. Next, we delve into the theoretical properties of the estimators presented in Equations \eqref{eq 2}-\eqref{eq 5}.

\section{Asymptotic Properties} \label{sec4}

We begin by examining the robustness properties of four estimators: $\hat{\beta}_{\text{mono}}$, $\tilde{\beta}_{\text{mono}}$, $\hat{\beta}_{\text{inde}}$, and $\tilde{\beta}_{\text{inde}}$ defined in subsection \ref{sec3-2}. Subsequently, we investigate their asymptotic normality.

\subsection{Double Robustness}

 We first summarize the robustness properties of the four proposed estimators. 
\begin{proposition} \label{double robustness}
We have that  

 (a) (Double Robustness) under Assumptions \ref{assump 1} and \ref{assump 2},  the proposed estimator $\hat{\beta}_{mono}$ given in equation  \eqref{eq 2} is a consistent estimator of $\beta$ if one of the following conditions is satisfied:     
    
     \begin{itemize}[leftmargin=40pt]
      \item[(i)] the propensity score model is correctly specified; 
        \item[(ii)] the outcome regression models are correctly specified.
    \end{itemize}
    
 (b) (Single Robustness) under Assumptions \ref{assump 1} and \ref{assump 3},  the proposed estimator $\hat{\beta}_{inde}$ given in equation \eqref{eq 3} is a consistent estimate of $\beta$ if the outcome regression models are correctly specified, regardless of whether the propensity score model is correctly specified or not.  
\end{proposition}


Proposition \ref{double robustness}(a) indicates that $\hat{\beta}_{mono}$ exhibits double robustness~\citep{Robins-Rotnitzky-Zhao-1994, Bang-Robins-2005}, that is,  $\hat{\beta}_{mono}$ remains a consistent estimator of $\beta$ if either the propensity score model $e(X)$ or the outcome regression model $\mu_a(X)$ for $a = 0, 1$ is correctly specified. This property provides additional protection against model misspecification on either the propensity score or the outcome regression functions.   
Proposition \ref{double robustness}(b) presents the single robustness property of $\hat{\beta}_{inde}$ under Assumptions \ref{assump 1} and \ref{assump 3}. It shows that $\hat{\beta}_{inde}$ is a consistent estimator of $\beta$ as long as the outcome regression model is correctly specified. This consistency does not rely on the correctness of model specification on the propensity score, providing a safeguard for it.

When the propensity score $e(X)$ is known, the proposed estimator $\tilde{\beta}_{mono}$ given in equation \eqref{eq 4} is always consistent, regardless of whether the outcome regression model $\hat{\mu}_a(X)$ for $a = 0, 1$ is specified correctly or not. 
In contrast, the consistency of the estimator $\tilde{\beta}_{inde}$ of $\beta$ given in equation $\eqref{eq 5}$ under Assumptions \ref{assump 1} and \ref{assump 3} depends entirely on whether the outcome regression model is correctly specified.


In addition to the proposed estimators, we can also construct the inverse probability weighting (IPW)~\citep{horvitz1952generalization}  and outcome regression (OR)~\citep{Tan-2007} estimators of $\beta$ under the corresponding identifiability assumptions.  Specifically, under Assumptions \ref{assump 1} and \ref{assump 2},  
\[  \beta = 1 - \frac{ \mathbb{E}[e(X)\mu_0(X)]}{\mathbb{E}[e(X)\mu_1(X)] } = 1 - \frac{ \mathbb{E}[e(X)  Y^0]}{\mathbb{E}[e(X) Y^1] }  =  1 - \frac{ \mathbb{E}[e(X) (1-A)Y / (1 - e(X)) ]}{\mathbb{E}[ A Y] }.   \]
Based on it, the IPW estimator of $\beta$ is given as 
\begin{equation}\label{ipw mono}
    \hat{\beta}_{mono, IPW} = 1 -\dfrac{\sum_{i = 1}^n \hat{e}(X_i)(1 - A_i)Y_i / (1 - \hat{e}(X_i))}{\sum_{i = 1}^n A_iY_i},
\end{equation}
which relies only on the nuisance parameter -- propensity score.   
Moreover, under Assumptions \ref{assump 1} and \ref{assump 2}, $\beta$ also can be reformulated as follows,   
\[  \beta = 1 - \frac{ \mathbb{E}[e(X)\mu_0(X)]}{\mathbb{E}[e(X)\mu_1(X)] } =1 - \frac{ \mathbb{E}[A\mu_0(X)]}{\mathbb{E}[A\mu_1(X)] }.   \]
Thus, the OR estimator is given by   
\begin{equation}\label{or mono}
    \hat{\beta}_{mono, OR} = 1 -\dfrac{\sum_{i = 1}^n A_i\hat \mu_0(X_i)}{\sum_{i = 1}^n A_i\hat \mu_1(X_i)},
\end{equation}
which relies only on the nuisance parameter -- outcome regression functions.  


The proposed estimator $\hat{\beta}_{mono}$ differs from the IPW estimator $\hat{\beta}_{mono, IPW}$ and the OR estimator $\hat{\beta}_{mono, OR}$ in that it relies on both the propensity score and the outcome regression functions. The consistency of $\hat{\beta}_{mono}$ is guaranteed as long as either the propensity score or the outcome regression function is correctly specified. In contrast, the IPW and OR estimators rely solely on the propensity score model and the outcome regression models, respectively, with their consistency depending on the correct specification of these models. 


\subsection{Asymptotic Normality and Efficiency}
In this subsection, we present the asymptotic normality and efficiency of the proposed estimators $\hat{\beta}_{\text{mono}}$, $\tilde{\beta}_{\text{mono}}$, $\hat{\beta}_{\text{inde}}$, and $\tilde{\beta}_{\text{inde}}$.  Let $|| \cdot ||_2$ represent the $L_2$-norm.

\begin{condition}\label{condition}
    For  $a \in \{0, 1\}$, 
    \begin{itemize}
        \item[(a)] $|| \hat{e} - e ||_2 \cdot || \hat{\mu}_a - \mu_a ||_2 = o_\mathbb{p}(n^{-1/2})$;
        \item[(b)] $|| \hat{\mu}_1 - \mu_1 ||_2 \cdot || \hat{\mu}_0 - \mu_0 ||_2 = o_\mathbb{p}(n^{-1/2})$.
    \end{itemize}  
\end{condition}

Condition \ref{condition} is a high-level condition involving the convergence rates of the propensity score estimator $\hat{e}(X)$ and the outcome regression function estimator $\hat{\mu}_a(X)$. This condition is widely adopted in machine-learning-aided causal inference~\citep{chernozhukov2018double, kennedy2023towards}. Condition \ref{condition} is mild as it imposes weak restrictions on the estimation of the nuisance parameter, requiring only that the $L_2$-error of the product of the convergence rates of $\hat{e}(X)$ and $\hat{\mu}_a(X)$ be smaller than $n^{-1/2}$. This means that the convergence rate of either $\hat{\mu}_a(X)$ or $\hat{e}(X)$ can be as slow as $n^{-1/4}$.  
The common parametric model, generalized linear model, lasso, and many flexible machine learning methods, such as random forests, satisfy this condition~\citep{chernozhukov2018double, Semenova-Chernozhukov}.  


\begin{thm}[Asymptotic Normality and Efficiency]\label{asy 1} Under Assumptions \ref{assump 1}-\ref{assump 2} and Condition \ref{condition},

(a) when the propensity score $e(X)$ is unknown,
    \begin{equation*}
        \sqrt{n}(\hat{\beta}_{mono} - \beta) \xrightarrow{d} \mathcal{N}(0, \sigma_1^2),
    \end{equation*}
    where $ \xrightarrow{d}$ denotes convergence in distribution and $\sigma_1^2 =\mathbb{V}(\phi_{\beta}(Z; \eta))$ is the semiparametric efficiency bound of $\beta$.  
    A consistent estimator of $\sigma_1^2$ is $\hat \sigma_1^2 = n^{-1}\sum_{i = 1}^n [\zeta_1(Z_i; \hat{\eta}) - \hat \beta_{mono}]^2$, where $\zeta_1(Z_i; \hat{\eta}) =  \big \{ A_i(Y_i - \hat \mu_0(X_i)) -  (1-A_i)(Y_i - \hat \mu_0(X_i)) \hat e(X_i)/(1 - \hat e(X_i))  \big \}  / (n^{-1}) \sum_{i=1}^n A_i Y_i$.

(b) when the propensity score $e(X)$ is known,
    \begin{equation*}
        \sqrt{n}(\tilde{\beta}_{mono} - \beta) \xrightarrow{d} \mathcal{N}(0, \sigma_2^2), 
    \end{equation*}
    where 
    $\sigma_2^2 = 
    \mathbb{V}(\tilde \phi_{\beta}(Z; \eta))$ is the semiparametric efficiency bound of $\beta$. A consistent estimator of $\sigma_2^2$ is $\hat\sigma_2^2 = n^{-1}\sum_{i = 1}^n [ \zeta_2(Z_i; \hat{\eta}) - \tilde \beta_{mono}]^2$, where $ \zeta_2(Z_i; \hat{\eta}) = \big \{ A_i(Y_i-\hat\mu_1(X_i)) - e(X_i)(1-A_i)(Y_i - \hat \mu_0(X_i))/1- e(X_i) + (\hat \mu_1(X_i) - \hat \mu_0(X_i)) e(X_i) \big \} / (n^{-1}) \sum_{i=1}^n \{A_i(Y_i-\hat\mu_1(X_i)) + \hat\mu_1(X_i) e(X_i) \}$. 
\end{thm}

Theorem \ref{asy 1} establishes the consistency and asymptotic normality of $\hat{\beta}_{\text{mono}}$ and $\tilde{\beta}_{\text{mono}}$ under the Assumptions \ref{assump 1} and \ref{assump 2}. Both estimators achieve asymptotic variances that attain the corresponding semiparametric efficient bounds, demonstrating the local efficiency of the proposed estimators. 
In addition, Theorem \ref{asy 1} provides the estimation of asymptotic variances using plug-in methods, which are shown to perform well in the simulation.
By comparing the asymptotic variances of $\hat{\beta}_{\text{mono}}$ and $\tilde{\beta}_{\text{mono}}$, we observe that the asymptotic variance of $\tilde{\beta}_{\text{mono}}$ is smaller than that of $\hat{\beta}_{\text{mono}}$ according to Proposition \ref{proposition 1}.

\begin{thm}[Asymptotic Normality and Efficiency] \label{asy 2} Under Assumptions \ref{assump 1} and \ref{assump 3} and Condition \ref{condition},

(a) when the propensity score $e(X)$ is unknown,
    \begin{equation*}
        \sqrt{n}(\hat{\beta}_{inde} - \beta) \xrightarrow{d} \mathcal{N}(0, \sigma_3^2),
    \end{equation*}
    where $\xrightarrow{d}$ denotes convergence in distribution and $\sigma_3^2 
    = \mathbb{V}(\varphi_{\beta}(Z; \eta))$. A consistent estimator of $\sigma_3^2$ is $\hat\sigma_3^2 = n^{-1}\sum_{i = 1}^n [\zeta_3(Z_i; \hat{\eta}) - \hat \beta_{inde}]^2$, where $\zeta_3(Z_i; \hat{\eta}) =  \big \{A_i \cdot (1 - \hat \mu_0(X_i))Y_i - \hat e(X_i)(1-A_i)(Y_i - \hat \mu_0(X_i))\hat \mu_1(X_i)/(1-\hat e(X_i))  \big \}  / (n^{-1}) \sum_{i=1}^n A_i Y_i $.\\ 
    
(b) when the propensity score $e(X)$ is known, 
    \begin{equation*}
        \sqrt{n}(\tilde{\beta}_{inde} - \beta) \xrightarrow{d} \mathcal{N}(0, \sigma_4^2),
    \end{equation*}
    where $\xrightarrow{d}$ denotes convergence in distribution and $\sigma_4^2 = \mathbb{V}(\tilde \varphi_{\beta}(Z; \eta))$. A consistent estimator of $\sigma_4^2$ is $\hat\sigma_4^2 = n^{-1}\sum_{i = 1}^n [ \zeta_{4}(Z_i; \hat{\eta}) - \tilde{\beta}_{inde}]^2$, where $ \zeta_{4}(Z_i; \hat{\eta}) =  \{ ( 1 - \hat \mu_0(X_i))A_i(Y_i - \hat \mu_1(X_i))+ ( 1 - \hat \mu_0(X_i)) \hat \mu_1(X_i)  e(X_i) - ((1-A_i)(Y_i-\hat \mu_0(X_i)) e(X_i) \hat \mu_1(X_i))/(1 - e(X_i)) \} / (n^{-1}) \sum_{i=1}^n \{A_i(Y_i-\hat\mu_1(X_i)) + \hat\mu_1(X_i) e(X_i) \}$.\\ 
\end{thm}


Similar to Theorem \ref{asy 1}, Theorem \ref{asy 2} establishes the asymptotic normality and local efficiency of the proposed estimators $\hat{\beta}_{\text{inde}}$ and $\tilde{\beta}_{\text{inde}}$. It also demonstrates the variance reduction property when the propensity scores are known compared to when they are unknown. 



\section{Simulation} \label{sec5}
In this section, we conduct extensive simulation studies to evaluate the performance of the proposed estimators with finite samples. 
We compare the proposed estimators with the competing estimators in terms of efficiency and double robustness under different simulation settings. To implement our proposed estimators, we employ logistic regression with main effects to estimate both the outcome regression functions and propensity score.   
Throughout the simulation, Treatment $A$ is obtained from logistic regression distribution $\mathbb{P}(A = 1| X) = \text{expit}(X^T\alpha)$, where $\text{expit}(x) = \exp(x)/\{1 + \exp(x)\}$ is the logistic function, and $\alpha$ is the corresponding parameter. 




The simulation studies each included 1000 replications with sample sizes of 500, 1000, and 2000, respectively. The results presented in the following tables include the metrics Bias, SSE, ESE, and CP95, where Bias and SSE are the mean and standard deviation of the Monte Carlo point estimates for 1000 repetitions, respectively,  ESE is the mean of the estimated asymptotic standard error given in Theorems \ref{asy 1} and \ref{asy 2}, and CP95 is the coverage proportions of 95\% confidence intervals based on 1000 repetitions. For each simulation case,
 We obtain the true value of $\beta$ based on a simulation with a sample size of 1,000,000.   

\subsection{Study I}
To investigate the performance of the four proposed estimators $\hat{\beta}_{mono}$, $\hat{\beta}_{inde}$, $\tilde{\beta}_{mono}$ and $\tilde{\beta}_{inde}$ in numerical simulations, we first consider the following four data-generation mechanisms:
\begin{itemize}[leftmargin=40pt]
    \item[Case 1] $X = (X_1, X_2)^T \sim N(0, 4I_2)$, $\mathbb{P}(A = 1| X) = \text{expit}((X_1 + X_2)/8)$, $\mathbb{P}(Y^0 = 1| X) = \text{expit}((X_1 - X_2) / 2)$, and $\mathbb{P}(Y^1 = 1| X) = \text{expit}((2X_1 + 3X_2)/3 + 1/2)$, where $I_2$ is an identity matrix. We further adjust the value of $Y^0$, setting $Y^0$ to 0 if $Y^1$ equals 0.
    
    \item[Case 2] $X = (X_1, X_2)^T \sim N(0, 4I_2)$, $\mathbb{P}(A = 1| X) = \text{expit}((X_1 + X_2)/8)$, $\mathbb{P}(Y^0 = 1| X) = \text{expit}((X_1 - X_2) / 2)$, and $\mathbb{P}(Y^1 = 1| X) = \text{expit}((2X_1 + 3X_2)/3 + 1/2)$, where $I_2$ is an identity matrix.
    
    \item[Case 3] $X = (X_1, X_2, X_3, X_4, X_5)^T \sim N(0, 4I_5)$, $\mathbb{P}(A = 1| X) = \text{expit}((X_1 + X_2 + X_3 + X_4 + X_5)/8)$, $\mathbb{P}(Y^0 = 1| X) = \text{expit}((X_1 - X_2 + X_3 - X_4 + X_5) / 2)$, and $\mathbb{P}(Y^1 = 1| X) = \text{expit}((2X_1 + 3X_2 + 2X_3 + 3X_4 + 2X_5)/3 + 1/2)$, where $I_5$ is an identity matrix. 
    We further adjust the value of $Y^0$, setting $Y^0$ to 0 if $Y^1$ equals 0. 
    
    \item[Case 4] $X = (X_1, X_2, X_3, X_4, X_5)^T \sim N(0, 4I_5)$, $\mathbb{P}(A = 1| X) = \text{expit}((X_1 + X_2 + X_3 + X_4 + X_5)/8)$, $\mathbb{P}(Y^0 = 1| X) = \text{expit}((X_1 - X_2 + X_3 - X_4 + X_5) / 2)$, and $\mathbb{P}(Y^1 = 1| X) = \text{expit}((2X_1 + 3X_2 + 2X_3 + 3X_4 + 2X_5)/3 + 1/2)$, where $I_5$ is an identity matrix.
\end{itemize}

Cases 1 and 3 satisfy Assumptions \ref{assump 1} and \ref{assump 2} and thus can be used to examine the numerical performance of $\hat \beta_{mono}$ and $\tilde \beta_{mono}$.  
On the other hand, Cases 2 and 4 satisfy Assumptions \ref{assump 1} and \ref{assump 3}. Consequently, they can be used to explore the numerical performance of $\hat \beta_{inde}$ and $\tilde \beta_{inde}$. 
In addition, the main difference between Cases 1-2 and Cases 3-4 is the dimension of covariates $X$; the former is two-dimensional, while the latter is five-dimensional. In Cases 1 and 3, to ensure monotonicity (Assumption \ref{assump 2}), we adjust the value of $Y^0$, setting $Y^0$ to 0 if $Y^1$ equals 0. 


\begin{table}[htbp]
  \centering
  \caption{Comparison of the proposed estimators with estimated or true propensity score for Cases 1--4.}
  \scalebox{0.88}{
    \begin{tabular}{cccccccccccccc}
    \toprule
          &       & \multicolumn{4}{c}{n = 500}   & \multicolumn{4}{c}{n = 1000}  & \multicolumn{4}{c}{n = 2000} \\
    \midrule
    Case  & Estimator & Bias  & SSE   & ESE   & CP95  & Bias  & SSE   & ESE   & CP95  & Bias  & SSE   & ESE   & CP95 \\
    \midrule
    \multicolumn{14}{c}{Estimated propensity score} \\
    \midrule
    Case 1 & $\hat \beta_{mono}$ & -0.001 & 0.054 & 0.053 & 0.939 & 0.000 & 0.038 & 0.037 & 0.940 & 0.000 & 0.027 & 0.026 & 0.952 \\
    Case 2 & $\hat \beta_{inde}$ & 0.001 & 0.042 & 0.042 & 0.945 & 0.000 & 0.029 & 0.029 & 0.947 & 0.001 & 0.021 & 0.021 & 0.945 \\
    Case 3 & $\hat \beta_{mono}$ & 0.000 & 0.058 & 0.055 & 0.940 & 0.000 & 0.040 & 0.039 & 0.939 & 0.001 & 0.027 & 0.028 & 0.951 \\
    Case 4 & $\hat \beta_{inde}$ & -0.002 & 0.047 & 0.045 & 0.934 & 0.001 & 0.032 & 0.032 & 0.950 & 0.003 & 0.021 & 0.022 & 0.950 \\
    \midrule
    \multicolumn{14}{c}{True propensity score} \\
    \midrule
    Case 1 & $\tilde \beta_{mono}$ & 0.000 & 0.051 & 0.051 & 0.946 & -0.002 & 0.037 & 0.036 & 0.947 & 0.002 & 0.025 & 0.026 & 0.952 \\
    Case 2 & $\tilde \beta_{inde}$ & 0.001 & 0.039 & 0.040 & 0.953 & 0.000 & 0.027 & 0.028 & 0.955 & 0.001 & 0.020 & 0.020 & 0.947 \\
    Case 3 & $\tilde \beta_{mono}$ & -0.002 & 0.054 & 0.052 & 0.931 & -0.001 & 0.039 & 0.037 & 0.930 & 0.000 & 0.026 & 0.026 & 0.949 \\
    Case 4 & $\tilde \beta_{inde}$ & 0.002 & 0.041 & 0.042 & 0.948 & 0.001 & 0.029 & 0.030 & 0.954 & 0.003 & 0.021 & 0.021 & 0.952 \\
    \bottomrule
    \end{tabular}}
     \begin{tablenotes} 
    \footnotesize
    \item Note: The Bias and SSE are the mean and standard deviation of the Monte Carlo point estimates for 1000 repetitions; ESE is the mean of the theoretical asymptotic variance given in Theorem \ref{asy 1} and \ref{asy 2} at 1000 repetitions; and CP95 is the coverage proportions of 95\% confidence intervals based on 1000 repetitions. 
    \end{tablenotes}
  \label{tab 1}%
\end{table}%

Table \ref{tab 1} summarizes the numerical results of the proposed estimators of $\beta$ for Cases 1--4. 
From Table \ref{tab 1}, we can draw the following observations: (1) The Bias of our proposed estimators is small across all cases and approaches zero as the sample size increases. This holds whether the propensity score is known or unknown, indicating the consistency of our proposed estimation method. (2) The value of ESE is remarkably close to the value of SSE, and the value of CP95 deviates very little from the nominal value of 0.95. These results clearly illustrate the asymptotic properties of Theorems \ref{asy 1} and \ref{asy 2}. (3) The values of SSE and ESE when the propensity score is known are lower than those when the propensity score is unknown, which is consistent with the description of Propositions \ref{proposition 1}. (4) As the sample size grows, the ESE and SSE become increasingly similar, and the CP95 value consistently converges to 0.95. This confirms the steady enhancement of our proposed estimator’s performance with increasing sample size.

\subsection{Study II} 
To illustrate the double robustness of $\hat \beta_{mono}$ as outlined in Proposition \ref{double robustness}, we provide the following six data-generation mechanisms:  
\begin{itemize}[leftmargin=40pt]
    \item[Case 5] $X = (X_1, X_2)^T \sim N(0, I_2)$, $\mathbb{P}(A = 1| X) = \text{expit}((X_1 + X_2)/2)$, $\mathbb{P}(Y^0 = 1| X) = \text{expit}((X_1 - X_2) / 2)$, and $\mathbb{P}(Y^1 = 1| X) = \text{expit}((2/5)X_1 + (3/5)X_2 + 1/2)$, where $I_2$ is an identity matrix. We further adjust the value of $Y^0$, setting $Y^0$ to 0 if $Y^1$ equals 0.
    \item[Case 6] $X = (X_1, X_2)^T \sim N(0, I_2)$, $\mathbb{P}(A = 1| X) = \text{expit}((sin(X_1) + log(1 + X_2^2))/2)$, $\mathbb{P}(Y^0 = 1| X) = \text{expit}((X_1 - X_2) / 2)$, and $\mathbb{P}(Y^1 = 1| X) = \text{expit}((2/5)X_1 + (3/5)X_2 + 1/2)$, where $I_2$ is an identity matrix. We further adjust the value of $Y^0$, setting $Y^0$ to 0 if $Y^1$ equals 0.
    \item[Case 7] $X = (X_1, X_2)^T \sim N(0, I_2)$, $\mathbb{P}(A = 1| X) = \text{expit}((X_1 + X_2)/2)$, $\mathbb{P}(Y^0 = 1| X) = \text{expit}((sin(X_1) - log(1 + X_2^2)) / 2)$, and $\mathbb{P}(Y^1 = 1| X) = \text{expit}((2/5)sin(X_1) + (3/5)log(1 + X_2^2) + 1/2)$, where $I_2$ is an identity matrix. We further adjust the value of $Y^0$, setting $Y^0$ to 0 if $Y^1$ equals 0.
    \item[Case 8] $X = (X_1, X_2)^T \sim N(0, I_2)$, $\mathbb{P}(A = 1| X) = \text{expit}((X_1 + X_2 + X_3 + X_4 + X_5)/2)$, $\mathbb{P}(Y^0 = 1| X) = \text{expit}((X_1 - X_2) / 2)$, and $\mathbb{P}(Y^1 = 1| X) = \text{expit}((2/5)X_1 + (3/5)X_2 + 1/2)$, where $I_2$ is an identity matrix. We further adjust the value of $Y^0$, setting $Y^0$ to 0 if $Y^1$ equals 0. 
    \item[Case 9] $X = (X_1, X_2, X_3, X_4, X_5)^T \sim N(0, I_5)$, $\mathbb{P}(A = 1| X) = \text{expit}((sin(X_1) + log(1 + X_2^2) + sin(X_1)cos(X_3) + exp(X_4) + X_4X_5)/2)$, $\mathbb{P}(Y^0 = 1| X) = \text{expit}((X_1 - X_2 + X_3 - X_4 + X_5) / 2)$, and $\mathbb{P}(Y^1 = 1| X) = \text{expit}((2/5)X_1 + (3/5)X_2 + (2/5)X_3 + (3/5)X_4 + (2/5)X_5 + 1/2)$, where $I_5$ is an identity matrix. We further adjust the value of $Y^0$, setting $Y^0$ to 0 if $Y^1$ equals 0. 
    \item[Case 10] $X = (X_1, X_2, X_3, X_4, X_5)^T \sim N(0, I_5)$, $\mathbb{P}(A = 1| X) = \text{expit}((X_1 + X_2 + X_3 + X_4 + X_5)/2)$, $\mathbb{P}(Y^0 = 1| X) = \text{expit}((sin(X_1) - log(1 + X_2^2) + sin(X_1)cos(X_3) - exp(X_4) + X_4X_5) / 2)$, and $\mathbb{P}(Y^1 = 1| X) = \text{expit}((2/5)sin(X_1) + (3/5)log(1 + X_2^2) + (2/5)sin(X_1)cos(X_3) + (3/5)exp(X_4) + (2/5)X_4X_5 + 1/2)$, where $I_5$ is an identity matrix. We further adjust the value of $Y^0$, setting $Y^0$ to 0 if $Y^1$ equals 0.  
\end{itemize}

All Cases 5--10 satisfy Assumptions \ref{assump 1} and \ref{assump 2}, and they can be classified as follows:
\begin{itemize}
    \item Cases 5 and 8: The outcome regression model (i.e. $\mu_a(x) \ for \ a = 0, 1$) and the propensity score model (i.e. $e(x)$) are both specified correctly.
    \item Cases 6 and 9:  The outcome regression model (i.e. $\mu_a(x) \ for \ a = 0, 1$) is specified correctly, but the propensity score model (i.e. $e(x)$) is specified incorrectly.
    \item Cases 7 and 10: The propensity score model (i.e. $e(x)$) is specified correctly, but the outcome regression model (i.e. $\mu_a(x) \ for \ a = 0, 1$) is specified incorrectly.
\end{itemize}
Cases 5--10 can be used to explore the double robustness property of $\hat \beta_{mono}$. The key distinction between Cases 5--7 and Cases 8--10 lies in the dimensionality of the covariates: Cases 5--7 have 2-dimensional covariates, while Cases 8--10 have 5-dimensional covariates. 

\begin{table}[h!]
  \centering
  \caption{Numerical performance of $\hat \beta_{mono}$}
  \scalebox{0.82}{
    \begin{tabular}{ccccccccccccccc}
    \toprule
          &   &     & \multicolumn{4}{c}{n = 500}   & \multicolumn{4}{c}{n = 1000}  & \multicolumn{4}{c}{n = 2000} \\
    \midrule
    Case  & $\hat e(X)$ & $\hat \mu_a(X)$  & Bias  & SSE   & ESE   & CP95  & Bias  & SSE   & ESE   & CP95  & Bias  & SSE   & ESE   & CP95 \\
    \midrule
    Case 5 & \Checkmark & \Checkmark & 0.002 & 0.060 & 0.061 & 0.942 & 0.000 & 0.046 & 0.043 & 0.949 & 0.001 & 0.031 & 0.031 & 0.950 \\
    Case 6 & \ding{55}  & \Checkmark & 0.000 & 0.055 & 0.054 & 0.946 & -0.003 & 0.038 & 0.038 & 0.952 & -0.001 & 0.028 & 0.027 & 0.950 \\
    Case 7 &  \Checkmark & \ding{55} & 0.001 & 0.050 & 0.051 & 0.951 & 0.001 & 0.036 & 0.035 & 0.948 & 0.000 & 0.025 & 0.025 & 0.947 \\
    \midrule
    Case 8 &\Checkmark & \Checkmark  & 0.000 & 0.061 & 0.061 & 0.944 & 0.000 & 0.042 & 0.043 & 0.957 & 0.001 & 0.030 & 0.031 & 0.954 \\
    Case 9 & \ding{55} & \Checkmark  & -0.006 & 0.062 & 0.061 & 0.951 & -0.003 & 0.044 & 0.043 & 0.947 & -0.003 & 0.030 & 0.030 & 0.948 \\
    Case 10 & \Checkmark & \ding{55}  & -0.001 & 0.043 & 0.043 & 0.946 & -0.001 & 0.031 & 0.030 & 0.950 & 0.000 & 0.022 & 0.021 & 0.948 \\
    \bottomrule
    \end{tabular}%
    }
    \begin{tablenotes} 
    \footnotesize
    \item[1] The symbols $\checkmark$ and $\times$ represent the corresponding model being specified correctly and incorrectly, respectively. 
    \item[2] The Bias and SSE are the mean and standard deviation of the Monte Carlo point estimates for 1000 repetitions; ESE is the mean of the theoretical asymptotic variance given in Theorem \ref{assump 1}and \ref{assump 2} at 1000 repetitions; and CP95 is the coverage proportions of 95\% confidence intervals based on 1000 repetitions. 
    \end{tablenotes}
  \label{tab 2}%
\end{table}%
Table \ref{tab 2} summarizes numerical results for $\hat \beta_{mono}$ in Cases 5--10. From the observations in Table \ref{tab 2} we can draw the following conclusions: (1) Regardless of the dimensionality of the covariates, the Bias remains small and very close to zero across all specified models. This indicates that $\hat{\beta}_{mono}$ is consistent. (2) In each Case, SSE and ESE are almost identical, while the value of CP95 is very close to its nominal value, which again verifies the large sample nature of $\hat \beta_{mono}$. (3) The Bias, ESE, SSE, and CP95 values are very close across the three models specified. As the sample size increases, these metrics gradually converge, showing that our proposed estimator performs increasingly well. This supports Proposition \ref{double robustness}(a), validating that our estimator is becoming more accurate.

 To verify the single robustness property of $\hat{\beta}_{inde}$ as described in Proposition \ref{double robustness}(b), we performed numerical simulations using Cases 11--16. The data generation mechanisms for Cases 11--16 are the same as those for Cases 5--10, respectively, except that there is no need to further adjust $Y^0$ based on the value of $Y^1$. 
The associated results for Cases 11--16  are displayed in Table \ref{tab 3}.

\begin{table}[htbp]
  \centering
  \caption{Numerical performance of $\hat \beta_{inde}$ }
  \scalebox{0.83}{
    \begin{tabular}{ccccccccccccccc}
    \toprule
          &       & \multicolumn{4}{c}{n = 500}   & \multicolumn{4}{c}{n = 1000}  & \multicolumn{4}{c}{n = 2000} \\
    \midrule
    Case  & $\hat e(X)$ & $\hat \mu_a(X)$ & Bias  & SSE   & ESE   & CP95  & Bias  & SSE   & ESE   & CP95  & Bias  & SSE   & ESE   & CP95 \\
    \midrule
    Case 11 & \Checkmark & \Checkmark & 0.001 & 0.047 & 0.045 & 0.942 & 0.002 & 0.032 & 0.032 & 0.943 & 0.001 & 0.022 & 0.022 & 0.948 \\
    Case 12 & \ding{55} & \Checkmark & 0.002 & 0.037 & 0.038 & 0.941 & 0.000 & 0.028 & 0.027 & 0.948 & 0.000 & 0.019 & 0.019 & 0.950 \\
    Case 13 & \Checkmark & \ding{55} & -0.002 & 0.042 & 0.042 & 0.938 & -0.004 & 0.030 & 0.029 & 0.935 & -0.004 & 0.021 & 0.021 & 0.931 \\
    \midrule
    Case 14 & \Checkmark & \Checkmark & 0.002 & 0.039 & 0.039 & 0.941 & 0.000 & 0.027 & 0.027 & 0.946 & 0.000 & 0.019 & 0.019 & 0.957 \\
    Case 15 & \ding{55} & \Checkmark & 0.000  & 0.039  & 0.039  & 0.953  & 0.000  & 0.028  & 0.027  & 0.946  & 0.000  & 0.020  & 0.019  & 0.949  \\
    Case 16 & \Checkmark & \ding{55} & 0.001 & 0.033 & 0.033 & 0.944 & 0.001 & 0.023 & 0.023 & 0.951 & 0.001 & 0.017 & 0.016 & 0.942 \\
    \bottomrule
    \end{tabular}%
    }
  \label{tab 3}%
\end{table}%

Based on the observations in Table \ref{tab 3}, we can draw the following conclusions: (1) In both Cases 11--12 and Cases 14--15, the Bias is small and very close to zero, suggesting that $\hat{\beta}_{mono}$ is consistent, in line with the conclusion of Proposition \ref{double robustness}(b). (2) 
The SSE and ESE values under Cases 11--12 and Cases 14--15 are almost identical, and the CP95 value is very close to its nominal value, further validating the large sample properties of $\hat \beta_{mono}$. (3) In terms of the metrics Bias, ESE, SSE, and CP95, we find that the results for Cases 13 and 16 are similar to those for Cases 11, 12, 14, and 15. This indicates that
 $\hat{\beta}_{mono}$ still performs well numerically, even the it is not a consistent estimator theoretically when the propensity scores are correctly specified and the outcome regressions are incorrectly specified. 
(4) The values of Bias, ESE, SSE, and CP95 are very close across all three model specifications. The gradual convergence of these metrics as the sample size increases highlights the superiority of our proposed estimator.


\subsection{Study III}
To evaluate the efficiency of our proposed estimators, we compare them with two commonly used methods -- the OR estimator and the IPW estimator defined in equations \eqref{or mono} and \eqref{ipw mono}.  
We further consider the following three data-generation mechanisms: 

\begin{itemize}[leftmargin=45pt]
    \item[Case 17] $X = (X_1, X_2, X_3, X_4, X_5)^T \sim N(0, 4I_5)$, $\mathbb{P}(A = 1| X) = \text{expit}((X_1 + X_2 + X_3 + X_4 + X_5)/2)$, $\mathbb{P}(Y^0 = 1| X) = \text{expit}((X_1 - X_2 + X_3 - X_4 + X_5) / 5)$, and $\mathbb{P}(Y^1 = 1| X) = \text{expit}((X_1 + 2X_2 + X_3 + 2X_4 + X_5)/5 + 1/2)$, where $I_5$ is an identity matrix. We further adjust the value of $Y^0$, setting $Y^0$ to 0 if $Y^1$ equals 0. 
    \item[Case 18] $X = (X_1, X_2, X_3, X_4, X_5)^T \sim N(0, 9I_5)$, $\mathbb{P}(A = 1| X) = \text{expit}((sin(X_1) + log(1 + X_2^2) + sin^2(X_3) + cos(X_2)sin(X_4) + X_5))$, $\mathbb{P}(Y^0 = 1| X) = \text{expit}((X_1 - X_2 + X_3 - X_4 + X_5) / 5)$, and $\mathbb{P}(Y^1 = 1| X) = \text{expit}((X_1 + X_2 + X_3 + X_4 + X_5)/5 + 1/2)$, where $I_5$ is an identity matrix. We further adjust the value of $Y^0$, setting $Y^0$ to 0 if $Y^1$ equals 0. 
    \item[Case 19] $X = (X_1, X_2, X_3, X_4, X_5)^T \sim N(0, 9I_5)$, $\mathbb{P}(A = 1| X) = \text{expit}((X_1 + X_2 + X_3 + X_4 + X_5)/2)$, $\mathbb{P}(Y^0 = 1| X) = \text{expit}(sin(X_1) - log(1 + X_2^2) + sin^2(X_3) - log(1 + \lvert X_5 \rvert)cos(X_4) + sin(X_5))$, and $\mathbb{P}(Y^1 = 1| X) = \text{expit}(sin(X_1) + log(1 + X_2^2) + sin^2(X_3) + log(1 + \lvert X_5 \rvert)cos(X_4) + sin(X_5) + 1)$, where $I_5$ is an identity matrix. We further adjust the value of $Y^0$, setting $Y^0$ to 0 if $Y^1$ equals 0. 
\end{itemize}

All Cases 17--19 satisfy Assumptions \ref{assump 1} and \ref{assump 2}. These Cases correspond to three types of model specification: (1) Case 17: All models are specified correctly. (2) Case 18: The propensity score model $\hat{e}(X)$ is incorrectly specified, but the outcome regression model $\hat{\mu}_a(X)$ is correctly specified. (3) Case 19: The propensity score model $\hat{e}(X)$ is correctly specified, but the outcome regression model $\hat{\mu}_a(X)$ is incorrectly specified.

\begin{table}[h!]
  \centering
  \caption{Comparison of the efficiency of different estimation methods.}
  \setlength{\tabcolsep}{3mm}{
  \small
    \begin{tabular}{cccccccccc}
    \toprule
          &       & \multicolumn{2}{c}{} & \multicolumn{2}{c}{n = 500} & \multicolumn{2}{c}{n = 1000} & \multicolumn{2}{c}{n = 2000} \\
    \midrule
    Case  & $\hat{e}(X)$ & $\hat{\mu}_a(X)$ & Approach & Bias  & SSE  & Bias  & SSE      & Bias  & SSE    \\
    \midrule
    \multirow{3}[0]{*}{Case 17} & \multirow{3}[0]{*}{\Checkmark} & \multirow{3}[0]{*}{\Checkmark} & proposed & -0.005 & 0.199  & -0.001 & 0.132  & 0.000 & 0.096  \\
          &       &       & OR    & -0.050 & 0.087  & -0.052 & 0.063  & -0.050 & 0.045  \\
          &       &       & IPW   & -0.002 & 0.244  & 0.002 & 0.182  & -0.001 & 0.183  \\
          \midrule
    \multirow{3}[0]{*}{Case 18} & \multirow{3}[0]{*}{\ding{55}} & \multirow{3}[0]{*}{\Checkmark} & proposed & -0.038 & 0.249  & -0.025 & 0.170  & -0.024 & 0.121  \\
          &       &       & OR    & -0.044 & 0.133  & -0.037 & 0.098  & -0.034 & 0.067  \\
          &       &       & IPW   & 0.058 & 0.390  & 0.078 & 0.211  & 0.075 & 0.237  \\
          \midrule
    \multirow{3}[0]{*}{Case 19} & \multirow{3}[0]{*}{\Checkmark} & \multirow{3}[0]{*}{\ding{55}} & proposed & -0.002 & 0.411  & 0.000 & 0.220  & 0.004 & 0.220  \\
          &       &       & OR    & -0.041 & 0.080  & -0.033 & 0.053  & -0.032 & 0.039  \\
          &       &       & IPW   & -0.002 & 0.496  & 0.010 & 0.235  & -0.002 & 0.327  \\
          \bottomrule
    \end{tabular}%
    \begin{tablenotes} 
    \footnotesize
    \item Note: The Bias and SSE are the mean and standard deviation of the Monte Carlo point estimates for 1000 repetitions.
    \end{tablenotes}
  \label{tab 4}%
  }
\end{table}%

The numerical results are summarized in Table \ref{tab 4}. From Table \ref{tab 4}, we can observe the following: (1)
When both the propensity score model and the outcome regression model are specified correctly, the Bias of our proposed estimator is lower than that of the OR estimator, while the SSE of our proposed estimator is not higher than that of the IPW method. Our proposed method performs comparably to the OR and IPW methods when all models are correctly specified. This similarity is demonstrated by the trade-off between Bias and SSE. When the models are specified correctly, our method maintains a balance between reducing Bias and minimizing SSE, indicating its effectiveness in providing accurate estimates. (2)When the propensity score model is specified incorrectly and the outcome regression model is specified correctly, the Bias and SSE of our proposed estimator are lower than those of the IPW method, indicating that the IPW method fails in this scenario. Additionally, the Bias of our proposed estimator is close to that of the OR method, and the SSE is only slightly larger than that of the OR method. This demonstrates the reliability of our proposed estimator under the Bias-SSE trade-off. (3) When the outcome regression model is misspecified, the OR method fails, which is evident from the significantly higher Bias compared to our proposed estimator. In this scenario, the Bias and SSE of our proposed estimator are close to those of the IPW method. This suggests that our proposed estimator remains efficient, further demonstrating the robustness of our proposed approach. (4) Combining the numerical simulation results under the three specified models, both the Bias and SSE of our proposed method decrease gradually with increasing sample size. Moreover, it outperforms the two comparative methods in different aspects, illustrating the advantages and effectiveness of our proposed estimation method.

\section{Application} \label{sec6}

In recent decades, stroke has been the second most common cause of death worldwide, with over 70\% of stroke-related deaths occurring in economically underdeveloped regions \citep{feigin2007stroke,feigin2017global}. As science advances, doctors are identifying the major underlying factors that lead to strokes, aiming to control these factors to effectively prevent stroke and similar diseases \citep{o2016global}. In this context, a multicenter case-control study with global coverage was conducted by \citet{o2010risk}, which explored the risk factors for stroke and their variations across different regions and populations. Many researchers have built on the data from \citet{o2010risk} to investigate the likelihood of various factors contributing to stroke through different methods. Notable studies include those by~\citet{o2017interstroke, smyth2022anger, smyth2023alcohol}. They analyze these factors using correlation-based metrics, while we analyze them from a perspective of attribution in causal inference. This allows us to rigorously examine the likelihood of these factors contributing to stroke, addressing the attribution problem. In this section, we continue to use data from the \citet{o2010risk} study to explore the risk factors contributing to stroke in different regions of the world. 

\subsection{Data Description and Setup}
The dataset from \citet{o2010risk}, referred to as INTERSTROKE, includes data from over 27,000 participants in 22 countries worldwide, collected from 2007 to 2010. This dataset comprises both stroke patients and matched controls, thus Assumption \ref{assump 1} is likely to hold. Following \citet{ferguson2023estimating}, we cleaned the data and removed observations with missing values. The final dataset included $ n = 13,712$ observations, with 6,856 stroke cases and 6,856 controls. For each observation, we used key indicators from the INTERSTROKE dataset as variables in our application study. The results are summarized in Table \ref{tab 5} below.

\begin{table}[htbp]
  \centering
  \caption{Indicators involved in application study}
    \begin{tabular}{cccrcc}
    \toprule
    No.   & indicators & values and meanings & \multicolumn{1}{c}{No.} & indicators & values and meanings \\
    \midrule
    1     & case & 0: no stroke; 1: stroke & \multicolumn{1}{c}{9} & sex   & 0: male; 1:female \\
    2     & region & 1-7, Geographic region\footnote{1: Western Europe, 2: Eastern/Central Europe/Middle East 3: Africa, 4: South Asia, 5: China, 6: South East Asia, 7: South America.} & \multicolumn{1}{c}{10} & age   & 0-100, age of individual \\
    3     & smoking & 0: Never, 1: Current & \multicolumn{1}{c}{11} & whr & 0-2, waist hip ratio \\
    4     & stress & 0: never; 1: sometimes & \multicolumn{1}{c}{12} & alcohol & 1:never; 2:low; 3: high intake \\
    5     & exercise &  0: active; 1: inactive & \multicolumn{1}{c}{13} & diet  & 0-100; healthy eating score \\
    6     & diabetes & 0: No; 1: Yes & \multicolumn{1}{c}{14} & lipids & 0-2, ratio of apolipoprotein B to A \\
    7     & heart disease & 0: No; 1: Yes & \multicolumn{1}{c}{15} & education & 1-4; education years\footnote{1: No education, 2: 1-8 years, 3: 9-12 years, 3: Technical college, 4: University.} \\
    8     & hypertension & 0: No; 1: Yes &       &       &  \\
    \bottomrule
    \end{tabular}%
  \label{tab 5}%
\end{table}%
We used the \texttt{case} indicator as the outcome $Y$, where $Y = 1$ indicates having a stroke and $Y = 0$ indicates not having a stroke. We investigated the effects of six factors individually—\texttt{stress}, \texttt{smoking}, \texttt{exercise}, \texttt{diabetes}, \texttt{hypertension}, and \texttt{heart disease}—on stroke. For instance, in studying the effect of stress on stroke, we defined the indicator \texttt{stress} as the cause variable $A$, where $A = 1$ denotes sometimes feeling stress and $A = 0$ denotes no stress. The covariates, denoted as $X$, consist of two components: a primitive set of variables (excluding \texttt{stress} and \texttt{case} as indicated in Table \ref{tab 5}) and a technical variable that represents the interaction terms between the continuous and discrete variables within the primitive set. 
Adding the technical variable allows for better capture of the complex relationships and dependencies between the variables, which can improve the estimation of the nuisance parameters $e(X)$ and $\mu_a(X)$, leading to a more accurate analysis \citep{smithson2013generalized,brauer2018linear,fan2022estimation}.
Similar procedures were applied to the other factors. Given the challenge of satisfying the independence assumption in observational data, we utilize $\hat{\beta}_{mono}$ as our proposed estimator in this section for application studies. To implement our proposed estimation method, we use LASSO regression with main effects to estimate the outcome regression functions $\mu_a(X)$ (for $a = 0,1$) and the propensity score $e(X)$ \citep{tibshirani1996regression,efron2004least,buhlmann2011statistics}. The regularization parameter is selected using 5-fold cross-validation.

\subsection{Different Factors Attributed to Influence Stroke} \label{sec 6.2}

We are currently exploring the global impact of six factors on the likelihood of stroke: \texttt{stress}, \texttt{smoking}, \texttt{exercise}, \texttt{diabetes}, \texttt{hypertension}, and \texttt{heart disease}. Each factor's contribution to stroke risk is being carefully examined. To confirm the efficiency of our proposed estimators $\hat{\beta}_{mono}$, we give the likelihood of stroke caused by these six factors under the IPW and OR methods with our proposed estimators, respectively. The standard error of estimate (ESE) for the proposed method is derived from the estimated asymptotic variance formula of Theorem \ref{asy 1}, while for the other two methods, it is derived from 200 bootstraps.
Also, we give the $p$-values of the corresponding estimators by $H_0 : \beta = 0$ against $H_1 : \beta \neq 0$, a two-sided test.

\begin{table}[htbp]
  \centering
  \caption{Six different factors may contribute to the probability of stroke}
   \begin{tabular}{cccccccccc}
    \toprule
          & \multicolumn{3}{c}{hypertension} & \multicolumn{3}{c}{diabetes} & \multicolumn{3}{c}{heart disease} \\
    \midrule
          & pn.est & ESE   & p-value & pn.est & ESE   & p-value & pn.est & ESE   & p-value \\
    \midrule
    proposed & 0.320 & 0.013 &  $<$ 0.001 & 0.142 & 0.017 & $<$ 0.001 & 0.321 & 0.017 & $<$ 0.001 \\
    OR    & 0.314 & 0.031 & $<$ 0.001 & 0.142 & 0.020 & $<$ 0.001 & 0.325 & 0.026 & $<$ 0.001 \\
    IPW   & 0.330 & 0.013 & $<$ 0.001 & 0.153 & 0.022 & $<$ 0.001 & 0.328 & 0.021 & $<$ 0.001 \\
    \midrule
          & \multicolumn{3}{c}{smoke} & \multicolumn{3}{c}{stress} & \multicolumn{3}{c}{exercise} \\
    \midrule
          & pn.est & ESE   & p-value & pn.est & ESE   & p-value & pn.est & ESE   & p-value \\
    \midrule
    proposed & 0.249 & 0.015 & $<$ 0.001 & 0.169 & 0.017 & $<$ 0.001 & 0.204 & 0.028 & $<$ 0.001 \\
    OR    & 0.244 & 0.025 & $<$ 0.001 & 0.172 & 0.021 & $<$ 0.001 & 0.202 & 0.064 & $<$ 0.001 \\
    IPW   & 0.262 & 0.016 & $<$ 0.001 & 0.202 & 0.022 & $<$ 0.001 & 0.297 & 0.031 & $<$ 0.001 \\
    \bottomrule
    \end{tabular}%
    \begin{tablenotes} 
     \centering
    \footnotesize
    \item Note: $p$-values are obtained by bilateral tests, that is, $H_0 : \beta = 0$ against $H_1 : \beta \neq 0$.
    \end{tablenotes}
  \label{tab 6}%
\end{table}%


Table \ref{tab 6} displays the probabilities of six different factors contributing to stroke across three methods. Based on the data presented in Table \ref{tab 6}, the following conclusions can be drawn: (1) In all scenarios, the $p$-value was less than 0.001, confirming that all six factors play a significant role in causing stroke. (2) In all three methods, the probability values for the factors contributing to stroke are similar, with the ESE of our proposed method slightly smaller, demonstrating the superiority of our approach. (3) Hypertension and heart disease are considered to be the most influential factors leading to stroke. In addition, smoking status and physical activity levels were also identified as significant influencing factors leading to stroke. On the other hand, diabetes and stress levels were determined to be less likely to lead to stroke. These conclusions are essentially similar to the results of the analyses by \citet{o2010risk}, \citet{o2016global}, and \citet{o2017interstroke}.

\subsection{Subgroup Analysis}

In this subsection, we will further explore the differences in the main causes of stroke across different regions through subgroup analysis.    
We maintain the same regional divisions as in the original dataset, with regions 1-7 representing Western Europe, Eastern/Central Europe/Middle East, Africa, South Asia, China, South East Asia, and South America, respectively. In keeping with the previous section, we continue to consider the impact of six factors - hypertension, diabetes, heart disease, smoking status, stress, and exercise - on stroke.

\begin{table}[htbp]
  \centering
  \caption{Importance of six different factors contributing to stroke in different regions}
   \scalebox{0.92}{
       \begin{tabular}{ccccccccccccc}
    \toprule
          & \multicolumn{2}{c}{hypertension} & \multicolumn{2}{c}{diabetes} & \multicolumn{2}{c}{heart disease} & \multicolumn{2}{c}{smoke} & \multicolumn{2}{c}{stress} & \multicolumn{2}{c}{exercise} \\
    \midrule
    region & pn.est & ESE   & pn.est & ESE   & pn.est & ESE   & pn.est & ESE   & pn.est & ESE   & pn.est & ESE \\
    \midrule
    1     & 0.291 & 0.031 & 0.123 & 0.042 & 0.298 & 0.032 & 0.193 & 0.044 & 0.169 & 0.036 & 0.206 & 0.032 \\
    2     & 0.352 & 0.050 & 0.184 & 0.043 & 0.332 & 0.039 & 0.197 & 0.061 & 0.193 & 0.045 & 0.281 & 0.068 \\
    3     & 0.323 & 0.043 & 0.076 & 0.089 & 0.279 & 0.105 & 0.283 & 0.061 & 0.119 & 0.060 & 0.103 & 0.090 \\
    4     & 0.329 & 0.029 & 0.086 & 0.052 & 0.325 & 0.068 & 0.202 & 0.034 & 0.158 & 0.049 & 0.185 & 0.078 \\
    5     & 0.285 & 0.021 & 0.125 & 0.038 & 0.317 & 0.040 & 0.280 & 0.024 & 0.115 & 0.047 & 0.254 & 0.055 \\
    6     & 0.347 & 0.061 & 0.112 & 0.061 & 0.376 & 0.068 & 0.333 & 0.051 & 0.232 & 0.061 & 0.478 & 0.137 \\
    7     & 0.393 & 0.044 & 0.230 & 0.041 & 0.357 & 0.035 & 0.267 & 0.043 & 0.230 & 0.042 & 0.110 & 0.058 \\
    \bottomrule
    \end{tabular}%
    }
    \begin{tablenotes}[flushleft]
    \footnotesize
    \item Note: The standard error of estimate (ESE) for the proposed method is derived from the estimated asymptotic variance formula of Theorem \ref{asy 1}, while for the other two methods, it is derived from 200 bootstraps.
    \end{tablenotes}
  \label{tab 7}%
\end{table}%

Table \ref{tab 7} illustrates the impact of various factors on stroke incidence across different regions. Hypertension and heart disease consistently emerge as the predominant contributors to stroke in all regions, according to the numerical findings. In Western Europe, Eastern/Central Europe/Middle East, and China, lack of exercise and smoking are identified as significant risk factors associated with increased stroke incidence. These lifestyle-related behaviors markedly elevate the risk of stroke within these populations. In Africa and South Asia, alongside hypertension and heart disease, smoking assumes a prominent role in contributing to stroke incidence, often correlated with regional economic development indicators. Moreover, in Southeast Asia, smoking and stress emerge as substantial contributors to stroke incidence, highlighting their significant prevalence and impact on public health in the region. In South America, in addition to hypertension and heart disease, factors such as smoking, high-stress levels, and diabetes are notable contributors to stroke incidence. These factors are prevalent in the region and exert considerable influence on the heightened stroke risk observed. In summary, our conclusions about the primary influencing factors across different regions align closely with the findings from the analyses by \citep{odonnell2013stroke, o2017interstroke,yusuf2020modifiable}.


\section{Extension to the Probability of Sufficient Causation} \label{sec7}

The methods developed in the previous sections primarily focus on PN. In this section, we extend these methods to estimation and inference on the probability of sufficient causation (PS). 

\subsection{Identification and Estimation of PS}

For ease of presentation, we let $\bar{\mu}_a = \mathbb{E}[\mu_a(X) \cdot (1 - e(X))]$, $\bar{\bar{\mu}}_a = \mathbb{E}[(1 - \mu_a(X)) \cdot (1 - e(X))]$ for a = 0, 1, and $\bar{\mu} = \mathbb{E}[\mu_1(X)\mu_0(X)(1 - e(X))]$.  We first present the identifiability of PS, defined by $\gamma:= \mathbb{P}(Y^1 = 1| A = 0, Y = 0)$, as described in Lemma \ref{lemma2}.

\begin{lemma}[Identifiability of PS] \label{lemma2} It follows that 

(a) under Assumptions \ref{assump 1} and \ref{assump 2}, $\gamma$ is identified as 
\begin{align*}
    \gamma &= \frac{\mathbb{E}[(1 - e(X))(\mu_1(X) - \mu_0(X))]}{\mathbb{E}[(1 - e(X))(1 - \mu_0(X))]} = \frac{\bar\mu_1 - \bar\mu_0}{\bar{\bar\mu}_0};
\end{align*}

(b)  under Assumptions  \ref{assump 1} and \ref{assump 3}, $\gamma$ is identified as 
\begin{align*}
    \gamma = \frac{\mathbb{E}[(1 - e(X))(1 - \mu_0(X))\mu_1(X)]}{\mathbb{E}[(1 - e(X))(1 - \mu_0(X))]} = \frac{\bar{\mu}_1 - \bar{\mu}}{\bar{\bar \mu}_0}.
\end{align*}

\end{lemma}

Lemma \ref{lemma2} provides the identification results for $\gamma$ under two sets of assumptions, similar to Lemma \ref{lemma1}. After establishing identifiability, we then present the efficient influence function for $\gamma$ under the corresponding assumptions. 
Here we only present results for cases where the propensity score is unknown; results for which the propensity score is known are provided in the Supplementary Material. We also let $\eta = \eta(X) := (e(X), \mu_0(X), \mu_1(X)) $ be the nuisance parameters.

\begin{thm}[Efficiency Bounds of PS]  \label{thm5} 
We have that 

(a) under Assumption \ref{assump 1} and \ref{assump 2} the efficient influence function of $\gamma$ is 
\begin{align*}
   &\phi_{\gamma} (Y, A, X; \eta) =  \frac{1}{\bar{\bar{\mu}}_0}\frac{A}{e(X)}(1 - e(X))(Y - \mu_1(X)) - \frac{\bar{\bar{\mu}}_1}{\bar{\bar{\mu}}_0^2}(1 - A)(Y - \mu_0(X)) \\
    &+ [\frac{1}{\bar{\bar{\mu}}_0}(1 - \mu_1(X)) - \frac{\bar{\bar{\mu}}_1}{\bar{\bar{\mu}}_0^2}(1 - \mu_0(X))](A - e(X)) + [ \frac{\bar{\bar{\mu}}_1}{\bar{\bar{\mu}}_0^2}(1 - \mu_0(X)) - \frac{1}{\bar{\bar{\mu}}_0}(1 - \mu_1(X))] (1 - e(X)) \\
    &= \frac{1}{\bar{\bar{\mu}}_0}\frac{A}{e(X)}(1 - e(X))(Y - \mu_1(X))
     - \frac{(1 - \gamma)}{\bar{\bar{\mu}}_0} (1-A)(Y -\mu_0(X)) \\
     &+ \frac{ \mu_0(X) - \mu_1(X) + \gamma(1 -\mu_0(X)) }{ \bar{\bar{\mu}}_0 } (A - 1).
\end{align*}
The associated semiparametric efficiency bound of $\gamma$ is $\mathbb{V}(\phi_{\gamma}(Y, A, X; \eta)) = \mathbb{E}[\phi^2_{\gamma}(Y, A, X; \eta)]$. \\

(b) under Assumption \ref{assump 1} and \ref{assump 3} the efficient influence function of $\gamma$ is
\begin{align*}
   \varphi_{\gamma}(Y, &A, X; \eta) = \ \frac{1}{\bar{\bar{\mu}}_0}\frac{A}{e(X)}(1 - e(X))(Y-\mu_1(X))(1 - \mu_0(X)) + (1 - A)(\frac{\bar{\mu}_1 - \bar{\mu}}{\bar{\bar{\mu}}_0^2} - \frac{1}{\bar{\bar{\mu}}_0}\mu_1(X))(Y - \mu_0(X)) \\
  &+ (\frac{\bar{\mu}_1 - \bar{\mu}}{\bar{\bar{\mu}}_0^2} - \frac{1}{\bar{\bar{\mu}}_0}\mu_1(X))(1 - \mu_0(X))(A - e(X)) + (\frac{1}{\bar{\bar{\mu}}_0}\mu_1(X) - \frac{\bar{\mu}_1 - \bar{\mu}}{\bar{\bar{\mu}}_0^2})(1 - \mu_0(X))(1 - e(X)) \\
  &= \frac{1}{\bar{\bar{\mu}}_0}\frac{A}{e(X)}(1 - e(X))(Y-\mu_1(X))(1 - \mu_0(X)) + (\frac{\gamma}{\bar{\bar{\mu}}_0} - \frac{1}{\bar{\bar{\mu}}_0}\mu_1(X))(1 - Y)(A - 1).  
\end{align*}
The associated semiparametric efficiency bound of $\gamma$ is $\mathbb{V}(\varphi_{\gamma}(Y, A, X; \eta)) = \mathbb{E}[\varphi^2_{\gamma}(Y, A, X; \eta)]$. \\
\end{thm}

Theorem \ref{thm5} presents the specific form of the EIFs and the associated semiparametric efficiency bounds for $\gamma$ under two sets of identifiability assumptions. The EIF of $\gamma$ under Assumptions \ref{assump 1} and \ref{assump 2} takes a different form compared to that under Assumptions \ref{assump 1} and \ref{assump 3}, thus reaffirming that Assumptions \ref{assump 2} and \ref{assump 3} convey different pieces of information on the efficiency of $\gamma$.

Next, based on the EIFs derived in Theorem \ref{thm5}, we construct the proposed estimators for $\gamma$ following procedures similar to those outlined in Algorithm \ref{algorithm1}. 
Specifically, under Assumptions \ref{assump 1} and \ref{assump 2}, the proposed estimator of $\gamma$ is given by 
   \begin{align}\label{ps mono}
    \hat \gamma_{mono} = \frac{ \dfrac{1}{n}\sum\limits_{i=1}^n \Big \{ (1-A_i)Y_i  - \dfrac{A_i}{\hat e(X_i)}(1 - \hat e(X_i))(Y_i - \hat \mu_1(X_i)) +  \hat \mu_1(X_i) (A_i - 1) \Big \}}{ \dfrac{1}{n} \sum\limits_{i=1}^n (1 - A_i)(Y_i - 1)}.  
   \end{align}
Similarly, the proposed estimator under Assumptions \ref{assump 1} and \ref{assump 3} of $\gamma$ is given as 
   \begin{align}\label{ps inde}
    \hat \gamma_{inde} = \frac{ \dfrac{1}{n}\sum\limits_{i=1}^n \Big \{ \hat \mu_1(X_i)(1 - Y_i)(A_i - 1) - \dfrac{A_i}{\hat e(X_i)}(1 - \hat e(X_i))(Y_i-\hat \mu_1(X_i))(1 - \hat \mu_0(X_i)) \Big \}}{ \dfrac{1}{n} \sum\limits_{i=1}^n (1 - Y_i)(A_i - 1)}.
\end{align}
The estimator of $\gamma$ under Assumptions \ref{assump 1} and \ref{assump 3} is slightly more complex compared to those under Assumptions \ref{assump 1} and \ref{assump 2}.

\subsection{Asymptotic Properties}

In this subsection, we present the large sample properties of the proposed estimators $\hat \gamma_{mono}$ and $\hat \gamma_{inde}$. 
We first give the robustness property of them.  

\begin{proposition} We have that 
\label{robustness PS} 

 (a) (Double Robustness) Under Assumptions \ref{assump 1} and \ref{assump 2}, the proposed estimator $\hat{\gamma}_{mono}$ given in equation \eqref{ps mono} is a consistent estimator of $\gamma$ if one of the following conditions is satisfied: (i) the propensity score model is correctly specified; (ii) the outcome regression model is correctly specified.
    
 (b) (Single Robustness) Under Assumptions \ref{assump 1} and \ref{assump 3}, the estimator $\hat{\gamma}_{inde}$ given in equation \eqref{ps inde} is a consistent estimate of $\gamma$, regardless of whether the propensity score model is correctly specified, provided that the outcome regression model is correctly specified.
 
\end{proposition}

Proposition \ref{robustness PS}(a) demonstrates that $\hat{\gamma}_{mono}$ is a doubly robust estimator under Assumptions \ref{assump 1} and \ref{assump 2}. This means that $\hat{\gamma}_{mono}$ provides a consistent estimate of $\gamma$ as long as either the propensity score model $\hat{e}(X)$ or the outcome regression model $\hat{\mu}_a(X)$ for $a = 0, 1$ is correctly specified, a property similar to that of $\hat{\beta}_{mono}$ described in Proposition \ref{double robustness}(a). 
Proposition \ref{robustness PS}(b) gives the single robustness property of $\hat{\gamma}_{inde}$ under Assumptions \ref{assump 1} and \ref{assump 3}. This property is similar to that of $\hat{\beta}_{mono}$ described $\hat \beta_{inde}$ described in Proposition \ref{double robustness}(b). 
Upon comparing the results in Proposition \ref{robustness PS} with those in Proposition \ref{double robustness}, we note that the properties of PN and PS under the two sets of assumptions are highly similar.  

\begin{thm}[Asymptotic Normality and Efficiency]
\label{asy 3} 
We have that 

(a) under Assumptions \ref{assump 1}--\ref{assump 2} and Condition \ref{condition}, 
    \begin{equation*}
        \sqrt{n}(\hat{\gamma}_{mono} - \gamma) \xrightarrow{d} \mathcal{N}(0, \sigma_5^2)
    \end{equation*}
    where  $\sigma_5^2 = \mathbb{V}(\phi_{\gamma}(Z; \eta))$ is the semiparametric efficiency
    bound of $\gamma$. A consistent estimator of $\sigma_5^2$ is $\hat\sigma_5^2 = n^{-1}\sum_{i = 1}^n [\zeta_5(Z_i; \hat{\eta}) - \hat \gamma_{mono}]^2$, where $\zeta_5(Z_i; \hat{\eta}) =  \big \{  (1-A_i)Y_i  - A_i(1 - \hat e(X_i))(Y_i - \hat \mu_1(X_i))/\hat e(X_i) +  \hat \mu_1(X_i) (A_i - 1) \big \}  / (n^{-1}) \sum_{i=1}^n \{(1 - A_i)(Y_i - 1) \} $.\\ 
    
(b) under Assumptions \ref{assump 1}--\ref{assump 3} and Condition \ref{condition}, 
    \begin{equation*}
        \sqrt{n}(\hat{\gamma}_{inde} - \gamma) \xrightarrow{d}  \mathcal{N}(0, \sigma_6^2)
    \end{equation*}
    where $\sigma_6^2 = \mathbb{V}(\varphi_{\gamma}(Z; \eta))$ is the semiparametric efficiency
    bound of $\gamma$. A consistent estimator of $\sigma_6^2$ is $\hat\sigma_6^2 = n^{-1}\sum_{i = 1}^n [\zeta_6(Z_i; \hat{\eta}) - \hat \gamma_{inde}]^2$, where $\zeta_6(Z_i; \hat{\eta}) = \big \{ \hat \mu_1(X_i)(1 - Y_i)(A_i - 1) - A_i(1 - \hat e(X_i))(Y_i-\hat \mu_1(X_i))(1 - \hat \mu_0(X_i))/\hat e(X_i) \big \} / (n^{-1}) \sum_{i=1}^n \{ (1 - Y_i)(A_i - 1)\}$.\\ 
\end{thm}

Theorem \ref{asy 3} shows that our proposed estimates $\hat \gamma_{mono}$ and $\hat \gamma_{inde}$ are consistent and asymptotically normal under mild conditions and the corresponding identifiability assumptions. Also, their asymptotic variances attain the corresponding semiparametric efficient bounds, which indicates the local efficiency. This validates the similarity between PS and PN in terms of theoretical properties. 


\section{Conclusion}  \label{sec8}

In this paper, we focus on the estimation and inference of the probability of necessary causation (PN) and the probability of sufficient causation (PS) under two sets of identifiability assumptions. For each set of identifiability assumptions, we derive the efficient influence functions (EIFs) and the semiparametric efficiency bounds for PN and PS, respectively. Based on these, we propose semiparametrically efficient estimators for PN and PS, and establish their consistency and asymptotic normality. Through extensive numerical simulations and real data applications, we demonstrate the superior performance of the proposed estimators in finite samples. 

The identifiability of PN and PS in the main text relies on either the monotonicity assumption (Assumption \ref{assump 2}) or the conditional independence assumption (Assumption \ref{assump 3}). Both of these assumptions may not hold in practice. To apply to a wider range of scenarios, future research is desirable by weakening these identifiability assumptions. One possible direction is to introduce multiple datasets~\citep{Jiang-etal2016}. When identifiability is difficult to achieve, we might consider improving the bounds under strongly ignorability (Assumption \ref{assump 1}) by introducing additional assumptions, such as multiple outcomes~\citep{mealli2013using, Ying-etal2024} or positive correlation between two potential outcomes~\citep{wu2024quantifying}.
Another interesting extension is to consider multiple cause variables. In the presence of multiple causes, methods that account for the complexity and interactions among these variables would be necessary \citep{tran2017implicit,ranganath2018multiple,kong2022identifiability}.  
\bibliography{ref}

\section*{Appendix}

\section*{Proof of Lemma 1}
Proof of Lemma 1(a). Under Assumptions 1 and 2, $\beta$ is identified as
\begin{align*}
    \beta &= \mathbb{P}(Y^0 = 0|A = 1, Y = 1) \\
    &= \frac{\mathbb{P}(Y^0 = 0, Y^1 = 1| A = 1)}{\mathbb{P}(Y^1 = 1| A = 1)} \\
    &= \frac{\sum_x \mathbb{P}(X = x|A = 1) \cdot \mathbb{P}(Y^0 = 0, Y^1 = 1| A = 1, X = x)}{\sum_x \mathbb{P}(X = x|A = 1) \cdot \mathbb{P}(Y^1 = 1| A = 1, X = x)} \\
    &= \frac{\sum_x e(x)\mathbb{P}(X = x) \cdot \mathbb{P}(Y^0 = 0, Y^1 = 1|X = x)}{\sum_x e(x)\mathbb{P}(X = x) \cdot \mathbb{P}(Y^1 = 1|X = x)} \\
    &=  \frac{\sum_x e(x)\mathbb{P}(X = x) \cdot [\mathbb{P}(Y^1 = 1|X = x) - \mathbb{P}(Y^0 = 1|X = x)]}{\sum_x e(x)\mathbb{P}(X = x) \cdot \mathbb{P}(Y^1 = 1|X = x)} \\
    &= 1 - \frac{ \mathbb{E}[e(X)\mu_0(X)]}{\mathbb{E}[e(X)\mu_1(X)] } \\
    &= 1 - \frac{\mu_0}{\mu_1}.
\end{align*}
Where the fourth equation holds because Assumption 1 and the fifth equation holds because Assumption 2, so $\beta$ can be identified.

Proof of Lemma 1(b). Under Assumptions 1 and 3, $\beta$ is identified as
\begin{align*}
    \beta &= \mathbb{P}(Y^0 = 0|A = 1, Y = 1) \\
    &= \frac{\mathbb{P}(Y^0 = 0, Y^1 = 1| A = 1)}{\mathbb{P}(Y^1 = 1| A = 1)} \\
    &= \frac{\sum_x \mathbb{P}(X = x|A = 1) \cdot \mathbb{P}(Y^0 = 0, Y^1 = 1| A = 1, X = x)}{\sum_x \mathbb{P}(X = x|A = 1) \cdot \mathbb{P}(Y^1 = 1| A = 1, X = x)} \\
    &= \frac{\sum_x e(x)\mathbb{P}(X = x) \cdot \mathbb{P}(Y^0 = 0, Y^1 = 1|X = x)}{\sum_x e(x)\mathbb{P}(X = x) \cdot \mathbb{P}(Y^1 = 1|X = x)} \\
    &=  \frac{\sum_x e(x)\mathbb{P}(X = x) \cdot [\mathbb{P}(Y^1 = 1|X = x)(1 - \mathbb{P}(Y^0 = 1|X = x))]}{\sum_x e(x)\mathbb{P}(X = x) \cdot \mathbb{P}(Y^1 = 1|X = x)} \\
    &= 1 - \frac{\mathbb{E}[e(X)\mu_0(X)\mu_1(X)]}{\mathbb{E}[e(X)\mu_1(X)]} \\
    &= 1 - \frac{\mu}{\mu_1}.
\end{align*}
Where the fourth equation holds because Assumption 1 and the fifth equation holds because Assumption 3, so $\beta$ can be identified.

\section*{Proof of Theorem 1}

Let $f(x)$ be the density function of $X$ and $f(y^0, y^1| x)$ be the joint distribution of $(Y^0, Y^1)$ conditional on $X = x$, then the density of $(Y^0, Y^1, A, X)$ is given by
\begin{align*}
    f(y^0, y^1, a, x) &= f(y^0, y^1, a| x)f(x) \\ 
    &= f(y^0, y^1| x)f(a| x)f(x) \\
    &= [f_1(y| x)e(x)]^a[f_0(y| x)(1 - e(x))]^{1 - a}f(x)
\end{align*}
where $f_1( \cdot | x) = \int f(y^0, \cdot| x)dy^0$ and $f_0( \cdot | x) = \int f(\cdot, y^1| x)dy^1$ are the marginal distribution function of $Y^1$ and $Y^0$ given $X = x$ respectively, $e(x)$ is the propensity score.

When the propensity score $e(X)$ is unknown.Consider a regular parametric submodel indexed by $\theta$,
\begin{equation*}
    f(y, a, x; \theta) = [f_1(y| x,\theta)e(x, \theta)]^a[f_0(y| x, \theta)(1 - e(x, \theta))]^{1 - a}f(x, \theta)
\end{equation*}
which equals $f(y ,a, x)$ when $\theta = \theta_0$. Then the score function for the parametric submodel is given by 
\begin{align*}
    s(y, a, x; \theta) = a \cdot s_1(y|x, \theta) + (1 - a) \cdot s_0(y|x, \theta) + \frac{a - e(x, \theta)}{e(x, \theta)(1 - e(x, \theta))}\dot{e}(x, \theta) + s(x, \theta)
\end{align*}
where $\dot{e}(x, \theta) = \partial e(x, \theta) / \partial \theta$, $s_1(y|x, \theta) = \partial \log f_1(y|x, \theta) / \partial \theta$, $s_0(y|x, \theta) = \partial \log f_0(y|x, \theta) / \partial \theta$, and $s(x, \theta) = \partial \log f(x, \theta) / \partial \theta$.

Thus, the tangent space is 
\begin{equation*}
    \mathcal{T} = \{a \cdot s_1(y|x) + (1 - a) \cdot s_0(y|x) + \alpha(x) \cdot (a - e(x)) + s(x) \}
\end{equation*}
where $\mathbb{E}[s_a(Y| X)| X = x] = \int s_a(y| x)f_a(y| x)dy = 0$ for a = 0,1, $\int s(x)f(x)dx = 0$, and $\alpha(x)$ is an arbitrary square-integrable measurable function of x.

\subsection*{Proof of Theorem 1(a)} 
Proof of Theorem 1(a). Under Assumptions 1 and 2, the parametric submodel indexed by $\theta$, the estimand $\beta$ can be written as 
    \begin{equation*}
        \beta(\theta) = \frac{\int (\int yf_1(y| x, \theta)dy - \int yf_0(y| x, \theta)dy)e(x, \theta)f(x, \theta)dx}{\iint yf_1(y|x, \theta)e(x, \theta)f(x, \theta)dydx }.
    \end{equation*}
The pathwise derivative of $\beta(\theta)$ at $\theta = \theta_0$ is given as

    \begin{align*}
        \frac{\partial \beta(\theta)}{\partial \theta} \bigg|_{\theta = \theta_0} &= \frac{\int (\int ys_1(y| x, \theta_0)f_1(y| x, \theta_0)dy - \int ys_0(y| x, \theta_0)f_0(y| x, \theta_0)dy)e(x, \theta_0)f(x, \theta_0)dx}{\iint yf_1(y|x, \theta_0)e(x, \theta_0)f(x, \theta_0)dydx} \\
         &+\frac{\int (\int yf_1(y| x, \theta_0)dy - \int yf_0(y| x, \theta_0)dy)\dot{e}(x, \theta_0)f(x, \theta_0)dx}{\iint yf_1(y|x, \theta_0)e(x, \theta_0)f(x, \theta_0)dydx} \\
        &+\frac{\int (\int yf_1(y| x, \theta_0)dy - \int yf_0(y| x, \theta_0)dy)e(x, \theta_0)s(x, \theta_0)f(x, \theta_0)dx}{\iint yf_1(y|x, \theta_0)e(x, \theta_0)f(x, \theta_0)dydx} \\
        &- \frac{(\int (\int yf_1(y| x, \theta_0)dy - \int yf_0(y| x, \theta_0)dy)e(x, \theta_0)f(x, \theta_0)dx)(\iint ys_1(y|x, \theta_0)f_1(y|x, \theta_0)e(x, \theta_0)f(x, \theta_0)dydx)}{[\iint yf_1(y|x, \theta_0)e(x, \theta_0)f(x, \theta_0)dydx]^2} \\
        &- \frac{(\int (\int yf_1(y| x, \theta_0)dy - \int yf_0(y| x, \theta_0)dy)e(x, \theta_0)f(x, \theta_0)dx)(\iint y f_1(y|x, \theta_0)\dot{e}(x, \theta_0)f(x, \theta_0)dydx)}{[\iint yf_1(y|x, \theta_0)e(x, \theta_0)f(x, \theta_0)dydx]^2} \\
        &- \frac{(\int (\int yf_1(y| x, \theta_0)dy - \int yf_0(y| x, \theta_0)dy)f(x, \theta_0)dx)(\iint yf_1(y|x, \theta_0)e(x, \theta_0)s(x, \theta_0)f(x, \theta_0)dydx)}{[\iint yf_1(y|x, \theta_0)e(x, \theta_0)f(x, \theta_0)dydx]^2} \\
        &= \frac{\mu_0}{\mu_1^2}\mathbb{E}[e(X)Y^1S_1(Y^1| X)] - \frac{1}{\mu_1}\mathbb{E}[e(X)Y^0S_0(Y^0| X)] \\
        &+ \frac{1}{(\mu_1)^2}\mathbb{E}[(\mu_0\mu_1(X) - \mu_1\mu_0(X))\dot{e}(X)] + \frac{1}{\mu_1^2}\mathbb{E}[(\mu_0\mu_1(X) - \mu_1\mu_0(X))e(X)S(X)]
    \end{align*}

    we first show that $\phi_\beta(Y, A, X; \eta)$ is an influence function for $\beta$, which suffices to verify that
    \begin{equation}\label{unknown case 1}
        \frac{\partial \beta(\theta)}{\partial \theta} \bigg|_{\theta = \theta_0} = \mathbb{E}[\phi_\beta(Y, A, X; \eta) \cdot s(Y, A, X; \theta_0)]
    \end{equation}

    The right side of equation (\ref{unknown case 1}) can be decomposed as follows,

    \begin{equation*}
       \mathbb{E}[\phi_\beta(Y, A, X; \eta) \cdot s(Y, A, X; \theta_0)] = B_1 + B_2 + B_3 + B_4,
    \end{equation*}
    
    where 

    \begin{align*}
        B_1 &= \mathbb{E}[\phi_\beta(Y, A, X; \eta) \cdot A \cdot S_1(Y|X)] \\
        B_2 &= \mathbb{E}[\phi_\beta(Y, A, X; \eta) \cdot (1 - A) \cdot S_0(Y|X)] \\
        B_3 &= \mathbb{E}[\phi_\beta(Y, A, X; \eta) \cdot \frac{A - e(X)}{e(X)(1 - e(X))}\dot{e}(X)] \\
        B_4 &= \mathbb{E}[\phi_\beta(Y, A, X; \eta) \cdot S(X)]
    \end{align*}

    We analyze $B_1, B_2, B_3$, and $B_4$ one by one. Since $\mathbb{E}[s_a(Y|X)|X = x] = 0$ for a = 0, 1, and \ $\mathbb{E}[S(X)] = 0$ we have

    \begin{align*}
        B_1 &= \mathbb{E}[\phi_\beta(Y, A, X; \eta) \cdot A \cdot S_1(Y|X)] \\
        &= \mathbb{E}[\{ \frac{\mu_0}{\mu_1^2}A(Y-\mu_1(X)) - \frac{1}{\mu_1}\frac{1-A}{1-e(X)}(Y - \mu_0(X))e(X) + \frac{1}{\mu_1^2}[\mu_0\mu_1(X) - \mu_1\mu_0(X)](A - e(X)) \\
        &+ \frac{1}{\mu_1^2}[\mu_0\mu_1(X) - \mu_1\mu_0(X)]e(X) \} \times A \cdot S_1(Y|X)] \\
        &= \mathbb{E}[\frac{\mu_0}{\mu_1^2}A(Y-\mu_1(X)) S_1(Y|X)] \\
        &= \mathbb{E}[\frac{\mu_0}{\mu_1^2}e(X)Y^1
        S_1(Y|X)] \\
        &= the \ first \ term \ of \ \frac{\partial \beta(\theta)}{\partial \theta} \bigg|_{\theta = \theta_0}.
    \end{align*}

    Likewise,
    \begin{align*}
        B_2 &= \mathbb{E}[\phi_\beta(Y, A, X; \eta) \cdot (1 - A) \cdot S_0(Y|X)] \\
        &= \mathbb{E}[\{ \frac{\mu_0}{\mu_1^2}A(Y-\mu_1(X)) - \frac{1}{\mu_1}\frac{1-A}{1-e(X)}(Y - \mu_0(X))e(X) + \frac{1}{\mu_1^2}[\mu_0\mu_1(X) - \mu_1\mu_0(X)](A - e(X)) \\
        &+ \frac{1}{\mu_1^2}[\mu_0\mu_1(X) - \mu_1\mu_0(X)]e(X) \} \times (1 - A) \cdot S_0(Y|X)] \\
        &= \mathbb{E}[- \frac{1}{\mu_1}\frac{1-A}{1-e(X)} e(X) \cdot Y \cdot S_0(Y|X)] \\
        &= \mathbb{E}[- \frac{1}{\mu_1} \cdot e(X) \cdot Y^0 S_0(Y^0|X)] \\
        &= the \ second \ term \ of \ \frac{\partial \beta(\theta)}{\partial \theta} \bigg|_{\theta = \theta_0}.
    \end{align*}

    In addition,
    \begin{align*}
        B_3 &= \mathbb{E}[\phi_\beta(Y, A, X; \eta) \cdot \frac{A - e(X)}{e(X)(1 - e(X))}\dot{e}(X)] \\
        &= \mathbb{E}[\{ \frac{\mu_0}{\mu_1^2}A(Y-\mu_1(X)) - \frac{1}{\mu_1}\frac{1-A}{1-e(X)}(Y - \mu_0(X))e(X) + \frac{1}{\mu_1^2}[\mu_0\mu_1(X) - \mu_1\mu_0(X)](A - e(X)) \\
        &+ \frac{1}{\mu_1^2}[\mu_0\mu_1(X) - \mu_1\mu_0(X)]e(X) \} \times \frac{A - e(X)}{e(X)(1 - e(X))}\dot{e}(X)] \\
        &= \mathbb{E}\{ [\frac{1}{\mu_1^2}[\mu_1(X) - \mu_1\mu_0(X)](A - e(X))] \times \frac{A - e(X)}{e(X)(1 - e(X))}\dot{e}(X)\} \\
        &= \mathbb{E}[\frac{1}{\mu_1^2}[\mu_0\mu_1(X) - \mu_1\mu_0(X)]A \times \frac{A - e(X)}{e(X)(1 - e(X))}\dot{e}(X)] \\
        &= \mathbb{E}[\frac{1}{\mu_1^2}(\mu_0\mu_1(X) - \mu_1\mu_0(X)) \cdot \dot{e}(X)] \\
        &= the \ third \ term \ of \ \frac{\partial \beta(\theta)}{\partial \theta} \bigg|_{\theta = \theta_0}.
    \end{align*}

    where the last equation follows from the law of iterated expectations. Also,

    \begin{align*}
        B_4 &= \mathbb{E}[\phi_\beta(Y, A, X; \eta) \cdot S(X)] \\
        &= \mathbb{E}[\frac{1}{\mu_1^2}[\mu_0\mu_1(X) - \mu_1\mu_0(X)]e(X) \cdot S(X)] \\
        &= the \ fourth \ term \ of \ \frac{\partial \beta(\theta)}{\partial \theta} \bigg|_{\theta = \theta_0}.
    \end{align*}

    Combing the results of $B_1, B_2, B_3$, and $B_4$ leads to equation (\ref{unknown case 1}). In addition, let $s_1(Y|X) = \dfrac{\mu_0}{\mu_1^2}(Y-\mu_1(X))$, $s_0(Y|X) = - \dfrac{1}{\mu_1}\dfrac{1}{1-e(X)}(Y - \mu_0(X))e(X)$, $\alpha(x) = \dfrac{1}{\mu_1^2}(\mu_0\mu_1(X) - \mu_1\mu_0(X))$ and $s(X) = \dfrac{1}{\mu_1^2}[\mu_0\mu_1(X) - \mu_1\mu_0(X)]e(X)$, then 
    \begin{align*}
        \phi_\beta(Y, A, X; \eta) = A \cdot s_1(Y|X) + (1 - A) \cdot s_0(Y|X) +\alpha(X)(A - e(X)) + s(X),
    \end{align*}

    which implies that $\phi_\beta(Y, A, X; \eta) \in \mathcal{T}$. Thus, the influence function $\phi_\beta(Y, A, X; \eta)$ is efficient.

\subsection*{Proof of Theorem 1(b)} 
Proof of Theorem 1(b). Under Assumptions 1 and 3, the parametric submodel indexed by $\theta$, the estimand $\beta$ can be written as
\begin{equation*}
        \beta(\theta) = 1 - \frac{\int (\int yf_1(y| x, \theta)dy)(\int yf_0(y| x, \theta)dy)e(x, \theta)f(x, \theta)dx}{\iint yf_1(y|x, \theta)e(x, \theta)f(x, \theta)dydx }
    \end{equation*}

    The pathwise derivative of $\beta(\theta)$ at $\theta = \theta_0$ is given as

    \begin{align*}
        \frac{\partial \beta(\theta)}{\partial \theta}& \bigg|_{\theta = \theta_0} = \ - \frac{\int (\int ys_1(y |x, \theta_0)f_1(y| x, \theta_0)dy)(\int yf_0(y| x, \theta_0)dy)e(x, \theta_0)f(x, \theta_0)dx}{\iint yf_1(y|x, \theta_0)e(x, \theta_0)f(x, \theta_0)dydx } \\
        & - \frac{\int (\int yf_1(y| x, \theta_0)dy)(\int ys_0(y|x, \theta_0)f_0(y| x, \theta_0)dy)e(x, \theta_0)f(x, \theta_0)dx}{\iint yf_1(y|x, \theta_0)e(x, \theta_0)f(x, \theta_0)dydx } \\
        & - \frac{\int (\int yf_1(y| x, \theta_0)dy)(\int yf_0(y| x, \theta_0)dy)\dot{e}(x, \theta_0)f(x, \theta_0)dx}{\iint yf_1(y|x, \theta_0)e(x, \theta_0)f(x, \theta_0)dydx } \\
        & - \frac{\int (\int yf_1(y| x, \theta_0)dy)(\int yf_0(y| x, \theta_0)dy)e(x, \theta_0)s(x, \theta_0)f(x, \theta_0)dx}{\iint yf_1(y|x, \theta_0)e(x, \theta_0)f(x, \theta_0)dydx } \\
        &+ \frac{[\int (\int yf_1(y| x, \theta_0)dy)(\int yf_0(y| x, \theta_0)dy)e(x, \theta_0)f(x, \theta_0)dx][\iint ys_1(y|x, \theta_0)f_1(y|x, \theta_0)e(x, \theta_0)f(x, \theta_0)dydx]}{[\iint yf_1(y|x, \theta_0)e(x, \theta_0)f(x, \theta_0)dydx]^2 } \\
        &+ \frac{[\int (\int yf_1(y| x, \theta_0)dy)(\int yf_0(y| x, \theta_0)dy)e(x, \theta_0)f(x, \theta_0)dx][\iint yf_1(y|x, \theta_0)\dot{e}(x, \theta_0)f(x, \theta_0)dydx]}{[\iint yf_1(y|x, \theta_0)e(x, \theta_0)f(x, \theta_0)dydx]^2 } \\
        &+ \frac{[\int (\int yf_1(y| x, \theta_0)dy)(\int yf_0(y| x, \theta_0)dy)e(x, \theta_0)f(x, \theta_0)dx][\iint yf_1(y|x, \theta_0)e(x, \theta_0)s(x, \theta_0)f(x, \theta_0)dydx]}{[\iint yf_1(y|x, \theta_0)e(x, \theta_0)f(x, \theta_0)dydx]^2 } \\
        &= \mathbb{E}[\frac{\mu - \mu_1\mu_0(X)}{(\mu_1)^2}e(X)Y^1S_1(Y^1| X)] - \frac{1}{\mu_1}\mathbb{E}[\mu_1(X)e(X)Y^0S_0(Y^0| X)] \\
        &+ \mathbb{E}[\frac{\mu - \mu_1\mu_0(X)}{(\mu_1)^2} \mu_1(X)\dot{e}(X)] + \mathbb{E}[\frac{\mu - \mu_1\mu_0(X)}{(\mu_1)^2}\mu_1(X)e(X)S(X)]
    \end{align*}

    we first show that $\varphi_{\beta}(Y, A, X; \eta)$ is an influence function for $\beta$, which suffices to verify that
    \begin{equation}\label{unknown case 2}
        \frac{\partial \beta(\theta)}{\partial \theta} \bigg|_{\theta = \theta_0} = \mathbb{E}[\varphi_{\beta}(Y, A, X; \eta) \cdot s(Y, A, X; \theta_0)]
    \end{equation}

    The right side of equation (\ref{unknown case 2}) can be decomposed as follows,

    \begin{equation*}
       \mathbb{E}[\varphi_{\beta}(Y, A, X; \eta) \cdot s(Y, A, X; \theta_0)] = B_5 + B_6 + B_7 + B_8,
    \end{equation*}
    
    where 

    \begin{align*}
        B_5 &= \mathbb{E}[\varphi_{\beta}(Y, A, X; \eta) \cdot A \cdot S_1(Y|X)] \\
        B_6 &= \mathbb{E}[\varphi_{\beta}(Y, A, X; \eta) \cdot (1 - A) \cdot S_0(Y|X)] \\
        B_7 &= \mathbb{E}[\varphi_{\beta}(Y, A, X; \eta) \cdot \frac{A - e(X)}{e(X)(1 - e(X))}\dot{e}(X)] \\
        B_8 &= \mathbb{E}[\varphi_{\beta}(Y, A, X; \eta) \cdot S(X)]
    \end{align*}

    We analyze $B_5, B_6, B_7$, and $B_8$ one by one. Since $\mathbb{E}[s_a(Y|X)|X = x] = 0$ for a = 0, 1, and \ $\mathbb{E}[s(X)] = 0$ we have

    \begin{align*}
        B_5 &= \mathbb{E}[\varphi_{\beta}(Y, A, X; \eta) \cdot A \cdot S_1(Y|X)] \\
        &= \mathbb{E}[\{ A \cdot \dfrac{\mu - \mu_1\mu_0(X)}{\mu_1^2}(Y-\mu_1(X)) - \dfrac{e(X)}{\mu_1}\dfrac{1-A}{1-e(X)}(Y - \mu_0(X))\mu_1(X) \\
        &+ \dfrac{\mu - \mu_1\mu_0(X)}{\mu_1^2}\mu_1(X)(A - e(X)) + \dfrac{\mu - \mu_1\mu_0(X)}{\mu_1^2}\mu_1(X)e(X)\} \times A \cdot S_1(Y|X)]
        \\
        &= \mathbb{E}[ A \cdot \dfrac{\mu - \mu_1\mu_0(X)}{\mu_1^2}(Y-\mu_1(X)) S_1(Y|X)] \\
        &= \mathbb{E}[\dfrac{\mu - \mu_1\mu_0(X)}{\mu_1^2}e(X)Y^1
        S_1(Y|X)] \\
        &= the \ first \ term \ of \ \frac{\partial \beta(\theta)}{\partial \theta} \bigg|_{\theta = \theta_0}.
    \end{align*}

    Likewise,
    \begin{align*}
        B_6 &= \mathbb{E}[\varphi_{\beta}(Y, A, X; \eta) \cdot (1 - A) \cdot S_0(Y|X)] \\
        &= \mathbb{E}[\{ A \cdot \dfrac{\mu - \mu_1\mu_0(X)}{\mu_1^2}(Y-\mu_1(X)) - \dfrac{e(X)}{\mu_1}\dfrac{1-A}{1-e(X)}(Y - \mu_0(X))\mu_1(X) \\
        &+ \dfrac{\mu - \mu_1\mu_0(X)}{\mu_1^2}\mu_1(X)(A - e(X)) + \dfrac{\mu - \mu_1\mu_0(X)}{\mu_1^2}\mu_1(X)e(X) \} \times (1 - A) \cdot S_0(Y|X)]
        \\
        &= \mathbb{E}[- \dfrac{e(X)}{\mu_1}\dfrac{1-A}{1-e(X)} (Y - \mu_0(X)) \mu_1(X) S_0(Y|X)] \\
        &= \mathbb{E}[- \dfrac{e(X)}{\mu_1}\mu_1(X) Y^0 S_0(Y|X)] \\
        &= the \ second \ term \ of \ \frac{\partial \beta(\theta)}{\partial \theta} \bigg|_{\theta = \theta_0}.
    \end{align*}

    In addition,
    \begin{align*}
        B_7 &= \mathbb{E}[\varphi_{\beta}(Y, A, X; \eta) \cdot \frac{A - e(X)}{e(X)(1 - e(X))}\dot{e}(X)] \\
        &= \mathbb{E}[\{ A \cdot \dfrac{\mu - \mu_1\mu_0(X)}{\mu_1^2}(Y-\mu_1(X)) - \dfrac{e(X)}{\mu_1}\dfrac{1-A}{1-e(X)}(Y - \mu_0(X))\mu_1(X) \\
        &+ \dfrac{\mu - \mu_1\mu_0(X)}{\mu_1^2}\mu_1(X)(A - e(X)) + \dfrac{\mu - \mu_1\mu_0(X)}{\mu_1^2}\mu_1(X)e(X) \} \times \frac{A - e(X)}{e(X)(1 - e(X))}\dot{e}(X)] \\
        &= \mathbb{E}[\dfrac{\mu - \mu_1\mu_0(X)}{\mu_1^2}\mu_1(X)A \cdot \frac{A - e(X)}{e(X)(1 - e(X))}\dot{e}(X)] \\
        &= \mathbb{E}[\dfrac{\mu - \mu_1\mu_0(X)}{\mu_1^2}\mu_1(X) \dot{e}(X)] \\
        &= the \ third \ term \ of \ \frac{\partial \beta(\theta)}{\partial \theta} \bigg|_{\theta = \theta_0}.
    \end{align*}

    where the last equation follows from the law of iterated expectations. Also,

    \begin{align*}
        B_8 &= \mathbb{E}[\varphi_{\beta}(Y, A, X; \eta) \cdot S(X)] \\
        &= \mathbb{E}[\{ A \cdot \dfrac{\mu - \mu_1\mu_0(X)}{\mu_1^2}(Y-\mu_1(X)) - \dfrac{e(X)}{\mu_1}\dfrac{1-A}{1-e(X)}(Y - \mu_0(X))\mu_1(X) \\
        &+ \dfrac{\mu - \mu_1\mu_0(X)}{\mu_1^2}\mu_1(X)(A - e(X)) + \dfrac{\mu - \mu_1\mu_0(X)}{\mu_1^2}\mu_1(X)e(X) \} \times S(X)] \\
        &= \mathbb{E}[\dfrac{\mu - \mu_1\mu_0(X)}{\mu_1^2}\mu_1(X) \cdot A \times S(X)] \\
        &= \mathbb{E}[\dfrac{\mu - \mu_1\mu_0(X)}{\mu_1^2}\mu_1(X) \cdot e(X) S(X)] \\
        &= the \ fourth \ term \ of \ \frac{\partial \beta(\theta)}{\partial \theta} \bigg|_{\theta = \theta_0}.
    \end{align*}

    Combing the results of $B_5, B_6, B_7$, and $B_8$ leads to equation (\ref{unknown case 2}). In addition, let $s_1(Y|X) = \dfrac{\mu - \mu_1\mu_0(X)}{\mu_1^2}(Y-\mu_1(X))$, $s_0(Y|X) = - \dfrac{e(X)}{\mu_1}\dfrac{1}{1-e(X)}(Y - \mu_0(X))\mu_1(X)$, $\alpha(X) = \dfrac{\mu - \mu_1\mu_0(X)}{\mu_1^2}\mu_1(X)$ and $s(X) = \dfrac{\mu - \mu_1\mu_0(X)}{\mu_1^2}\mu_1(X)e(X)$, then 
    \begin{align*}
        \varphi_{\beta}(Y, A, X; \eta) = A \cdot s_1(Y|X) + (1 - A) \cdot s_0(Y|X) + \alpha(X)(A - e(X)) + s(X),
    \end{align*}

    which implies that $\varphi_{\beta}(Y, A, X; \eta) \in \mathcal{T}$. Thus, the influence function $\varphi_{\beta}(Y, A, X; \eta)$ is efficient.

\section*{Proof of Corollary 1}
Under Assumptions 1 and 2, the semiparametric efficiency bound of $\beta$ is
\begin{align*}
    \mathbb{V}(\phi_{\beta}(Y, A, X; \eta)) &= \mathbb{E}[\phi_{\beta}^2(Y, A, X; \eta) ] \\ &= \mathbb{E}[\frac{(1-\beta)^2}{\mu_1^2} A Y^2 + \frac{e^2(X)}{\mu_1^2} \frac{(1-A)}{(1 - e(X))^2}(Y-\mu_0(X))^2  +   \frac{  \mu_0^2(X) }{\mu_1^2} A - 2\frac{(1-\beta)\mu_0(X)}{\mu_1^2} A Y ]. 
\end{align*}

Under Assumptions 1 and 3, the semiparametric efficiency bound of $\beta$ is
\begin{align*}
    \mathbb{V}(\varphi_{\beta}(Y, A, X; \eta)) &= \mathbb{E}[\varphi_{\beta}^2(Y, A, X; \eta) ] \\
    &= \mathbb{E}[A \cdot \dfrac{(1 - \beta - \mu_0(X))^2}{\mu_1^2}Y^2 + \dfrac{e^2(X)}{\mu^2_1}\dfrac{1-A}{(1-e(X))^2}(Y - \mu_0(X))^2\mu^2_1(X)].
\end{align*}

The difference between $\mathbb{V}(\phi_{\beta}(Y, A, X; \eta))$ and $\mathbb{V}(\varphi_{\beta}(Y, A, X; \eta))$ is 
\begin{align*}
    \mathbb{V}(\phi_{\beta}(Y, &A, X; \eta)) - \mathbb{V}(\varphi_{\beta}(Y, A, X; \eta)) \\ &= \mathbb{E}[\frac{(1-\beta)^2}{\mu_1^2} A Y^2 + \frac{e^2(X)}{\mu_1^2} \frac{(1-A)}{(1 - e(X))^2}(Y-\mu_0(X))^2  +   \frac{  \mu_0^2(X) }{\mu_1^2} A - 2\frac{(1-\beta)\mu_0(X)}{\mu_1^2} A Y ] - \\ & \quad \ \mathbb{E}[A \cdot \dfrac{(1 - \beta - \mu_0(X))^2}{\mu_1^2}Y^2 + \dfrac{e^2(X)}{\mu^2_1}\dfrac{1-A}{(1-e(X))^2}(Y - \mu_0(X))^2\mu^2_1(X)] \\
    &= \mathbb{E}[\frac{A\mu_0^2(X)}{\mu_1^2}(1 - Y^2) + \frac{(1 - A)e^2(X)}{\mu_1^2(1 - e(X))^2}(Y - \mu_0(X))^2(1 - \mu_1^2(X)) + 2\frac{A(1 - \beta)}{\mu_1^2}\mu_0(X)Y(Y - 1)] \\
    &= \mathbb{E}[\frac{A\mu_0^2(X)}{\mu_1^2}(1 - Y^2) + \frac{(1 - A)e^2(X)}{\mu_1^2(1 - e(X))^2}(Y - \mu_0(X))^2(1 - \mu_1^2(X))]
\end{align*}

\section*{Proof of Theorem 2}

Let $f(x)$ be the density function of $X$ and $f(y^0, y^1| x)$ be the joint distribution of $(Y^0, Y^1)$ conditional on $X = x$, then the density of $(Y^0, Y^1, A, X)$ is given by
\begin{align*}
    f(y^0, y^1, a, x) &= f(y^0, y^1, a| x)f(x) \\ 
    &= f(y^0, y^1| x)f(a| x)f(x) \\
    &= [f_1(y| x)e(x)]^a[f_0(y| x)(1 - e(x))]^{1 - a}f(x),
\end{align*}
where $f_1( \cdot | x) = \int f(y^0, \cdot| x)dy^0$ and $f_0( \cdot | x) = \int f(\cdot, y^1| x)dy^1$ are the marginal distribution function of $Y^1$ and $Y^0$ given $X = x$ respectively, $e(x)$ is the propensity score. When the true propensity score $e(X)$ is known, the parametric submodel indexed by $\theta$ is, 
    \begin{equation*}
        f(y, a, x; \theta) = [f_1(y| x,\theta)e(x)]^a[f_0(y| x, \theta)(1 - e(x))]^{1 - a}f(x, \theta),
    \end{equation*}
    which equals $f(y ,a, x)$ when $\theta = \theta_0$. The associated score function for the parametric submodel becomes
    \begin{equation*}
        s(y, a, x; \theta) = a \cdot s_1(y|x, \theta) + (1 - a) \cdot s_0(y|x, \theta) + s(x, \theta).
    \end{equation*}
    Thus, the tangent space is 
    \begin{equation*}
    \mathcal{T} = \{a \cdot s_1(y|x) + (1 - a) \cdot s_0(y|x) + s(x) \},
    \end{equation*}

    where $\mathbb{E}[s_a(Y| X)| X = x] = \int s_a(y| x)f_a(y| x)dy = 0$ for a = 0,1, $\int s(x)f(x)dx = 0$.

\subsection*{Proof of Theorem 2(a)}
Proof of Theorem 2(a). Under Assumptions 1 and 2, when the propensity score $e(X)$ is known, the parametric submodel indexed by $\theta$, the
estimand $\beta$ can be written as 
\begin{equation*}
        \beta(\theta) = \frac{\int (\int yf_1(y| x, \theta)dy - \int yf_0(y| x, \theta)dy)e(x)f(x, \theta)dx}{\iint yf_1(y|x, \theta)e(x)f(x, \theta)dydx }
    \end{equation*}

    The pathwise derivative of $\beta(\theta)$ at $\theta = \theta_0$ is given as

    \begin{align*}
        \frac{\partial \beta(\theta)}{\partial \theta} &\bigg|_{\theta = \theta_0} =\frac{\int (\int ys_1(y| x, \theta_0)f_1(y| x, \theta_0)dy - \int ys_0(y| x, \theta_0)f_0(y| x, \theta_0)dy)e(x)f(x, \theta_0)dx}{\iint yf_1(y|x, \theta_0)e(x)f(x, \theta_0)dydx} \\
        &+\frac{\int (\int yf_1(y| x, \theta_0)dy - \int yf_0(y| x, \theta_0)dy)s(x, \theta_0)e(x)f(x, \theta_0)dx}{\iint yf_1(y|x, \theta_0)e(x)f(x, \theta_0)dydx} \\
        &- \frac{(\int (\int yf_1(y| x, \theta_0)dy - \int yf_0(y| x, \theta_0)dy)e(x)f(x, \theta_0)dx)(\iint ys_1(y|x, \theta_0)f_1(y|x, \theta_0)e(x)f(x, \theta_0)dydx)}{[\iint yf_1(y|x, \theta_0)e(x)f(x, \theta_0)dydx]^2} \\
        &- \frac{(\int (\int yf_1(y| x, \theta_0)dy - \int yf_0(y| x, \theta_0)dy)e(x)f(x, \theta_0)dx)(\iint yf_1(y|x, \theta_0)s(x, \theta_0)e(x)f(x, \theta_0)dydx)}{[\iint yf_1(y|x, \theta_0)e(x)f(x, \theta_0)dydx]^2} \\
        &= \frac{\mu_0}{(\mu_1)^2}\mathbb{E}[e(X)Y^1S_1(Y^1| X)] - \frac{1}{\mu_1}\mathbb{E}[e(X)Y^0S_0(Y^0| X)] \\
        &+ \frac{\mu_0}{(\mu_1)^2}\mathbb{E}[e(x)\mu_1(X)S(X)] - \frac{1}{\mu_1}\mathbb{E}[e(x)\mu_0(X)S(X)]
    \end{align*}

    we first show that $\tilde \phi_\beta(Y, A, X; \eta)$ is an influence function for $\beta$, which suffices to verify that
    \begin{equation}\label{ known case 1}
        \frac{\partial \beta(\theta)}{\partial \theta} \bigg|_{\theta = \theta_0} = \mathbb{E}[\tilde \phi_\beta(Y, A, X; \eta) \cdot s(Y, A, X; \theta_0)]
    \end{equation}

    The right side of equation (\ref{ known case 1}) can be decomposed as follows,

    \begin{equation*}
       \mathbb{E}[\tilde \phi_\beta(Y, A, X; \eta) \cdot s(Y, A, X; \theta_0)] = C_1 + C_2 + C_3,
    \end{equation*}
    
    where 

    \begin{align*}
        C_1 &= \mathbb{E}[\tilde \phi_\beta(Y, A, X; \eta) \cdot A \cdot S_1(Y|X)] \\
        C_2 &= \mathbb{E}[\tilde \phi_\beta(Y, A, X; \eta) \cdot (1 - A) \cdot S_0(Y|X)] \\
        C_3 &= \mathbb{E}[\tilde \phi_\beta(Y, A, X; \eta) \cdot S(X)]
    \end{align*}

    We analyze $C_1, C_2$, and $C_3$ one by one. Since $\mathbb{E}[s_a(Y|X)|X = x] = 0$ for a = 0, 1, and \ $\mathbb{E}[S(X)] = 0$ we have

    \begin{align*}
        C_1 &= \mathbb{E}[\tilde \phi_\beta(Y, A, X; \eta) \cdot A \cdot S_1(Y|X)] \\
        &= \mathbb{E}[\{ \frac{\mu_0}{\mu_1^2}A(Y-\mu_1(X)) - \frac{e(X)}{\mu_1}\frac{1-A}{1-e(X)}(Y - \mu_0(X)) + \frac{\mu_0}{\mu_1^2}\mu_1(X)e(X) \\
        &\quad \ - \frac{1}{\mu_1}\mu_0(X)e(X)\} \times A \cdot S_1(Y|X)] \\
        &= \mathbb{E}[\frac{\mu_0}{\mu_1^2}A(Y-\mu_1(X))Y S_1(Y|X)] \\
        &= \mathbb{E}[\frac{\mu_0}{\mu_1^2}e(X)Y^1
        S_1(Y|X)] \\
        &= the \ first \ term \ of \ \frac{\partial \beta(\theta)}{\partial \theta} \bigg|_{\theta = \theta_0}.
    \end{align*}

    Likewise,
    \begin{align*}
        C_2 &= \mathbb{E}[\tilde \phi_\beta(Y, A, X; \eta) \cdot (1 - A) \cdot S_0(Y|X)] \\
        &= \mathbb{E}[\{ \frac{\mu_0}{\mu_1^2}A(Y-\mu_1(X)) - \frac{e(X)}{\mu_1}\frac{1-A}{1-e(X)}(Y - \mu_0(X)) + \frac{\mu_0}{\mu_1^2}\mu_1(X)e(X) \\
        &\quad \ - \frac{1}{\mu_1}\mu_0(X)e(X)\} \times (1 - A) \cdot S_0(Y|X)] \\
        &= \mathbb{E}[- \frac{e(X)}{\mu_1}\frac{1-A}{1-e(X)} Y S_0(Y|X)] \\
        &= \mathbb{E}[- \frac{e(X)}{\mu_1} Y^0 S_0(Y^0|X)] \\
        &= the \ second \ term \ of \ \frac{\partial \beta(\theta)}{\partial \theta} \bigg|_{\theta = \theta_0}.
    \end{align*}

    where the last equation follows from the law of iterated expectations. Also,

    \begin{align*}
        C_3 &= \mathbb{E}[\tilde \phi_\beta(Y, A, X; \eta) \cdot S(X)] \\
        &= \mathbb{E}[\{ \frac{\mu_0}{\mu_1^2}A(Y-\mu_1(X)) - \frac{e(X)}{\mu_1}\frac{1-A}{1-e(X)}(Y - \mu_0(X)) + \frac{\mu_0}{\mu_1^2}\mu_1(X)e(X) \\
        &\quad \ - \frac{1}{\mu_1}\mu_0(X)e(X)\} \times S(X)] \\
        &= \mathbb{E}[\{ \frac{\mu_0}{\mu_1^2}\mu_1(X) - \frac{1}{\mu_1}\mu_0(X)\} \cdot e(X) \cdot S(X)] \\
        &= the \ third \ and \ fourth \ term \ of \ \frac{\partial \beta(\theta)}{\partial \theta} \bigg|_{\theta = \theta_0}.
    \end{align*}

    Combing the results of $C_1, C_2$, and $C_3$ leads to equation (\ref{ known case 1}). In addition, let $s_1(Y|X) = \dfrac{\mu_0}{\mu_1^2}(Y-\mu_1(X))$, $s_0(Y|X) = -\dfrac{e(X)}{\mu_1}\dfrac{1}{1-e(X)}(Y - \mu_0(X))$, and $s(X) = \dfrac{\mu_0}{\mu_1^2}\mu_1(X)e(X) - \dfrac{1}{\mu_1}\mu_0(X)e(X)$, then 
    \begin{align*}
        \phi_\beta(Y, A, X) = A \cdot s_1(Y|X) + (1 - A) \cdot s_0(Y|X) + s(X),
    \end{align*}

    which implies that $\tilde \phi_\beta(Y, A, X; \eta) \in \mathcal{T}$. Thus, the influence function $\tilde \phi_\beta(Y, A, X; \eta)$ is efficient.

\subsection*{Proof of Theorem 2(b)}
Proof of Theorem 2(b). Under Assumptions 1 and 3, when the propensity score $e(X)$ is known, the parametric submodel indexed by $\theta$, the
estimand $\beta$ can be written as
    \begin{equation*}
        \beta(\theta) = 1 - \frac{\int (\int yf_1(y| x, \theta)dy)(\int yf_0(y| x, \theta)dy)e(x)f(x, \theta)dx}{\iint yf_1(y|x, \theta)e(x)f(x, \theta)dydx }
    \end{equation*}

    The pathwise derivative of $\beta(\theta)$ at $\theta = \theta_0$ is given as

    \begin{align*}
         \frac{\partial \beta(\theta)}{\partial \theta}& \bigg|_{\theta = \theta_0} = \ - \frac{\int (\int ys_1(y |x, \theta_0)f_1(y| x, \theta_0)dy)(\int yf_0(y| x, \theta_0)dy)e(x)f(x, \theta_0)dx}{\iint yf_1(y|x, \theta_0)e(x)f(x, \theta_0)dydx } \\
        & - \frac{\int (\int yf_1(y| x, \theta_0)dy)(\int ys_0(y|x, \theta_0)f_0(y| x, \theta_0)dy)e(x)f(x, \theta_0)dx}{\iint yf_1(y|x, \theta_0)e(x)f(x, \theta_0)dydx } \\
        & - \frac{\int (\int yf_1(y| x, \theta_0)dy)(\int yf_0(y| x, \theta_0)dy)e(x)s(x, \theta_0)f(x, \theta_0)dx}{\iint yf_1(y|x, \theta_0)e(x)f(x, \theta_0)dydx } \\
        &+ \frac{[\int (\int yf_1(y| x, \theta_0)dy)(\int yf_0(y| x, \theta_0)dy)e(x)f(x, \theta_0)dx][\iint ys_1(y|x, \theta_0)f_1(y|x, \theta_0)e(x)f(x, \theta_0)dydx]}{[\iint yf_1(y|x, \theta_0)e(x)f(x, \theta_0)dydx]^2 } \\
        &+ \frac{[\int (\int yf_1(y| x, \theta_0)dy)(\int yf_0(y| x, \theta_0)dy)e(x)f(x, \theta_0)dx][\iint yf_1(y|x, \theta_0)e(x)s(x, \theta_0)f(x, \theta_0)dydx]}{[\iint yf_1(y|x, \theta_0)e(x)f(x, \theta_0)dydx]^2 } \\
        &= \mathbb{E}[(\dfrac{\mu}{\mu_1^2} - \dfrac{\mu_0(X)}{\mu_1})e(X)Y^1S_1(Y^1| X)] - \frac{1}{\mu_1}\mathbb{E}[\mu_1(X)e(X)Y^0S_0(Y^0| X)] \\
        &+ \mathbb{E}[(\dfrac{\mu}{\mu_1^2} - \dfrac{\mu_0(X)}{\mu_1})\mu_1(X)e(X)S(X)]
    \end{align*}

    we first show that $\tilde \varphi_{\beta}(Y, A, X; \eta)$ is an influence function for $\beta$, which suffices to verify that
    \begin{equation}\label{ known case 2}
        \frac{\partial \beta(\theta)}{\partial \theta} \bigg|_{\theta = \theta_0} = \mathbb{E}[\tilde \varphi_{\beta}(Y, A, X; \eta) \cdot s(Y, A, X; \theta_0)]
    \end{equation}

    The right side of equation (\ref{ known case 2}) can be decomposed as follows,

    \begin{equation*}
       \mathbb{E}[\tilde \varphi_{\beta}(Y, A, X; \eta) \cdot s(Y, A, X; \theta_0)] = C_4 + C_5 + C_6,
    \end{equation*}
    
    where 

    \begin{align*}
        C_4 &= \mathbb{E}[\tilde \varphi_{\beta}(Y, A, X; \eta) \cdot A \cdot S_1(Y|X)] \\
        C_5 &= \mathbb{E}[\tilde \varphi_{\beta}(Y, A, X; \eta) \cdot (1 - A) \cdot S_0(Y|X)] \\
        C_6 &= \mathbb{E}[\tilde \varphi_{\beta}(Y, A, X; \eta) \cdot S(X)]
    \end{align*}

    We analyze $C_4, C_5$, and $C_6$ one by one. Since $\mathbb{E}[s_a(Y|X)|X = x] = 0$ for a = 0, 1, and \ $\mathbb{E}[S(X)] = 0$ we have

    \begin{align*}
        C_4 &= \mathbb{E}[\tilde \varphi_{\beta}(Y, A, X; \eta) \cdot A \cdot S_1(Y|X)] \\
        &= \mathbb{E}[\{ (\dfrac{\mu}{\mu_1^2} - \dfrac{\mu_0(X)}{\mu_1})A(Y-\mu_1(X)) - \dfrac{e(X)}{\mu_1}\dfrac{1-A}{1-e(X)}\mu_1(X)(Y - \mu_0(X)) \\
        &+ (\dfrac{\mu}{\mu_1^2} - \dfrac{\mu_0(X)}{\mu_1})\mu_1(X)e(X) \} \times A \cdot S_1(Y|X)] \\
        &= \mathbb{E}[(\dfrac{\mu}{\mu_1^2} - \dfrac{\mu_0(X)}{\mu_1}) A Y S_1(Y|X)] \\
        &= \mathbb{E}[(\dfrac{\mu}{\mu_1^2} - \dfrac{\mu_0(X)}{\mu_1})e(X) Y^1 S_1(Y^1|X)] \\
        &= the \ first \ term \ of \ \frac{\partial \beta(\theta)}{\partial \theta} \bigg|_{\theta = \theta_0}.
    \end{align*}

    Likewise,
    \begin{align*}
        C_5 &= \mathbb{E}[\tilde \varphi_{\beta}(Y, A, X; \eta) \cdot (1 - A) \cdot S_0(Y|X)] \\
        &= \mathbb{E}[\{ (\dfrac{\mu}{\mu_1^2} - \dfrac{\mu_0(X)}{\mu_1})A(Y-\mu_1(X)) - \dfrac{e(X)}{\mu_1}\dfrac{1-A}{1-e(X)}\mu_1(X)(Y - \mu_0(X)) \\
        &+ (\dfrac{\mu}{\mu_1^2} - \dfrac{\mu_0(X)}{\mu_1})\mu_1(X)e(X) \} \times (1 - A) \cdot S_0(Y|X)] \\
        &= \mathbb{E}[- \dfrac{e(X)}{\mu_1}\dfrac{1-A}{1-e(X)}\mu_1(X) Y  S_0(Y|X)] \\
        &= \mathbb{E}[- \dfrac{e(X)}{\mu_1}\mu_1(X) Y^0 S_0(Y^0|X)] \\
        &= the \ second \ term \ of \ \frac{\partial \beta(\theta)}{\partial \theta} \bigg|_{\theta = \theta_0}.
    \end{align*}

    where the last equation follows from the law of iterated expectations. Also,

    \begin{align*}
        C_6 &= \mathbb{E}[\tilde \varphi_{\beta}(Y, A, X; \eta) \cdot S(X)] \\
        &= \mathbb{E}[\{ (\dfrac{\mu}{\mu_1^2} - \dfrac{\mu_0(X)}{\mu_1})A(Y-\mu_1(X)) - \dfrac{e(X)}{\mu_1}\dfrac{1-A}{1-e(X)}\mu_1(X)(Y - \mu_0(X)) \\
        &+ (\dfrac{\mu}{\mu_1^2} - \dfrac{\mu_0(X)}{\mu_1})\mu_1(X)e(X) \} \times S(X)] \\
        &= \mathbb{E}[(\dfrac{\mu}{\mu_1^2} - \dfrac{\mu_0(X)}{\mu_1})\mu_1(X)e(X) \cdot S(X)] \\
        &= the \ third \ term \ of \ \frac{\partial \beta(\theta)}{\partial \theta} \bigg|_{\theta = \theta_0}.
    \end{align*}

    Combing the results of $C_4, C_5$, and $C_6$ leads to equation (\ref{ known case 2}). In addition, let $s_1(Y|X) = (\dfrac{\mu}{\mu_1^2} - \dfrac{\mu_0(X)}{\mu_1})(Y-\mu_1(X))$, $s_0(Y|X) = - \dfrac{e(X)}{\mu_1}\dfrac{\mu_1(X)}{1-e(X)}(Y - \mu_0(X))$, and $s(X) = (\dfrac{\mu}{\mu_1^2} - \dfrac{\mu_0(X)}{\mu_1})\mu_1(X)e(X)$, then 
    \begin{align*}
        \tilde \varphi_{\beta}(Y, A, X; \eta) = A \cdot s_1(Y|X) + (1 - A) \cdot s_0(Y|X) + s(X),
    \end{align*}

    which implies that $\tilde \varphi_{\beta}(Y, A, X; \eta) \in \mathcal{T}$. Thus, the influence function $\tilde \varphi_{\beta}(Y, A, X; \eta)$ is efficient.

\section*{Proof of Proposition 1}

\subsection*{Proof of Proposition 1(a)}
Proof of Proposition 1(a). Under Assumptions 1 and 2, if the propensity score $e(X)$ is unknown, the semiparametric efficiency bound of $\beta$ is
\begin{align*}
    \mathbb{V}(\phi_{\beta}(&Y, A, X; \eta)) = \E[\phi_{\beta}^2(Y, A, X; \eta) ] \\
    &= \mathbb{E}[ \{\frac{\mu_0}{\mu_1^2}A(Y-\mu_1(X)) - \frac{1}{\mu_1}\frac{1-A}{1-e(X)}(Y - \mu_0(X))e(X) + \frac{1}{\mu_1^2}[\mu_0\mu_1(X) - \mu_1\mu_0(X)](A - e(X))\\
    &+ \frac{1}{\mu_1^2}[\mu_0\mu_1(X) - \mu_1\mu_0(X)]e(X) \}^2 ] \\
    &= \mathbb{E}[\frac{\mu^2_0}{\mu_1^4}A(Y-\mu_1(X))^2 + \frac{1}{\mu^2_1}\frac{1-A}{(1-e(X))^2}(Y - \mu_0(X))^2e^2(X) + \frac{1}{\mu_1^4}[\mu_0\mu_1(X) - \mu_1\mu_0(X)]^2(A - e(X))^2 \\
    &+ \frac{1}{\mu_1^4}[\mu_0\mu_1(X) - \mu_1\mu_0(X)]^2e^2(X) + 2\frac{1}{\mu_1^4}[\mu_0\mu_1(X) - \mu_1\mu_0(X)]^2(A - e(X))e(X)] \\
    &= \mathbb{E}[\frac{\mu^2_0}{\mu_1^4}A(Y-\mu_1(X))^2 + \frac{1}{\mu^2_1}\frac{1-A}{(1-e(X))^2}(Y - \mu_0(X))^2e^2(X) + \frac{1}{\mu_1^4}[\mu_0\mu_1(X) - \mu_1\mu_0(X)]^2e(X)].
\end{align*}

Under Assumptions 1 and 2, if the propensity score $e(X)$ is known, the semiparametric efficiency bound of $\beta$ is
\begin{align*}
    \mathbb{V}(\tilde \phi_{\beta}(&Y, A, X; \eta)) = \E[\tilde \phi_{\beta}^2(Y, A, X; \eta) ] \\
    &= \mathbb{E}[\{ \frac{\mu_0}{\mu_1^2}A(Y-\mu_1(X)) - \frac{e(X)}{\mu_1}\frac{1-A}{1-e(X)}(Y - \mu_0(X)) + \frac{\mu_0}{\mu_1^2}\mu_1(X)e(X) - \frac{1}{\mu_1}\mu_0(X)e(X) \}^2] \\
    &= \mathbb{E}[ \frac{\mu^2_0}{\mu_1^4}A(Y-\mu_1(X))^2 + \frac{e^2(X)}{\mu^2_1}\frac{1-A}{(1-e(X))^2}(Y - \mu_0(X))^2 + \frac{\mu^2_0}{\mu_1^4}\mu^2_1(X)e^2(X) \\
    &+ \frac{1}{\mu^2_1}\mu^2_0(X)e^2(X) - 2\frac{\mu_0}{\mu_1^3}\mu_0(X)\mu_1(X)e^2(X)] \\
    &= \mathbb{E}[\frac{\mu^2_0}{\mu_1^4}A(Y-\mu_1(X))^2 + \frac{e^2(X)}{\mu^2_1}\frac{1-A}{(1-e(X))^2}(Y - \mu_0(X))^2 + \frac{1}{\mu_1^4}(\mu_0\mu_1(X) - \mu_1\mu_0(X))^2e^2(X)].
\end{align*}

The magnitude of difference between the semiparametric efficient bounds when the propensity score is unknown than known is
\begin{align*}
    \mathbb{V}( \phi_{\beta}(&Y, A, X; \eta)) - \mathbb{V}(\tilde \phi_{\beta}(Y, A, X; \eta)) \\
    &= \mathbb{E}[\frac{\mu^2_0}{\mu_1^4}A(Y-\mu_1(X))^2 + \frac{1}{\mu^2_1}\frac{1-A}{(1-e(X))^2}(Y - \mu_0(X))^2e^2(X) + \frac{1}{\mu_1^4}[\mu_0\mu_1(X) - \mu_1\mu_0(X)]^2e(X)] \\
    &- \mathbb{E}[\frac{\mu^2_0}{\mu_1^4}A(Y-\mu_1(X))^2 + \frac{e^2(X)}{\mu^2_1}\frac{1-A}{(1-e(X))^2}(Y - \mu_0(X))^2 + \frac{1}{\mu_1^4}(\mu_0\mu_1(X) - \mu_1\mu_0(X))^2e^2(X)] \\
    &= \mathbb{E}[\frac{1}{\mu_1^4}(\mu_0\mu_1(X) - \mu_1\mu_0(X))^2e(X)(1 - e(X))].
\end{align*}

This demonstrates Proposition 1(a).

\subsection*{Proof of Proposition 1(b)}
Proof of Proposition 1(b). Under Assumptions 1 and 3, if the propensity score $e(X)$ is unknown, the semiparametric efficiency bound of $\beta$ is
\begin{align*}
    \mathbb{V}&(\varphi_{\beta}(Y, A, X; \eta)) = \E[\varphi_{\beta}^2(Y, A, X; \eta) ] \\
    &= \mathbb{E}[ \{ A \cdot \dfrac{\mu - \mu_1\mu_0(X)}{\mu_1^2}(Y-\mu_1(X)) - \dfrac{e(X)}{\mu_1}\dfrac{1-A}{1-e(X)}(Y - \mu_0(X))\mu_1(X) + \dfrac{\mu - \mu_1\mu_0(X)}{\mu_1^2}\mu_1(X)(A - e(X)) \\
    &+ \dfrac{\mu - \mu_1\mu_0(X)}{\mu_1^2}\mu_1(X)e(X) \}^2 ] \\
    &= \mathbb{E}[A \cdot \dfrac{(\mu - \mu_1\mu_0(X))^2}{\mu_1^4}(Y-\mu_1(X))^2 + \dfrac{e^2(X)}{\mu^2_1}\dfrac{1-A}{(1-e(X))^2}(Y - \mu_0(X))^2\mu^2_1(X) + \dfrac{(\mu - \mu_1\mu_0(X))^2}{\mu_1^4}\mu^2_1(X)e^2(X)  \\
    &+ \dfrac{(\mu - \mu_1\mu_0(X))^2}{\mu_1^4}\mu^2_1(X)(A - e(X))^2 + 2\dfrac{(\mu - \mu_1\mu_0(X))^2}{\mu_1^4}\mu^2_1(X)(A - e(X))e(X)] \\
    &= \mathbb{E}[A \cdot \dfrac{(\mu - \mu_1\mu_0(X))^2}{\mu_1^4}(Y-\mu_1(X))^2 + \dfrac{e^2(X)}{\mu^2_1}\dfrac{1-A}{(1-e(X))^2}(Y - \mu_0(X))^2\mu^2_1(X) + \dfrac{(\mu - \mu_1\mu_0(X))^2}{\mu_1^4}\mu^2_1(X)e(X)]  
\end{align*}

Under Assumptions 1 and 3, if the propensity score $e(X)$ is known, the semiparametric efficiency bound of $\beta$ is
\begin{align*}
    \mathbb{V}&( \tilde \varphi_{\beta}(Y, A, X; \eta)) = \E[ \tilde \varphi_{\beta}^2(Y, A, X; \eta) ] \\
    &= \mathbb{E}[ \{ (\dfrac{\mu}{\mu_1^2} - \dfrac{\mu_0(X)}{\mu_1})A(Y-\mu_1(X)) - \dfrac{e(X)}{\mu_1}\dfrac{1-A}{1-e(X)}\mu_1(X)(Y - \mu_0(X)) + (\dfrac{\mu}{\mu_1^2} - \dfrac{\mu_0(X)}{\mu_1})\mu_1(X)e(X) \}^2 ] \\
    &= \mathbb{E}[ (\dfrac{\mu}{\mu_1^2} - \dfrac{\mu_0(X)}{\mu_1})^2A(Y-\mu_1(X))^2 + \dfrac{e^2(X)}{\mu^2_1}\dfrac{1-A}{(1-e(X))^2}\mu^2_1(X)(Y - \mu_0(X))^2 + (\dfrac{\mu}{\mu_1^2} - \dfrac{\mu_0(X)}{\mu_1})^2\mu^2_1(X)e^2(X) ]  \\
    &= \mathbb{E}[A \dfrac{(\mu - \mu_1\mu_0(X))^2}{\mu_1^4}(Y-\mu_1(X))^2 + \dfrac{e^2(X)}{\mu^2_1}\dfrac{1-A}{(1-e(X))^2}(Y - \mu_0(X))^2\mu^2_1(X) + \dfrac{(\mu - \mu_1\mu_0(X))^2}{\mu_1^4}\mu^2_1(X)e^2(X)]
\end{align*}

The magnitude of difference between the semiparametric efficient bounds when the propensity score is unknown than known is
\begin{align*}
    \mathbb{V}( \varphi_{\beta}(&Y, A, X; \eta)) - \mathbb{V}(\tilde \varphi_{\beta}(Y, A, X; \eta)) \\
    &= \mathbb{E}[A \cdot \dfrac{(\mu - \mu_1\mu_0(X))^2}{\mu_1^4}(Y-\mu_1(X))^2 + \dfrac{e^2(X)}{\mu^2_1}\dfrac{1-A}{(1-e(X))^2}(Y - \mu_0(X))^2\mu^2_1(X) + \dfrac{(\mu - \mu_1\mu_0(X))^2}{\mu_1^4}\mu^2_1(X)e(X)] \\
    &- \mathbb{E}[A \dfrac{(\mu - \mu_1\mu_0(X))^2}{\mu_1^4}(Y-\mu_1(X))^2 + \dfrac{e^2(X)}{\mu^2_1}\dfrac{1-A}{(1-e(X))^2}(Y - \mu_0(X))^2\mu^2_1(X) + \dfrac{(\mu - \mu_1\mu_0(X))^2}{\mu_1^4}\mu^2_1(X)e^2(X)] \\
    &= \mathbb{E}[\frac{1}{\mu_1^4}(\mu\mu_1(X) - \mu_1\mu_1(X)\mu_0(X))^2e(X)(1 - e(X))].
\end{align*}

This demonstrates Proposition 1(b).

\section*{Proof of Proposition 2}

The proofs of Proposition 2(a) and Proposition 2(b) are similar, so we only provide the detailed proof of Proposition 2(a).

{\bf Proposition 2(a)}(Double Robustness). Under Assumptions 1 and 2, the proposed estimator $\hat{\beta}_{mono}$ given in equation (2) is a consistent estimator of $\beta$ if one of the following conditions is satisfied: 
    \begin{itemize}
        \item[(i)] the propensity score model is correctly specified;
        \item[(ii)] the outcome regression models are correctly specified.
    \end{itemize}

\begin{pf} 
    We note here that the propensity score model is correctly specified i.e. $\hat{e}(x) = \bar e(x)$ and the outcome regression models are correctly specified i.e. $\hat{\mu}_a(x) = \bar \mu_a(x)$ for $a = 0, 1$.
    For the denominator, according to the law of large numbers when $n \rightarrow \infty$, 
    \begin{equation*}
        \dfrac{1}{n} \sum\limits_{i=1}^n A_i Y_i \stackrel{p}{\longrightarrow} \mathbb{E}[A \cdot Y]
    \end{equation*}
    where $\mathbb{E}[A \cdot Y] = \mathbb{E}\{ \mathbb{E}[AY | X]\} = \mathbb{E}\{ e(X) \mathbb{E}[Y | X, A = 1]\} = \mathbb{E}\{ e(X) \mu_1(X) \} = \mu_1$, so we have $\dfrac{1}{n} \sum\limits_{i=1}^n A_i Y_i \stackrel{p}{\longrightarrow} \mu_1$.

    For the numerator, according to the law of large numbers when $n \rightarrow \infty$, 
    \begin{equation*}
        \dfrac{1}{n}\sum\limits_{i=1}^n \Big \{ A_i(Y_i - \hat \mu_0(X_i)) -  \dfrac{(1-A_i)(Y_i - \hat \mu_0(X_i)) \hat e(X_i) }{1 - \hat e(X_i)}  \Big \} \stackrel{p}{\longrightarrow} \mathbb{E}[ A(Y - \bar \mu_0(X)) -  \dfrac{(1-A)(Y - \bar \mu_0(X)) \bar e(X) }{1 - \bar e(X)} ]
    \end{equation*}
    
    If $\bar{e}(x) = e(x)$,  the right side of the above formula becomes to \\
    \begin{align*}
        &\mathbb{E}[ A(Y - \bar \mu_0(X)) -  \dfrac{(1-A)(Y - \bar \mu_0(X)) \bar e(X) }{1 - \bar e(X)} ] \\
        = \ &\mathbb{E}[ A(Y - \bar \mu_0(X)) -  \dfrac{(1-A)(Y - \bar \mu_0(X)) e(X) }{1 - e(X)} ] \\
        = \ &\mathbb{E}[ e(X)\mathbb{E}[(Y - \bar \mu_0(X))| X, A = 1] ] - \mathbb{E} [ (1 - e(X)) \mathbb{E}[\dfrac{(Y - \bar \mu_0(X)) e(X) }{1 - e(X)}| X, A = 0] ] \\
        = \ &\mathbb{E}[ e(X)(\mu_1(X) - \bar \mu_0(X)) ] - \mathbb{E} [ (\mu_0(X) - \bar \mu_0(X)) e(X)  ] \\
        = \ &\mu_1 - \mu_0
    \end{align*}
    so we have 
    \begin{equation*}
        \dfrac{1}{n}\sum\limits_{i=1}^n \Big \{ A_i(Y_i - \hat \mu_0(X_i)) -  \dfrac{(1-A_i)(Y_i - \hat \mu_0(X_i)) \hat e(X_i) }{1 - \hat e(X_i)}  \Big \} \stackrel{p}{\longrightarrow} \mu_1 - \mu_0
    \end{equation*}
    Under slutsky's theorem, 
    \begin{align*}
    \hat{\beta}(Z,\hat e, \hat \mu_0, \mu_1) = \dfrac{\dfrac{1}{n}\sum\limits_{i=1}^n \Big \{ A_i(Y_i - \hat \mu_0(X_i)) -  \dfrac{(1-A_i)(Y_i - \hat \mu_0(X_i)) \hat e(X_i) }{1 - \hat e(X_i)}  \Big \}  }{ \dfrac{1}{n} \sum\limits_{i=1}^n A_i Y_i } \stackrel{p}{\longrightarrow} \frac{\mu_1 - \mu_0}{\mu_1} = \beta
   \end{align*}
    This proves the conclusion (i).
    
    If $\bar{\mu}_a(x) = \mu_a(x)$ for a = 0, 1, 
    \begin{align*}
        &\mathbb{E}[ A(Y - \bar \mu_0(X)) -  \dfrac{(1-A)(Y - \bar \mu_0(X)) \bar e(X) }{1 - \bar e(X)} ]\\
        = \ &\mathbb{E}[ A(Y - \mu_0(X)) -  \dfrac{(1-A)(Y - \mu_0(X)) \bar e(X) }{1 - \bar e(X)} ] \\
        = \ &\mathbb{E}[ e(X)\mathbb{E}[(Y - \mu_0(X))| X, A = 1] ] - \mathbb{E} [ (1 - e(X)) \mathbb{E}[\dfrac{(Y - \mu_0(X)) \bar e(X) }{1 - \bar e(X)}| X, A = 0] ] \\
        = \ &\mathbb{E}[ e(X)(\mu_1(X) - \mu_0(X)) ] \\
        = \ &\mu_1 - \mu_0
    \end{align*}
    so we have 
    \begin{equation*}
        \dfrac{1}{n}\sum\limits_{i=1}^n \Big \{ A_i(Y_i - \hat \mu_0(X_i)) -  \dfrac{(1-A_i)(Y_i - \hat \mu_0(X_i)) \hat e(X_i) }{1 - \hat e(X_i)}  \Big \} \stackrel{p}{\longrightarrow} \mu_1 - \mu_0
    \end{equation*}
    Under slutsky's theorem, 
    \begin{align*}
    \hat{\beta}(Z,\hat e, \hat \mu_0, \mu_1) = \dfrac{\dfrac{1}{n}\sum\limits_{i=1}^n \Big \{ A_i(Y_i - \hat \mu_0(X_i)) -  \dfrac{(1-A_i)(Y_i - \hat \mu_0(X_i)) \hat e(X_i) }{1 - \hat e(X_i)}  \Big \}  }{ \dfrac{1}{n} \sum\limits_{i=1}^n A_i Y_i } \stackrel{p}{\longrightarrow} \frac{\mu_1 - \mu_0}{\mu_1} = \beta
   \end{align*}
    This proves the conclusion (ii).
    
\end{pf}

\section*{Proof of Robustness of $\tilde \beta_{mono}$}

We have that $\tilde \beta_{mono}$ is an unbiased estimator of $\beta$ regardless of whether the model of $\mu_a(X)$ is specified correctly(i.e. $\bar{\mu}_a(x)$ estimates $\mu_a(x)$ accurately).

\begin{pf} 
     For the denominator, according to the law of large numbers when $n \rightarrow \infty$, 
    \begin{equation*}
        \dfrac{1}{n} \sum\limits_{i=1}^n A_i(Y_i-\hat\mu_1(X_i)) + \hat\mu_1(X_i) e(X_i) \stackrel{p}{\longrightarrow} \mathbb{E}[A (Y-\bar\mu_1(X)) + \bar\mu_1(X) e(X)]
    \end{equation*}

    For the numerator, according to the law of large numbers when $n \rightarrow \infty$, 
    \begin{align*}
        \dfrac{1}{n}\sum\limits_{i=1}^n \Big \{ A_i(Y_i-\hat\mu_1(X_i)) - e(X_i)\dfrac{1-A_i}{1- e(X_i)}(Y_i - \hat \mu_0(X_i)) + (\hat \mu_1(X_i) - \hat \mu_0(X_i)) e(X_i) \Big \} \stackrel{p}{\longrightarrow} \\
        \mathbb{E}[  A(Y-\bar \mu_1(X)) - e(X)\dfrac{1-A}{1- e(X)}(Y - \bar \mu_0(X)) + (\bar \mu_1(X) - \bar \mu_0(X)) e(X) ]
    \end{align*}
    
    If $\bar{\mu}_a(x) = \mu_a(x)$, for a = 0, 1, the right side of the above formula becomes to \\
    \begin{align*}
        &\mathbb{E}[A (Y-\bar\mu_1(X)) + \bar\mu_1(X) e(X)] \\
        = \ &\mathbb{E}[A (Y-\mu_1(X)) + \mu_1(X) e(X)] \\
        = \ &\mathbb{E}[e(X) (\mu_1(X) -\mu_1(X)) + \mu_1(X) e(X)] \\
        = \ &\mathbb{E}[\mu_1(X) e(X)] \\
        = \ & \mu_1
    \end{align*}
    where the second equality follows by the law of iterated expectations \\
    \begin{align*}
        & \mathbb{E}[  A(Y-\bar \mu_1(X)) -e(X)\dfrac{1-A}{1- e(X)}(Y - \bar \mu_0(X)) + (\bar \mu_1(X) - \bar \mu_0(X)) e(X) ] \\
        = \ &\mathbb{E}[ A(Y- \mu_1(X)) - e(X)\dfrac{1-A}{1- e(X)}(Y - \mu_0(X)) + (\mu_1(X) -\mu_0(X)) e(X) ] \\
       = \ &\mathbb{E}[ e(X)(\mu_1(X)- \mu_1(X)) - e(X)(\mu_0(X) - \mu_0(X)) + (\mu_1(X) - \mu_0(X)) e(X) ] \\
        = \ &\mathbb{E}[ e(X)(\mu_1(X) - \mu_0(X)) ] \\
        = \ &\mu_1 - \mu_0
    \end{align*}
     where the second equality follows by the law of iterated expectations, so we have 
    \begin{align*}
        \dfrac{1}{n} \sum\limits_{i=1}^n A_i(Y_i-\hat\mu_1(X_i)) + \hat\mu_1(X_i) e(X_i) \stackrel{p}{\longrightarrow} \mu_1 
        \end{align*}
        \begin{align*}
        \dfrac{1}{n}\sum\limits_{i=1}^n \Big \{ A_i(Y_i - \hat \mu_0(X_i)) -  \dfrac{(1-A_i)(Y_i - \hat \mu_0(X_i)) e(X_i) }{1 - e(X_i)}  \Big \} \stackrel{p}{\longrightarrow} \mu_1 - \mu_0
    \end{align*}
    Under slutsky's theorem, 
    \begin{align*}
    \hat{\beta}(Z,\hat e, \hat \mu_0, \mu_1) = \dfrac{\dfrac{1}{n}\sum\limits_{i=1}^n \Big \{ A_i(Y_i - \hat \mu_0(X_i)) -  \dfrac{(1-A_i)(Y_i - \hat \mu_0(X_i)) e(X_i) }{1 - e(X_i)}  \Big \}  }{ \dfrac{1}{n} \sum\limits_{i=1}^n A_i Y_i } \stackrel{p}{\longrightarrow} \frac{\mu_1 - \mu_0}{\mu_1} = \beta
   \end{align*}
    This proves the conclusion for $\bar \mu_a(X) = \mu_a(X)$.
    
   If $\bar{\mu}_a(x) \neq \mu_a(x)$ for a = 0, 1, the right side of the above formula becomes to \\
    \begin{align*}
        &\mathbb{E}[A (Y-\bar\mu_1(X)) + \bar\mu_1(X) e(X)] \\
        = \ &\mathbb{E}[e(X) (\mu_1(X) -   \bar \mu_1(X)) + \bar\mu_1(X) e(X)] \\
        = \ &\mathbb{E}[\mu_1(X) e(X)] \\
        = \ &\mu_1
    \end{align*}
    where the first equality follows by the law of iterated expectations \\
    \begin{align*}
        &\mathbb{E}[ A(Y-\bar \mu_1(X)) - e(X)\dfrac{1-A}{1- e(X)}(Y - \bar \mu_0(X)) + (\bar \mu_1(X) - \bar \mu_0(X)) e(X) ] \\
       = \ &\mathbb{E}[ e(X)(\mu_1(X)-\bar \mu_1(X)) - e(X)(\mu_0(X) - \bar \mu_0(X)) + (\bar \mu_1(X) - \bar \mu_0(X)) e(X) ] \\
        = \ &\mathbb{E}[ e(X)(\mu_1(X) - \mu_0(X)) ] \\
        = \ &\mu_1 - \mu_0
    \end{align*}
     where the first equality follows by the law of iterated expectations, so we have 
    \begin{align*}
        \dfrac{1}{n} \sum\limits_{i=1}^n A_i(Y_i-\hat\mu_1(X_i)) + \hat\mu_1(X_i) e(X_i) \stackrel{p}{\longrightarrow} \mu_1 
        \end{align*}
        \begin{align*}
        \dfrac{1}{n}\sum\limits_{i=1}^n \Big \{ A_i(Y_i - \hat \mu_0(X_i)) -  \dfrac{(1-A_i)(Y_i - \hat \mu_0(X_i)) e(X_i) }{1 - e(X_i)}  \Big \} \stackrel{p}{\longrightarrow} \mu_1 - \mu_0
    \end{align*}
    Under slutsky's theorem, 
    \begin{align*}
    \hat{\beta}(Z,\hat e, \hat \mu_0, \mu_1) = \dfrac{\dfrac{1}{n}\sum\limits_{i=1}^n \Big \{ A_i(Y_i - \hat \mu_0(X_i)) -  \dfrac{(1-A_i)(Y_i - \hat \mu_0(X_i)) e(X_i) }{1 - e(X_i)}  \Big \}  }{ \dfrac{1}{n} \sum\limits_{i=1}^n A_i Y_i } \stackrel{p}{\longrightarrow} \frac{\mu_1 - \mu_0}{\mu_1} = \beta
   \end{align*}
    This proves the conclusion for $\bar \mu_a(X) \neq \mu_a(X)$.
  
\end{pf}

\section*{Proof of Theorem 3}

\begin{thm}[Asymptotic Normality and Efficiency] Under Assumptions 1-2 and Condition 1,

(a) when the propensity score $e(X)$ is unknown,
    \begin{equation*}
        \sqrt{n}(\hat{\beta}_{mono} - \beta) \xrightarrow{d} \mathcal{N}(0, \sigma_1^2),
    \end{equation*}
    where $ \xrightarrow{d}$ denotes convergence in distribution and $\sigma_1^2 =\mathbb{V}(\phi_{\beta}(Y, A, X; \eta))$ is the semiparametric efficiency bound of $\beta$. 
    A consistent estimator of $\sigma_1^2$ is $\hat \sigma_1^2 = n^{-1}\sum_{i = 1}^n [\zeta_1(Z_i; \hat{\eta}) - \hat \beta_{mono}]^2$, where $\zeta_1(Z_i; \hat{\eta}) =  \big \{ A_i(Y_i - \hat \mu_0(X_i)) -  (1-A_i)(Y_i - \hat \mu_0(X_i)) \hat e(X_i)/(1 - \hat e(X_i))  \big \}  / (n^{-1}) \sum_{i=1}^n A_i Y_i$.

(b) when the propensity score $e(X)$ is known,
    \begin{equation*}
        \sqrt{n}(\tilde{\beta}_{mono} - \beta) \xrightarrow{d} \mathcal{N}(0, \sigma_2^2), 
    \end{equation*}
    where 
    $\sigma_2^2 =
    \mathbb{V}(\tilde \phi_{\beta}(Y, A, X; \eta))$ is the semiparametric efficiency bound of $\beta$. A consistent estimator of $\sigma_2^2$ is $\hat\sigma_2^2 = n^{-1}\sum_{i = 1}^n [ \zeta_2(Z_i; \hat{\eta}) - \tilde \beta_{mono}]^2$, where $ \zeta_2(Z_i; \hat{\eta}) = \big \{ A_i(Y_i-\hat\mu_1(X_i)) - e(X_i)(1-A_i)(Y_i - \hat \mu_0(X_i))/1- e(X_i) + (\hat \mu_1(X_i) - \hat \mu_0(X_i)) e(X_i) \big \} / (n^{-1}) \sum_{i=1}^n \{A_i(Y_i-\hat\mu_1(X_i)) + \hat\mu_1(X_i) e(X_i) \}$. 
\end{thm}

\subsection*{Proof of Theorem 3(a)}

    Recall that 
    \begin{align*}
    \hat{\beta}_{mono} = \dfrac{\dfrac{1}{n}\sum\limits_{i=1}^n \Big \{ A_i(Y_i - \hat \mu_0(X_i)) -  \dfrac{(1-A_i)(Y_i - \hat \mu_0(X_i)) \hat e(X_i) }{1 - \hat e(X_i)}  \Big \}  }{ \dfrac{1}{n} \sum\limits_{i=1}^n A_i Y_i }. 
    \end{align*}
    The term $\sqrt{n}(\hat{\beta}_{mono} - \beta)$ can be decomposed as follow, 
    \begin{equation*}
        \sqrt{n}(\hat{\beta}_{mono} - \beta) = H_{1n} + H_{2n},
    \end{equation*}
    where 
    \begin{align*}
        H_{1n} &= \sqrt{n}\Big \{ \dfrac{\dfrac{1}{n}\sum\limits_{i=1}^n \big \{ A_i(Y_i - \mu_0(X_i)) -  \dfrac{(1-A_i)(Y_i - \mu_0(X_i)) e(X_i) }{1 - e(X_i)}  \big \}  }{ \dfrac{1}{n} \sum\limits_{i=1}^n A_i Y_i } - \beta \Big \}, \\
        H_{2n} &= \sqrt{n}\Big \{ \dfrac{\dfrac{1}{n}\sum\limits_{i=1}^n \big \{ A_i(Y_i - \hat \mu_0(X_i)) -  \dfrac{(1-A_i)(Y_i - \hat \mu_0(X_i)) \hat e(X_i) }{1 - \hat e(X_i)}  \big \}  }{ \dfrac{1}{n} \sum\limits_{i=1}^n A_i Y_i } - \\ & \quad \quad \quad \quad \quad \dfrac{\dfrac{1}{n}\sum\limits_{i=1}^n \big \{ A_i(Y_i - \mu_0(X_i)) -  \dfrac{(1-A_i)(Y_i - \mu_0(X_i)) e(X_i) }{1 - e(X_i)}  \big \}  }{ \dfrac{1}{n} \sum\limits_{i=1}^n A_i Y_i } \Big \}.
    \end{align*}
    For $H_{1n}$, \\
    \begin{align*}
        H_{1n} = \dfrac{1}{\sqrt{n}} \sum\limits_{i=1}^n\dfrac{ \big \{ A_i(Y_i - \mu_0(X_i)) -  \dfrac{(1-A_i)(Y_i - \mu_0(X_i)) e(X_i) }{1 - e(X_i)} -A_iY_i\beta \big \}  }{ \dfrac{1}{n} \sum\limits_{i=1}^n A_i Y_i }.
    \end{align*}
    By the central limit theorem, it is easy to see that 
    \begin{equation*}
        H_{1n} \stackrel{d}{\longrightarrow} \mathcal{N}(0, \sigma^2),
    \end{equation*}
    where $\sigma^2 = \mu_1^{-2}Var( A(Y - \mu_0(X)) -  \frac{(1-A)(Y - \mu_0(X)) e(X) }{1 - e(X)} -AY\beta)$ because of $\dfrac{1}{n} \sum\limits_{i=1}^n A_i Y_i \stackrel{p}{\longrightarrow} \mu_1$ and $\stackrel{d}{\longrightarrow}$ denotes convergence in distribution.

    Next we focus on analyzing $H_{2n}$, which can be further decomposed as 
    \begin{equation*}
        H_{2n} = H_{2n} - \mathbb{E}[H_{2n}] + \mathbb{E}[H_{2n}].
    \end{equation*}
    Define the Gateaux derivative of the generic function $f(Z; e, \mu_0, \mu_1)$ in the direction $[\hat e - e, \hat \mu_0 - \mu_0, \hat \mu_1 - \mu_1]$ as 
    \begin{align*}
        &\partial_{[\hat{e}-e,\hat{\mu}_0-\mu_0,\hat{\mu}_1-\mu_1]}f(Z;e,\mu_0,\mu_1)\\
        &=\frac{\partial f(Z;e+\alpha_1(\hat{e}-e),\mu_0,\mu_1)}{\partial\alpha_1}\Bigg|_{\alpha_1=0} +\frac{\partial f(Z;e,\mu_0+\alpha_2(\hat{\mu}_0-\mu_0),\mu_1)}{\partial\alpha_2}\Bigg|_{\alpha_2=0}
        +\frac{\partial f(Z;e,\mu_0,\mu_1+\alpha_3(\hat{\mu}_1-\mu_1))}{\partial\alpha_3}\Bigg|_{\alpha_3=0}.
    \end{align*}
    By a Taylor expansion for $\mathbb{E}[H_{2n}]$ in direction  $[\hat e - e, \hat \mu_0 - \mu_0]$ yields that
    \begin{align*}
    \mathbb{E}[H_{2n}] & = \sqrt{n}\mathbb{E}[\hat \beta(Z;\hat{e},\hat{\mu}_0,\mu_1)- \beta(Z;e,\mu_0,\mu_1)] \\
    &=\sqrt{n}\partial_{[\hat{e}-e,\hat{\mu}_0-\mu_0]}\mathbb{E}[\beta(Z;e,\mu_0,\mu_1)]+\sqrt{n}\frac12\partial_{[\hat{e}-e,\hat{\mu}_0-\mu_0]}^2\mathbb{E}[\beta(Z;e,\mu_0,\mu_1)]+\cdots
    \end{align*}
    The first-order term 
    \begin{align*}
        &\sqrt{n}\partial_{[\hat{e}-e,\hat{\mu}_0-\mu_0]}\mathbb{E}[\beta(Z;e,\mu_0,\mu_1)] \\
        =&\sqrt{n}\mathbb{E}[\dfrac{ e(X)-A}{ (\dfrac{1}{n} \sum\limits_{i=1}^n A_i Y_i)(1 - e(X)) }](\hat \mu_0(X) - \mu_0(X)) - \sqrt{n}\mathbb{E}[\dfrac{ (1 - A)(Y - \mu_0(X))}{ (\dfrac{1}{n} \sum\limits_{i=1}^n A_i Y_i)(1 - e(X))^2 }](\hat e(X) - e(X)) \\
        = &0,
    \end{align*}
    where the last equation follows from $\mathbb{E}[A| X] - e(X) = 0$ and $\mathbb{E}[(1 - A)(Y - \mu_0(X))| X] = 0$. So the first-order term converges to $o_\mathbb{P}(1)$.\\
    For the second-order term, we get 
    \begin{align*}
        &\sqrt{n}\frac12\partial_{[\hat{e}-e,\hat{\mu}_0-\mu_0]}^2\mathbb{E}[\beta(Z;e,\mu_0,\mu_1)] \\
        =& 2\sqrt{n}\mathbb{E}[\dfrac{ 1 -A}{ (\dfrac{1}{n} \sum\limits_{i=1}^n A_i Y_i)(1 - e(X))^2 }](\hat \mu_0(X) - \mu_0(X))(\hat e(X) - e(X)) - \sqrt{n}\mathbb{E}[\dfrac{ (1 - A)(Y - \mu_0(X))}{ (\dfrac{1}{n} \sum\limits_{i=1}^n A_i Y_i)(1 - e(X))^3 }](\hat e(X) - e(X))^2  \\
        =&2\sqrt{n}\mathbb{E}[\dfrac{ 1 -A}{ (\dfrac{1}{n} \sum\limits_{i=1}^n A_i Y_i)(1 - e(X))^2 }](\hat \mu_0(X) - \mu_0(X))(\hat e(X) - e(X)) \\
        =&\sqrt{n}O_\mathbb{P}\left(||\hat{e}(X)-e(X)||_2\cdot||\hat{\mu}_0(X)-\mu_0(X)||_2\right) \\
        =&\sqrt{n}o_\mathbb{P}(n^{-1/2}) \\
        =&o_\mathbb{P}(1),
    \end{align*}
    where the  second equation because of $\mathbb{E}[(1 - A)(Y - \mu_0(X))| X] = 0$. All higher-order terms can be shown to be dominated by the second-order term. Therefore, $\mathbb{E}[H_{2n}] = o_\mathbb{P}(1)$. In addition, we get that $H_{2n} - \mathbb{E}[H_{2n}] = o_\mathbb{P}(1)$ by calculating $Var\{H_{2n} - \mathbb{E}[H_{2n}]\} =  o_\mathbb{P}(1)$. This proves the conclusion of Theorem 3(a).

\subsection*{Proof of Theorem 3(b)}

    Recall that 
    \begin{align*}
    \tilde {\beta}_{mono} = \dfrac{\dfrac{1}{n}\sum\limits_{i=1}^n \Big \{ A_i(Y_i-\hat\mu_1(X_i)) - e(X_i)\dfrac{1-A_i}{1- e(X_i)}(Y_i - \hat \mu_0(X_i)) + (\hat \mu_1(X_i) - \hat \mu_0(X_i)) e(X_i) \Big \}  }{ \dfrac{1}{n} \sum\limits_{i=1}^n A_i(Y_i-\hat\mu_1(X_i)) + \hat\mu_1(X_i) e(X_i) }. 
    \end{align*}
    The term $\sqrt{n}(\tilde {\beta}_{mono} - \beta)$ can be decomposed as follow, 
    \begin{equation*}
        \sqrt{n}(\tilde {\beta}_{mono} - \beta) = U_{1n} + U_{2n} + U_{3n},
    \end{equation*}
    where 
    \begin{align*}
        U_{1n} &= \sqrt{n}\Big \{ \dfrac{\dfrac{1}{n}\sum\limits_{i=1}^n \Big \{ A_i(Y_i-\mu_1(X_i)) - e(X_i)\dfrac{1-A_i}{1- e(X_i)}(Y_i - \mu_0(X_i)) + (\mu_1(X_i) - \mu_0(X_i)) e(X_i) \Big \}  }{ \dfrac{1}{n} \sum\limits_{i=1}^n A_i(Y_i-\mu_1(X_i)) + \mu_1(X_i) e(X_i) } - \beta \Big \} \\
        U_{2n} &= \sqrt{n}\Big \{ \dfrac{\dfrac{1}{n}\sum\limits_{i=1}^n \Big \{ A_i(Y_i-\mu_1(X_i)) - e(X_i)\dfrac{1-A_i}{1- e(X_i)}(Y_i - \mu_0(X_i)) + (\mu_1(X_i) - \mu_0(X_i)) e(X_i) \Big \}  }{ \dfrac{1}{n} \sum\limits_{i=1}^n A_i(Y_i-\hat\mu_1(X_i)) + \hat\mu_1(X_i) e(X_i) } - \\ 
        & \quad \quad \quad \quad \quad \dfrac{\dfrac{1}{n}\sum\limits_{i=1}^n \Big \{ A_i(Y_i-\mu_1(X_i)) - e(X_i)\dfrac{1-A_i}{1- e(X_i)}(Y_i - \mu_0(X_i)) + (\mu_1(X_i) - \mu_0(X_i)) e(X_i) \Big \}  }{ \dfrac{1}{n} \sum\limits_{i=1}^n A_i(Y_i-\mu_1(X_i)) + \mu_1(X_i) e(X_i) } \Big \} \\
        U_{3n} &= \sqrt{n}\Big \{ \dfrac{\dfrac{1}{n}\sum\limits_{i=1}^n \Big \{ A_i(Y_i-\hat\mu_1(X_i)) - e(X_i)\dfrac{1-A_i}{1- e(X_i)}(Y_i - \hat \mu_0(X_i)) + (\hat \mu_1(X_i) - \hat \mu_0(X_i)) e(X_i) \Big \}  }{ \dfrac{1}{n} \sum\limits_{i=1}^n A_i(Y_i-\hat\mu_1(X_i)) + \hat\mu_1(X_i) e(X_i) } - \\ 
        & \quad \quad \quad \quad \quad \dfrac{\dfrac{1}{n}\sum\limits_{i=1}^n \Big \{ A_i(Y_i-\mu_1(X_i)) - e(X_i)\dfrac{1-A_i}{1- e(X_i)}(Y_i - \mu_0(X_i)) + (\mu_1(X_i) - \mu_0(X_i)) e(X_i) \Big \}  }{ \dfrac{1}{n} \sum\limits_{i=1}^n A_i(Y_i-\hat\mu_1(X_i)) + \hat\mu_1(X_i) e(X_i) } \Big \}
    \end{align*}
    For $U_{1n}$, \\
    \begin{align*}
        U_{1n} = \dfrac{1}{\sqrt{n}} \sum\limits_{i=1}^n\dfrac{  (1 - \beta)A_i(Y_i-\mu_1(X_i)) - e(X_i)\dfrac{1-A_i}{1- e(X_i)}(Y_i - \mu_0(X_i)) + (1 - \beta)\mu_1(X_i)e(X_i) - \mu_0(X_i)e(X_i) }{ \dfrac{1}{n} \sum\limits_{i=1}^n A_i(Y_i-\mu_1(X_i)) + \mu_1(X_i) e(X_i) }
    \end{align*}
    By the central limit theorem, it is easy to see that 
    \begin{equation*}
        U_{1n} \stackrel{d}{\longrightarrow} \mathcal{N}(0, \sigma^2)
    \end{equation*}
    where $\sigma^2 = \mu_1^{-2}Var[( 1 - \beta)A(Y-\mu_1(X)) - e(X)\frac{1-A}{1- e(X)}(Y - \mu_0(X)) + (1 - \beta)\mu_1(X)e(X) - \mu_0(X)e(X)]$ because of $\dfrac{1}{n} \sum\limits_{i=1}^n A_i(Y_i-\mu_1(X_i)) + \mu_1(X_i) e(X_i) \stackrel{p}{\longrightarrow} \mu_1$ and $\stackrel{d}{\longrightarrow}$ denotes convergence in distribution.

    Next we focus on analyzing $U_{2n}$,
    \begin{align*}
        U_{2n} &= \sqrt{n}\Big \{ \dfrac{\dfrac{1}{n}\sum\limits_{i=1}^n \Big \{ A_i(Y_i-\mu_1(X_i)) - e(X_i)\dfrac{1-A_i}{1- e(X_i)}(Y_i - \mu_0(X_i)) + (\mu_1(X_i) - \mu_0(X_i)) e(X_i) \Big \}  }{ \dfrac{1}{n} \sum\limits_{i=1}^n A_i(Y_i-\hat\mu_1(X_i)) + \hat\mu_1(X_i) e(X_i) } - \\ 
        & \quad \quad \quad \quad \quad \dfrac{\dfrac{1}{n}\sum\limits_{i=1}^n \Big \{ A_i(Y_i-\mu_1(X_i)) - e(X_i)\dfrac{1-A_i}{1- e(X_i)}(Y_i - \mu_0(X_i)) + (\mu_1(X_i) - \mu_0(X_i)) e(X_i) \Big \}  }{ \dfrac{1}{n} \sum\limits_{i=1}^n A_i(Y_i-\mu_1(X_i)) + \mu_1(X_i) e(X_i) } \Big \} \\
        = &\sqrt{n}(\mu_1 - \mu_0)\Big \{ \dfrac{1}{ \dfrac{1}{n} \sum\limits_{i=1}^n A_i(Y_i-\hat\mu_1(X_i)) + \hat\mu_1(X_i) e(X_i) } - \dfrac{1}{ \dfrac{1}{n} \sum\limits_{i=1}^n A_i(Y_i-\mu_1(X_i)) + \mu_1(X_i) e(X_i) } \Big \} + o_\mathbb{P}(1) \\
        = &\sqrt{n}(\mu_0 - \mu_1)\Big \{ \dfrac{\dfrac{1}{n} \sum\limits_{i=1}^n A_i(Y_i-\hat\mu_1(X_i)) + \hat\mu_1(X_i) e(X_i) - \dfrac{1}{n} \sum\limits_{i=1}^n A_i(Y_i-\mu_1(X_i)) + \mu_1(X_i) e(X_i)}{ [\dfrac{1}{n} \sum\limits_{i=1}^n A_i(Y_i-\hat\mu_1(X_i)) + \hat\mu_1(X_i) e(X_i)][\dfrac{1}{n} \sum\limits_{i=1}^n A_i(Y_i-\mu_1(X_i)) + \mu_1(X_i) e(X_i)] } \Big \} + o_\mathbb{P}(1)\\
        = &\sqrt{n}\frac{\mu_0 - \mu_1}{\mu_1^2}\Big \{ \dfrac{1}{n} \sum\limits_{i=1}^n A_i(Y_i-\hat\mu_1(X_i)) + \hat\mu_1(X_i) e(X_i) - \dfrac{1}{n} \sum\limits_{i=1}^n A_i(Y_i-\mu_1(X_i)) + \mu_1(X_i) e(X_i) \Big \} + o_\mathbb{P}(1)\\
    \end{align*}
    where the second equation because of $\frac{1}{n}\sum\limits_{i=1}^n \Big \{ A_i(Y_i-\mu_1(X_i)) - e(X_i)\frac{1-A_i}{1- e(X_i)}(Y_i - \mu_0(X_i)) + (\mu_1(X_i) - \mu_0(X_i)) e(X_i) \Big \} \stackrel{p}{\longrightarrow} \mu_1 - \mu_0$ by law of large numbers when $n \rightarrow \infty$. The last equation because of $\frac{1}{n} \sum\limits_{i=1}^n A_i(Y_i-\mu_1(X_i)) + \mu_1(X_i) e(X_i) \stackrel{p}{\longrightarrow} \mu_1$ by the law of large numbers when $n \rightarrow \infty$, and the term $\frac{1}{n} \sum\limits_{i=1}^n A_i(Y_i-\hat \mu_1(X_i)) + \hat \mu_1(X_i) e(X_i)$ can expand as 
    \begin{align*}
        &\frac{1}{n} \sum\limits_{i=1}^n \{ A_i(Y_i-\hat \mu_1(X_i)) + \hat \mu_1(X_i) e(X_i) \} \\
        ={}& \frac{1}{n} \sum\limits_{i=1}^n \{ A_i(Y_i- \mu_1(X_i)) +  \mu_1(X_i) e(X_i) \} + \frac{1}{n} \sum\limits_{i=1}^n (A_i - e(X_i))(\mu_1(X_i) - \hat \mu_1(X_i)) \\
        ={}& \frac{1}{n} \sum\limits_{i=1}^n \{ A_i(Y_i- \mu_1(X_i)) +  \mu_1(X_i) e(X_i) \} + o_\mathbb{P}(1) \\
        ={}&\mu_1 + o_\mathbb{P}(1)
    \end{align*}
    where the second equation because of Cauchy Schwarz inequality and the condtion. So $\frac{1}{n} \sum\limits_{i=1}^n \{ A_i(Y_i-\hat \mu_1(X_i)) + \hat \mu_1(X_i) e(X_i) \} \stackrel{p}{\longrightarrow} \mu_1$. \\
    $U_{2n}$ can be further decomposed as 
    \begin{equation*}
        U_{2n} = U_{2n} - \mathbb{E}[U_{2n}] + \mathbb{E}[U_{2n}]
    \end{equation*}
    By a Taylor expansion for $\mathbb{E}[U_{2n}]$ in direction  $[ \hat \mu_1 - \mu_1]$ yields that
    \begin{align*}
    \mathbb{E}[H_{2n}] & = \sqrt{n}\frac{\mu_0 - \mu_1}{\mu_1^2}\mathbb{E}[\hat g_2(Z; e, \hat \mu_1)- g_2(Z;e,\mu_1)] \\
    &=\sqrt{n}\frac{\mu_0 - \mu_1}{\mu_1^2}\partial_{[\hat{\mu}_0-\mu_0]}\mathbb{E}[g_2(Z;e,\mu_1)]+\sqrt{n}\frac{\mu_0 - \mu_1}{\mu_1^2}\frac12\partial_{[\hat{\mu}_0-\mu_0]}^2\mathbb{E}[g_2(Z;e,\mu_1)]+\cdots \\
    &= \mathbb{E}[e(X) - A](\hat \mu_1(X) - \mu_1(X)) + 0 + \cdots \\
    &=0
    \end{align*}
    where the last equation follows from $\mathbb{E}[A| X] - e(X) = 0$ . All higher-order terms can be shown to be dominated by the first-order term. Therefore, $\mathbb{E}[U_{2n}] = o_\mathbb{P}(1)$. In addition, we get that $U_{2n} - \mathbb{E}[U_{2n}] = o_\mathbb{P}(1)$ by calculating $Var\{U_{2n} - \mathbb{E}[U_{2n}]\} =  o_\mathbb{P}(1)$. This proves the $u_{2n} = o_\mathbb{P}(1)$. \\
    
    For $U_{3n}$ the analysis is similar to $U_{2n}$,
    \begin{align*}
        U_{3n} &= \sqrt{n}\Big \{ \dfrac{\dfrac{1}{n}\sum\limits_{i=1}^n \Big \{ A_i(Y_i-\hat\mu_1(X_i)) - e(X_i)\dfrac{1-A_i}{1- e(X_i)}(Y_i - \hat \mu_0(X_i)) + (\hat \mu_1(X_i) - \hat \mu_0(X_i)) e(X_i) \Big \}  }{ \dfrac{1}{n} \sum\limits_{i=1}^n A_i(Y_i-\hat\mu_1(X_i)) + \hat\mu_1(X_i) e(X_i) } - \\ 
        & \quad \quad \quad \quad \quad \dfrac{\dfrac{1}{n}\sum\limits_{i=1}^n \Big \{ A_i(Y_i-\mu_1(X_i)) - e(X_i)\dfrac{1-A_i}{1- e(X_i)}(Y_i - \mu_0(X_i)) + (\mu_1(X_i) - \mu_0(X_i)) e(X_i) \Big \}  }{ \dfrac{1}{n} \sum\limits_{i=1}^n A_i(Y_i-\hat\mu_1(X_i)) + \hat\mu_1(X_i) e(X_i) } \Big \} \\
        & = \sqrt{n}\frac{1}{\mu_1}\Big \{ \dfrac{1}{n}\sum\limits_{i=1}^n \big \{ A_i(Y_i-\hat\mu_1(X_i)) - e(X_i)\dfrac{1-A_i}{1- e(X_i)}(Y_i - \hat \mu_0(X_i)) + (\hat \mu_1(X_i) - \hat \mu_0(X_i)) e(X_i) \big \}  - \\ 
        & \quad \quad \quad \quad \quad \dfrac{1}{n}\sum\limits_{i=1}^n \big \{ A_i(Y_i-\mu_1(X_i)) - e(X_i)\dfrac{1-A_i}{1- e(X_i)}(Y_i - \mu_0(X_i)) + (\mu_1(X_i) - \mu_0(X_i)) e(X_i) \big \}   \Big \} \\
    \end{align*}
     $U_{3n}$ can be further decomposed as 
    \begin{equation*}
        U_{3n} = U_{3n} - \mathbb{E}[U_{3n}] + \mathbb{E}[U_{3n}]
    \end{equation*}
    By a Taylor expansion for $\mathbb{E}[U_{3n}]$ in direction  $[ \hat \mu_0 - \mu_0, \hat \mu_1 - \mu_1]$ yields that
    \begin{align*}
    \mathbb{E}[H_{3n}] & = \sqrt{n}\frac{1}{\mu_1}\mathbb{E}[\hat g_3(Z; e,\hat \mu_0, \hat \mu_1)- g_3(Z;e,\mu_0, \mu_1)] \\
    &=\sqrt{n}\frac{1}{\mu_1}\partial_{[\hat{\mu}_0-\mu_0, \hat{\mu}_1-\mu_1]}\mathbb{E}[g_3(Z;e,\mu_0, \mu_1)]+\sqrt{n}\frac{1}{\mu_1}\frac12\partial_{[\hat{\mu}_0-\mu_0, \hat{\mu}_1-\mu_1]}^2\mathbb{E}[g_3(Z;e,\mu_0, \mu_1)]+\cdots \\
    &= \sqrt{n}\frac{1}{\mu_1}\mathbb{E}[e(X) - A](\hat \mu_1(X) - \mu_1(X)) + \sqrt{n}\frac{1}{\mu_1}\mathbb{E}[\frac{e(X)}{1 - e(X)}(e(X) - A)](\hat \mu_0(X) - \mu_0(X)) + 0 + \cdots \\
    &=0
    \end{align*}
    where the last equation follows from $\mathbb{E}[A| X] - e(X) = 0$ . All higher-order terms can be shown to be dominated by the first-order term. Therefore, $\mathbb{E}[U_{3n}] = o_\mathbb{P}(1)$. In addition, we get that $U_{3n} - \mathbb{E}[U_{3n}] = o_\mathbb{P}(1)$ by calculating $Var\{U_{3n} - \mathbb{E}[U_{3n}]\} =  o_\mathbb{P}(1)$. This proves the conclusion of Theorem 3(b).

\section*{Proof of Lemma 2}
Proof of Lemma 2(a). Under Assumptions 1 and 2, $\gamma$ is identified as
\begin{align*}
    \gamma &= \mathbb{P}(Y^1 = 1|A = 0, Y = 0) \\
    &= \frac{\mathbb{P}(Y^0 = 0, Y^1 = 1| A = 0)}{\mathbb{P}(Y^0 = 0| A = 0)} \\
    &= \frac{\sum_x \mathbb{P}(X|A = 0) \cdot \mathbb{P}(Y^0 = 0, Y^1 = 1| A = 0, X)}{\sum_x \mathbb{P}(X|A = 0) \cdot \mathbb{P}(Y^0 = 0| A = 0, X)} \\
    &= \frac{\sum_x (1 - e(X))\mathbb{P}(X) \cdot \mathbb{P}(Y^0 = 0, Y^1 = 1|X)}{\sum_x (1 - e(X))\mathbb{P}(X) \cdot \mathbb{P}(Y^0 = 0|X)} \\
    &=  \frac{\sum_x (1 - e(X))\mathbb{P}(X) \cdot [\mathbb{P}(Y^1 = 1|X) - \mathbb{P}(Y^0 = 1|X)]}{\sum_x (1 - e(X))\mathbb{P}(X) \cdot \mathbb{\mathbb{P}}(Y^0 = 0|X)} \\
    &= \frac{\mathbb{E}[(1 - e(X))(\mu_1(X) - \mu_0(X))]}{\mathbb{E}[(1 - e(X))(1 - \mu_0(X))]} \\
    &= \frac{\bar\mu_1 - \bar\mu_0}{\bar{\bar\mu}_0}.
\end{align*}
Where the fourth equation holds because Assumption 1 and the fifth equation holds because Assumption 2, so $\gamma$ can be identified.

Proof of Lemma 2(b). Under Assumptions 1 and 3, $\gamma$ is identified as
\begin{align*}
    PS &= \mathbb{P}(Y^1 = 1|A = 0, Y = 0) \\
    &= \frac{\mathbb{P}(Y^0 = 0, Y^1 = 1| A = 0)}{\mathbb{P}(Y^0 = 0| A = 0)} \\
    &= \frac{\sum_x \mathbb{P}(X|A = 0) \cdot \mathbb{P}(Y^0 = 0, Y^1 = 1| A = 0, X)}{\sum_x \mathbb{P}(X|A = 0) \cdot \mathbb{P}(Y^0 = 0| A = 0, X)} \\
    &= \frac{\sum_x (1 - e(X))\mathbb{P}(X) \cdot \mathbb{P}(Y^0 = 0, Y^1 = 1|X)}{\sum_x (1 - e(X))\mathbb{P}(X) \cdot \mathbb{P}(Y^0 = 0|X)} \\
    &=  \frac{\sum_x (1 - e(X))\mathbb{P}(X) \cdot [\mathbb{P}(Y^0 = 0|X) \cdot \mathbb{P}(Y^1 = 1|X)]}{\sum_x (1 - e(X))\mathbb{P}(X) \cdot \mathbb{P}(Y^0 = 0|X)} \\
    &= \frac{\mathbb{E}[(1 - e(X))(1 - \mu_0(X))\mu_1(X)]}{\mathbb{E}[(1 - e(X))(1 - \mu_0(X))]} \\
    &= \frac{\bar{\mu}_1 - \bar{\mu}}{\bar{\bar \mu}_0}.
\end{align*}
Where the fourth equation holds because Assumption 1 and the fifth equation holds because Assumption 3, so $\gamma$ can be identified.

\section*{EIFs for $\gamma$ when $e(X)$ is Known}

The efficient influence function for $\gamma$ when the propensity score is known is shown below under two sets of assumptions.
\subsection*{Assumptions 1 and 2}
When the propensity score $e(X)$ is known, under the Assumptions 1 and 2,
the EIF of $\gamma$ is
\begin{align*}
  \tilde \phi_\gamma(Y, A, X; \eta) &= \ \dfrac{1}{\bar{\bar{\mu}}_0}\frac{A}{e(X)}(1 - e(X))(Y-\mu_1(X)) + \dfrac{\bar{\mu}_1 - \bar{\mu}_0 - \bar{\bar{\mu}}_0}{\bar{\bar{\mu}}_0^2}(1-A)(Y - \mu_0(X)) \\
  &+ [\dfrac{1}{\bar{\bar{\mu}}_0}(\mu_1(X) - \mu_0(X)) - \dfrac{\bar{\mu}_1 - \bar{\mu}_0}{\bar{\bar{\mu}}_0^2}(1 - \mu_0(X))](1 - e(X)) \\
  &= \dfrac{1}{\bar{\bar{\mu}}_0}\frac{A}{e(X)}(1 - e(X))(Y-\mu_1(X)) + \dfrac{\gamma - 1}{\bar{\bar{\mu}}_0}(1-A)(Y - \mu_0(X)) \\
  &+ [\dfrac{1}{\bar{\bar{\mu}}_0}(\mu_1(X) - \mu_0(X)) - \dfrac{\gamma}{\bar{\bar{\mu}}_0}(1 - \mu_0(X))](1 - e(X)). 
\end{align*}

The associated semiparametric efficiency bound of $\gamma$ is $\mathbb{V}(\tilde \phi_{\gamma}(Y, A, X; \eta)) = \mathbb{E}[\tilde \phi^2_{\gamma}(Y, A, X; \eta)]$.

\subsection*{Assumptions 1 and 3}
When the propensity score $e(X)$ is known, under the Assumptions 1 and 3,
the EIF of $\gamma$ is
\begin{align*}
  \tilde \varphi_{\gamma}(Y, A, X; \eta) &= \ \frac{1}{\bar{\bar{\mu}}_0}\frac{A}{e(X)}(1 - e(X))(Y-\mu_1(X))(1 - \mu_0(X)) + (1 - A)(\frac{\bar{\mu}_1 - \bar{\mu}}{\bar{\bar{\mu}}_0^2} - \frac{1}{\bar{\bar{\mu}}_0}\mu_1(X))(Y - \mu_0(X)) \\
  &+ (\frac{1}{\bar{\bar{\mu}}_0}\mu_1(X) - \frac{\bar{\mu}_1 - \bar{\mu}}{\bar{\bar{\mu}}_0^2})(1 - \mu_0(X))(1 - e(X)) \\
  &= \frac{1}{\bar{\bar{\mu}}_0}\frac{A}{e(X)}(1 - e(X))(Y-\mu_1(X))(1 - \mu_0(X)) + (1 - A)(\frac{\gamma}{\bar{\bar{\mu}}_0} - \frac{1}{\bar{\bar{\mu}}_0}\mu_1(X))(Y - \mu_0(X)) \\
  &+ (\frac{1}{\bar{\bar{\mu}}_0}\mu_1(X) - \frac{\gamma}{\bar{\bar{\mu}}_0})(1 - \mu_0(X))(1 - e(X)). 
\end{align*}

The associated semiparametric efficiency bound of $\gamma$ is $\mathbb{V}(\tilde \varphi_{\gamma}(Y, A, X; \eta)) = \mathbb{E}[\tilde \varphi^2_{\gamma}(Y, A, X; \eta)]$.

\section*{Estimator of $\gamma$ when $e(X)$ is known}

When the propensity score $e(X)$ is known, under the Assumptions 1 and 2,
the Estimator of $\gamma$ is
\begin{align*}
    \hat \gamma = \frac{ \dfrac{1}{n}\sum\limits_{i=1}^n \Big \{ \dfrac{A_i}{e(X_i)}(1 - e(X_i))(Y_i-\hat \mu_1(X_i)) -(1-A_i)(Y_i - \hat \mu_0(X_i)) + (\hat \mu_1(X_i) - \hat \mu_0(X_i))(1 - e(X_i)) \Big \}}{ \dfrac{1}{n} \sum\limits_{i=1}^n \Big \{ (1 - \hat \mu_0(X_i))(1 - e(X_i)) - (1 - A_i)(Y_i - \hat \mu_0(X_i)) \Big \}}. 
\end{align*}

When the propensity score $e(X)$ is known, under the Assumptions 1 and 3,
the Estimator of $\gamma$ is
\begin{small}
\begin{align*}
   \hat \gamma = \frac{ \dfrac{1}{n}\sum\limits_{i=1}^n \Big \{ \frac{A_i}{e(X_i)}(1 - e(X_i))(Y_i-\hat \mu_1(X_i))( 1 - \hat \mu_0(X_i)) -(1-A_i)(Y_i - \hat \mu_0(X_i))\hat \mu_1(X_i) + \hat \mu_1(X_i)(1 - \hat \mu_0(X_i))(1 - e(X_i)) \Big \}}{ \dfrac{1}{n} \sum\limits_{i=1}^n \Big \{ (1 - \hat \mu_0(X_i))(1 - e(X_i)) - (1 - A_i)(Y_i - \hat \mu_0(X_i)) \Big \}}.
\end{align*}
\end{small}

\section*{Note}
Many Propositions and Theorems in this paper are parallel, and we omit the proofs of Proposition 3, Theorem 4, Theorem 5 and Theorem 6. The proof of Proposition 3 can be referred to Proposition 2. The proofs of Theorem 4 and Theorem 6 can be referred to Theorem 3.  The proof of Theorem 5 can be referred to Theorem 1 and Theorem 2.

\end{document}